\documentclass[11pt, epsfig]{article}
\usepackage{epsfig, amsmath, amssymb, amsthm, times}
\usepackage{graphicx}
\def\theequation{\arabic{section}.\arabic{equation}}
\renewcommand{\theequation}{\thesection.\arabic{equation}}

\newtheorem {thm}{Theorem}[section]
\newtheorem {lem}[thm]{Lemma}

\newtheorem {cor}[thm]{Corollary}

\theoremstyle{defintion}

\theoremstyle{remark}
\newtheorem{rem}[thm]{Remark}

\theoremstyle{example}
\newtheorem{ex}[thm]{Example}

\theoremstyle{assumption}

\def\pf{{\it Proof.\;}}
\def\var{\mathrm{var}~}
\def\cov{\mathrm{cov}~}
\def\E{{\mathbb E~}}
\def\P{{\mathbb P~}}
\def\R{{\mathbb R}}
\def\N{{\mathbb N}}

\def\lbl{\label}
\def\be{\begin{equation}}
\def\ee{\end{equation}}

\def\qed{\square}
\def\tr{\mathrm{Tr}}

\def\t{\mathsf{T}}
\def\tl{\tilde}

\title{Shaping bursting by electrical coupling and noise}
\author{Georgi S. Medvedev and Svitlana Zhuravytska
\thanks{
Department of Mathematics, Drexel University, 3141 Chestnut Street,
Philadelphia, PA 19104, {\tt $\{$medvedev, sz38$\}$ @drexel.edu} 
}
}

\begin{document}
\maketitle
\begin{abstract} 
Gap-junctional coupling is an important way of communication between 
neurons and other excitable cells. Strong electrical coupling synchronizes
activity across cell ensembles. Surprisingly, in the presence of noise
synchronous oscillations generated by an electrically coupled network may
differ qualitatively from the oscillations produced by uncoupled individual 
cells forming the network. A prominent example of such behavior is the synchronized
bursting in islets of Langerhans formed by pancreatic $\beta-$cells, which in 
isolation are known to exhibit irregular spiking \cite{SRK88, SR91}.
At the heart of this intriguing phenomenon lies denoising, a remarkable ability
of electrical coupling to diminish the effects of noise acting on individual
cells. 

In this paper, building on an earlier analysis of denoising in networks
of integrate-and-fire neurons \cite{medvedev09}
and our recent study of spontaneous activity in a closely related model
of the Locus Coeruleus network \cite{MZ}, we derive quantitative estimates
characterizing denoising in electrically coupled networks of conductance-based
models of square wave bursting cells. Our analysis reveals the interplay
of the intrinsic properties of the individual cells and network topology
and their respective contributions to this important effect. In particular, we show that 
networks on graphs with large algebraic connectivity \cite{Fiedler73} or 
small total effective resistance \cite{Bollobas98} are better equipped for 
implementing denoising. As a by-product of the analysis of denoising,
we analytically estimate  the rate with which trajectories converge
to the synchronization subspace and the    
stability of the latter to random perturbations. These estimates reveal
the role of the network topology in synchronization. 
The analysis is complemented by numerical simulations
of electrically coupled conductance-based networks.
Taken together, these results explain the mechanisms underlying synchronization
and denoising in an important class of biological models.
\end{abstract}

\section{Introduction}
\lbl{intro}
\setcounter{equation}{0}
Cells in the nervous system are organized in complex interconnected
networks, which feature a rich variety of electrical activity. 
There is abundant experimental evidence linking spatio-temporal features
of the firing patterns generated by neuronal networks to various 
physiological and cognitive processes.
Therefore, elucidating dynamical principles underlying 
pattern-formation in neuronal networks is an important problem of
mathematical physiology.

Square wave bursting, one of the most common firing patterns, is characterized
by alternating periods of fast oscillations of the membrane potential
and quiescence \cite{RIN87, IZH00}. Typically, cells that generate 
bursting, under different conditions exhibit other firing patterns
such as periodic or aperiodic spiking or reside in the excitable regime
\cite{chay85, M05, M06, WR95}. Transitions between different dynamical regimes
in excitable cells often signal important physiological or cognitive 
events such as changes in the rate of hormone secretion or
neurotransmitter release as in the cases of pancreatic $\beta-$cells
\cite{SRK88} and midbrain dopamine neurons \cite{gra4}; or changes in
respiratory rhythm or attentional state as in the cases of 
neurons in the Pre-Botzinger complex \cite{BRS99} or Locus Coeruleus 
\cite{UCS99} respectively. Not surprisingly, mathematical models
have been used extensively to explain the origins of different firing patterns. 
For single cell models, mechanisms underlying various modes of
electrical activity have been thoroughly studied 
\cite{LT, IZH00, M05, RIN87, RE89, TR92}.
Some of the techniques developed for single cell models extend 
to cover small networks \cite{MMR10, RT00}. However, mathematical analysis of 
large conductance-based networks without special 
assumptions on network topology is an outstanding problem.
Furthermore, there is a growing body of experimental and
theoretical studies indicating the  importance of noise in shaping
neuronal dynamics
\cite{CCI, DVS00, GP06, HM, KB09, LR10, Lon97, MS95, MZ, PS07, SRT, wrk}.
In the presence of noise in the network dynamics, its analysis
becomes even a more challenging problem.
The goal of this paper is to elucidate  principal factors shaping
synchronous activity in large electrically 
coupled networks of bursting capable cells forced by small white noise. 

Many ingredients contribute to the output of neuronal networks.
Among them, intrinsic properties of the individual cells (local
dynamics) and the type and the structure of connections between cells 
(network topology) are probably the most important ones.
Direct electrical coupling through gap-junctions is a common way 
of communication between neurons as well as between cells of the
heart, pancreas, and other physiological systems \cite{CL04}.
The role of  electrical coupling in shaping firing patterns
generated by neuronal networks has been studied using many different
techniques: the theory for weakly connected networks 
\cite{HI97, KE88, PMG}, Poincare maps \cite{GH07, CK, LR03, MC}, and
Lyapunov functions \cite{MK}, to name a few.  In the present study, 
we consider  a relatively less studied case of strong electrical
coupling \cite{COO08, MK}, for which we develop two complementary
approaches based on center-manifold reduction \cite{Kuz98} and fast-slow 
decomposition \cite{BG, MKKR}.
Importantly, our method covers networks with arbitrary topology,
which allows us to study a large class of models
and to reveal the role of the network topology in shaping network
dynamics.

Under fairly general conditions, strong electrical coupling synchronizes 
activity across the network \cite{medvedev10a, medvedev10}.
Therefore, one might expect that dynamics of electrically coupled networks
of bursting cells will closely  resemble that of a single cell provided
the coupling is strong enough.
This is true in general for deterministic models. However, in the presence of 
noise network dynamics, while still synchronous, can be qualitatively
different from that of a single cell uncoupled from its neighbors.
For instance, single cell models, which  exhibit irregular spiking in isolation 
(Fig. \ref{f.intro}a) can generate very regular synchronous bursting when they are 
coupled electrically (Fig. \ref{f.intro}b). Likewise, a coupled network exhibiting
synchronous spiking for extremely long period of time (Fig. \ref{f.intro}c)
may be formed from bursting cells (Fig. \ref{f.intro}d). 
The first scenario illustrated by Fig.~\ref{f.intro}a,b was 
proposed in \cite{SRK88, SR91} to
explain why  pancreatic $\beta-$cells burst within
electrically coupled islets of Langerhans, but in isolation
exhibit irregular spiking. Numerical experiments and formal analysis
in \cite{SRK88, SR91} show that noise shaping the dynamics of individual
$\beta-$cells becomes less effective when these cells are coupled electrically.
It is certainly intuitive (albeit not obvious mathematically) that 
in a coupled network
the effects of uncorrelated stochastic processes acting on individual cells  
may become weaker due to averaging. However, a remarkable property of electrical
coupling, which as the analysis below shows should not be taken for granted, is
that the variability of the coupled system can be fully controlled by varying 
two parameters: the coupling strength and the size of the network. We identify 
analytical conditions, which guarantee that the variability of the coupled
systems (the meaning of variability will be explained below) goes to zero
as the network size and the coupling strength tend to infinity.

\begin{figure}
\begin{center}
{\bf a}\epsfig{figure=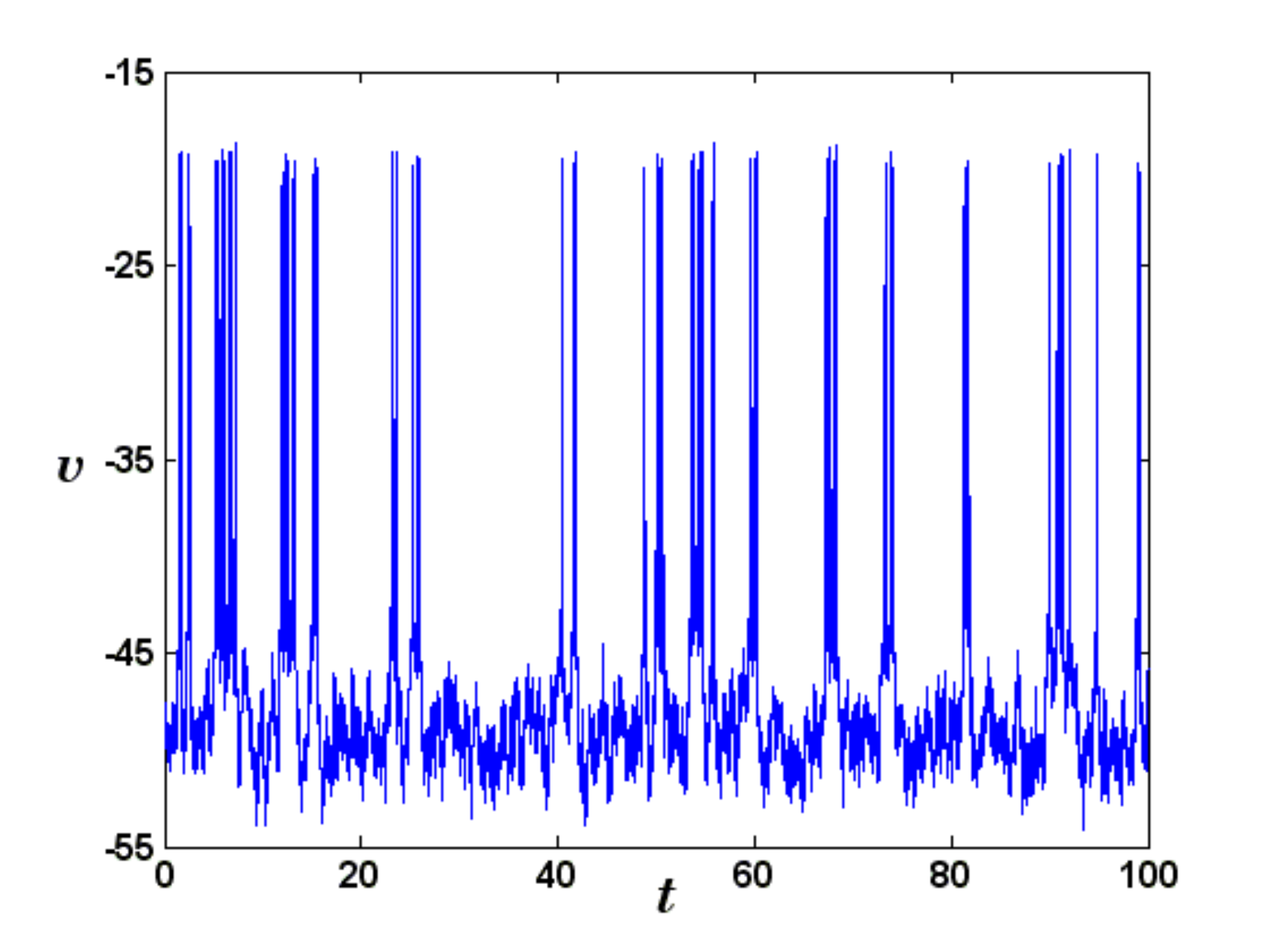, height=1.6in, width=2.05in}
{\bf b}\epsfig{figure=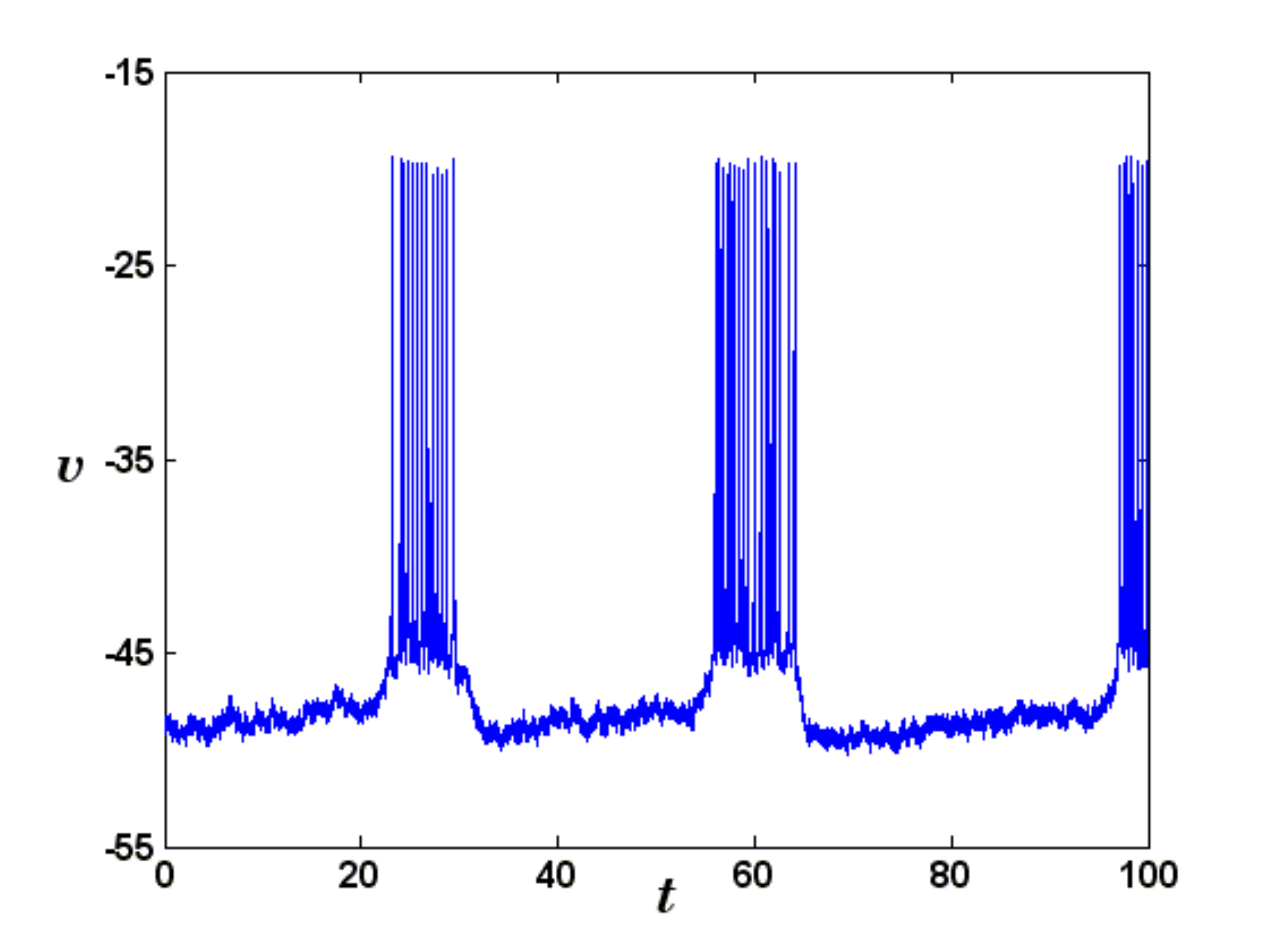, height=1.6in, width=2.05in}\\
{\bf c}\epsfig{figure=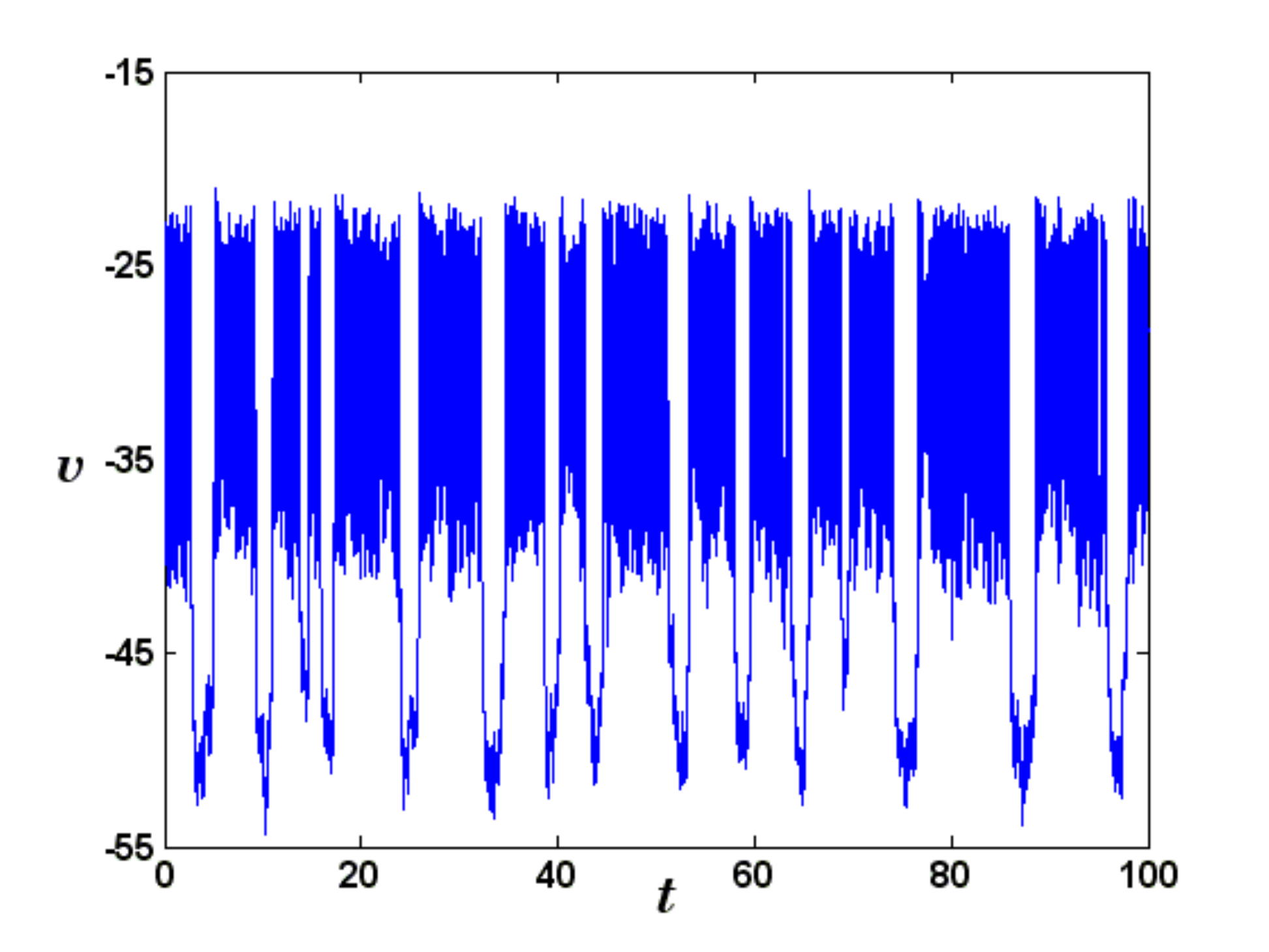, height=1.6in, width=2.05in}
{\bf d}\epsfig{figure=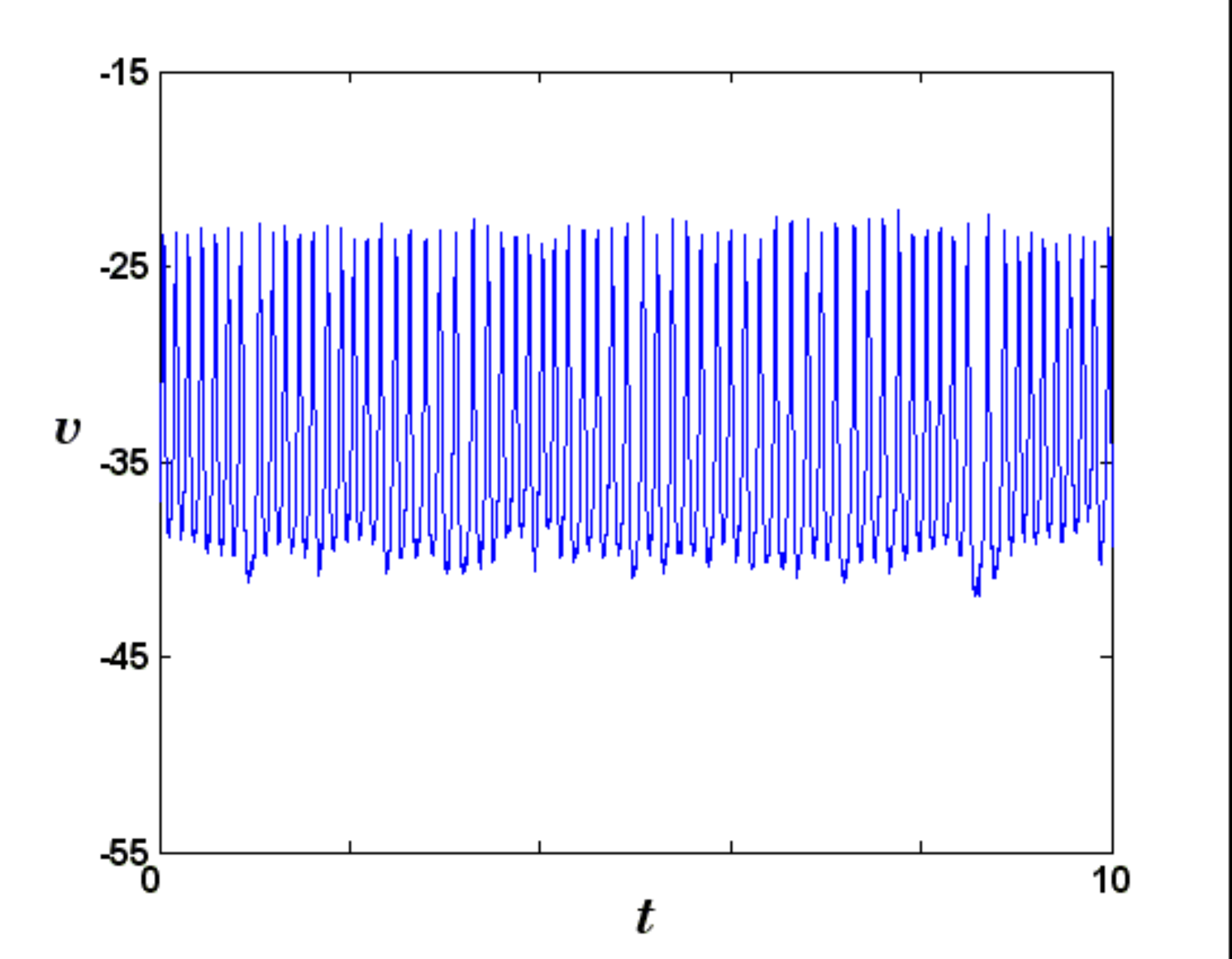, height=1.6in, width=2.05in}
\end{center}
\caption{ Scenario A: Irregular spiking generated by the single models 
(\ref{ch.1})-(\ref{ch.3}) shown in 
(a) is transformed into bursting  as shown in (b) when the cells are coupled 
electrically. A complementary Scenario B is illustrated in (b) and (c). 
The single cell model generates irregular bursting (c).
After the coupling is turned on the pattern of firing is effectively switched to 
spiking (d).
}
\lbl{f.intro}
\end{figure}

Besides bursting in electrically coupled islets of Langerhans, other phenomena
where denoising plays a central role, include episodes of phasic firing 
in the Locus Coeruleus network \cite{UCS99} and enhanced reliability of
neural responses in gap-junctionally coupled networks \cite{medvedev09, TSP10}.
In this paper, building on an earlier analysis of denoising in networks
of integrate-and-fire neurons \cite{medvedev09}
and our recent study of spontaneous activity in a closely related model
of the Locus Coeruleus network \cite{MZ}, we derive quantitative estimates
characterizing denoising in electrically coupled networks of conductance-based
models. We find that the results obtained for integrate-and-fire
models for individual cells do not extend automatically to 
conductance-based models with higher-dimensional state phase. We identify additional 
features of the local dynamics and coupling architecture
that are needed to guarantee denoising. In particular, our analysis highlights
the role of the bifurcation structure of the bursting cell model for denoising.
It also elucidates the contribution of the network topology to 
this important effect. We show that networks on the graphs with large algebraic
connectivity \cite{Fiedler73} or small total effective resistance \cite{Bollobas98}
are better equipped for implementing denoising. 
As a by-product of the analysis of denoising,
we analytically estimate  the rate with which trajectories converge
to the synchronization subspace and the    
stability of the latter to random perturbations.
Taken together, these results explain the mechanisms underlying synchronization
and denoising in an important class of biological models.

The organization of the paper is as follows. In the next section, we formulate
our assumptions on the single cell model (\S\ref{single-cell}) and explain how
the network is formed (\S\ref{coupled-network}). In \S\S\ref{graph},\ref{examples}, 
we collect
necessary information from the algebraic graph theory \cite{Biggs}, which will be used 
for describing the role of the network topology in dynamical phenomena analyzed in this paper. 
In Section~\ref{transitions}, we introduce two scenarios (A and B) leading to distinct 
firing patterns produced by the single cell models and by synchronized networks of these models.
In the remainder of this section, we analyze the first of these scenarios illustrated
in Fig.~\ref{f.intro} (a,b). 
First, in Lemma~\ref{easy}, we show that in the single cell model, bursting can be destroyed 
with small noise (Figure \ref{f.intro}a). This counter-intuitive result
relies on the presence of a slow timescale in the dynamics of bursting. 
Even very small noise can have significant qualitative effects on the
dynamics of a slow-fast system, because during the (long) periods of slow evolution
there is sufficient time for large deviations (which are extremely unlikely
on time intervals of order $O(1)$) to develop. The proof of the lemma uses large 
deviation estimates 
and is adapted from \cite{FR01}. Thus, given a deterministic model of a bursting cell,
one can switch its dynamics into irregular spiking by adding noise. We then show that 
when many such cells are coupled together, the effects of noise weaken and 
the bursting of the underlying deterministic model reemerges (Figure \ref{f.intro}b). 

The analysis of the coupled system proceeds in two steps. First, we use the
center-manifold reduction \cite{Kuz98}  to approximate  the coupled system near an
excitable equilibrium by a simpler lower-dimensional system of ordinary differential
equations. Second, thanks to the gradient
structure of the reduced problem, we can accurately estimate expected time that a trajectory
of the fast subsystem of the coupled system spends near the excitable equilibrium. We show that this
time is much longer for the network model than for the single cell one. 
This analysis (based on large deviation estimates \cite{FW}) yields one way of quantitative 
description of denoising. 

In Section~\ref{synchronization}, we present a complementary method based on a 
slow-fast decomposition. We show that when the coupling is strong, network dynamics near 
the excitable equilibrium splits into two modes: fast synchronization and ultra-slow 
noise-driven escape from the basin    
of attraction of the equilibrium along  a low-dimensional synchronization subspace.
The results of this section yield valuable insights into synchronization properties 
of the coupled system. In particular, we estimate the rate of convergence of trajectories to the 
synchronization subspace and the stability of the latter against random perturbations.
The estimates show explicitly the contribution of the structural properties of the network
to stability properties of the synchronization subspace.

In Section~\ref{denoising}, we take a look at denoising from a slightly different
angle. Specifically, we study the linearization of the coupled system near an excitable 
equilibrium directly without invoking center manifold reduction. The analysis shows that 
when the dynamics of the individual cells lives in multidimensional phase space 
(unlike integrate-and-fire models or those of excitable cells near a saddle-node bifurcation), 
unless the coupling is full rank (see \cite{medvedev10a} for the definition
of full versus partial rank coupling) denoising should not be expected.
These results show that the proximity to a saddle-node bifurcation in the fast subsystem 
of a square wave bursting neuron model, which makes the dynamics near the excitable
equilibrium essentially one-dimensional, is critical for observing 
distinct dynamics generated by the single cell and network models.
Therefore, the bifurcation structure of the bursting cell models
like those used in \cite{SRK88, SR91} and in this paper is  important 
for the interplay between electrical coupling and noise in shaping
the dynamics of the coupled network.
Finally, the results of this study are summarized in Section~\ref{discuss}.

\section{The model}
\lbl{the-model}
In this section, we introduce a model that will be studied
in the remainder of this paper. First, we formulate the 
assumptions on the single cell model and describe its main dynamical regimes. 
Next, we introduce the 
coupled system. At the end of this section, we review some notions 
and facts from the algebraic graph theory \cite{Biggs}, which will be 
useful for characterizing the contribution of the network 
topology to dynamics.   
\begin{figure}
\begin{center}
{\bf a}\epsfig{figure=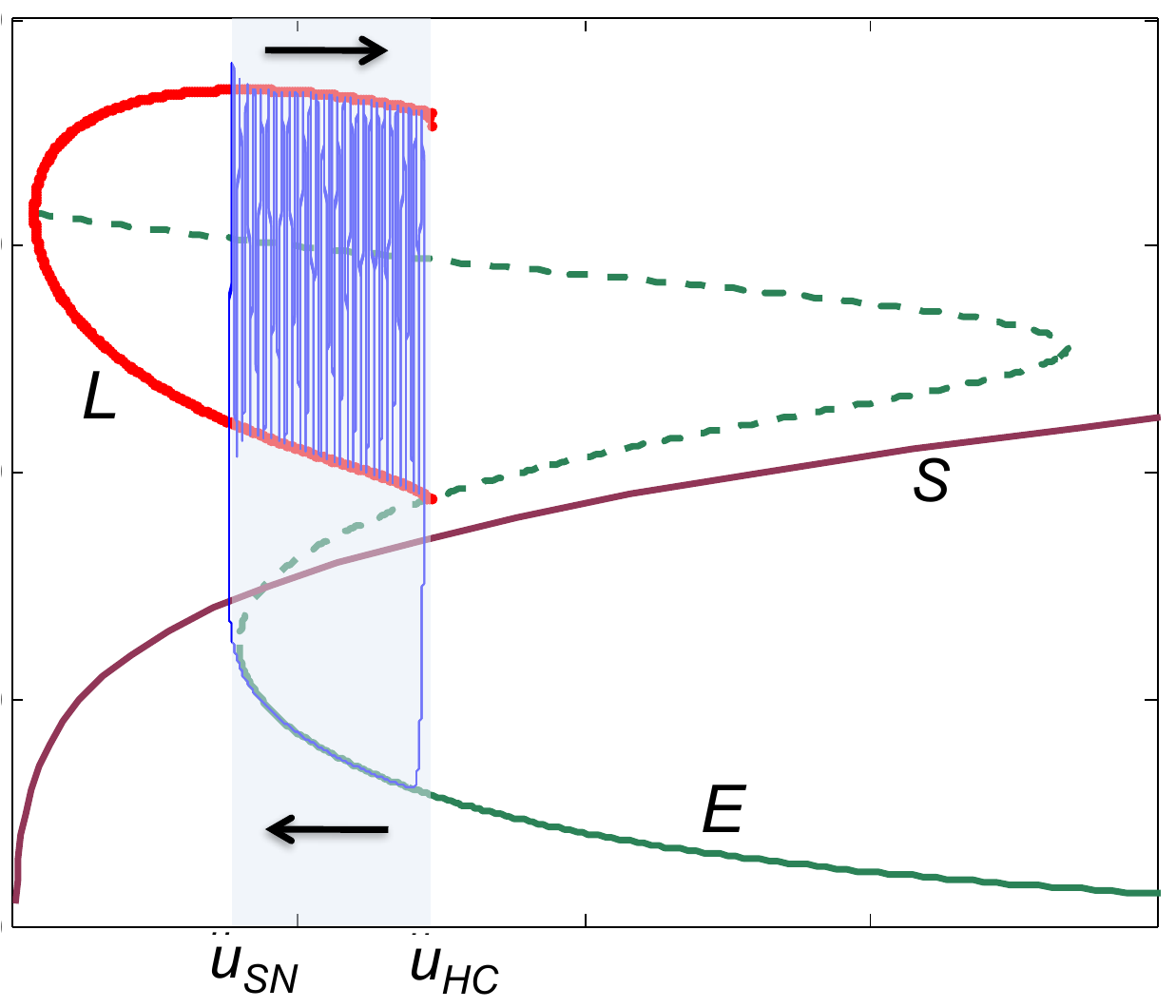, height=1.6in, width=2in}
{\bf b}\epsfig{figure=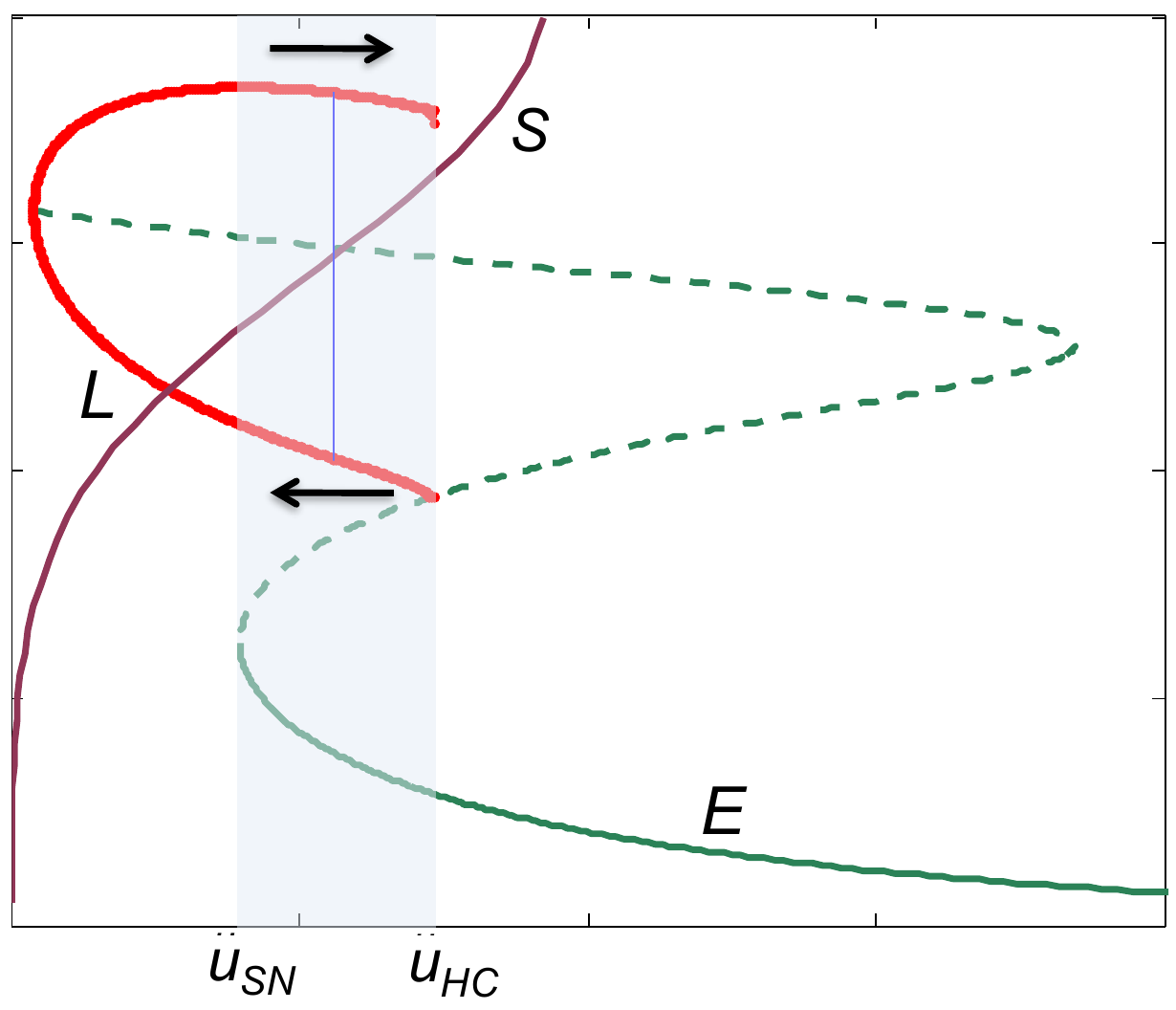, height=1.6in, width=2in}
{\bf c}\epsfig{figure=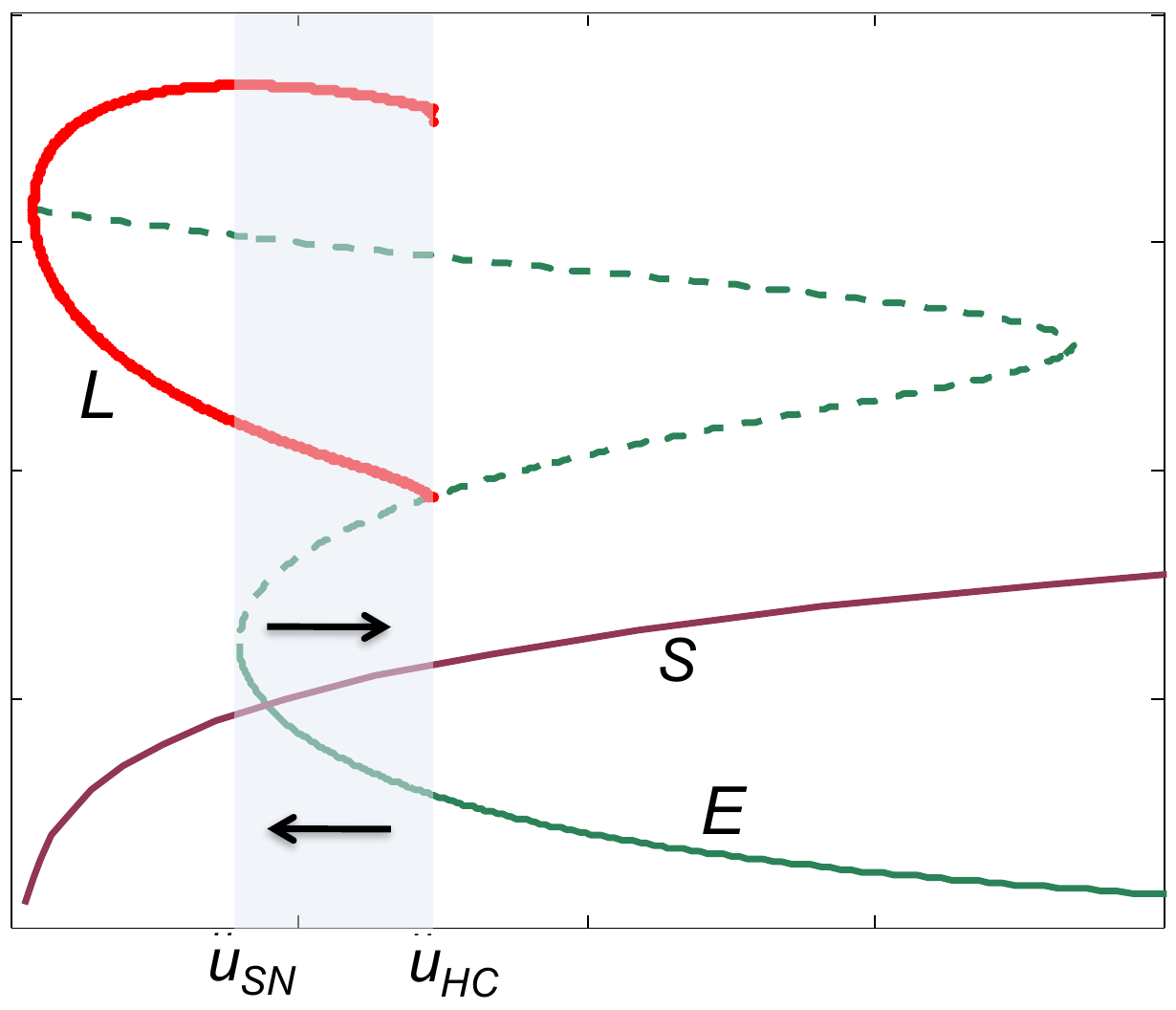, height=1.6in, width=2in}\\
{\bf d}\epsfig{figure=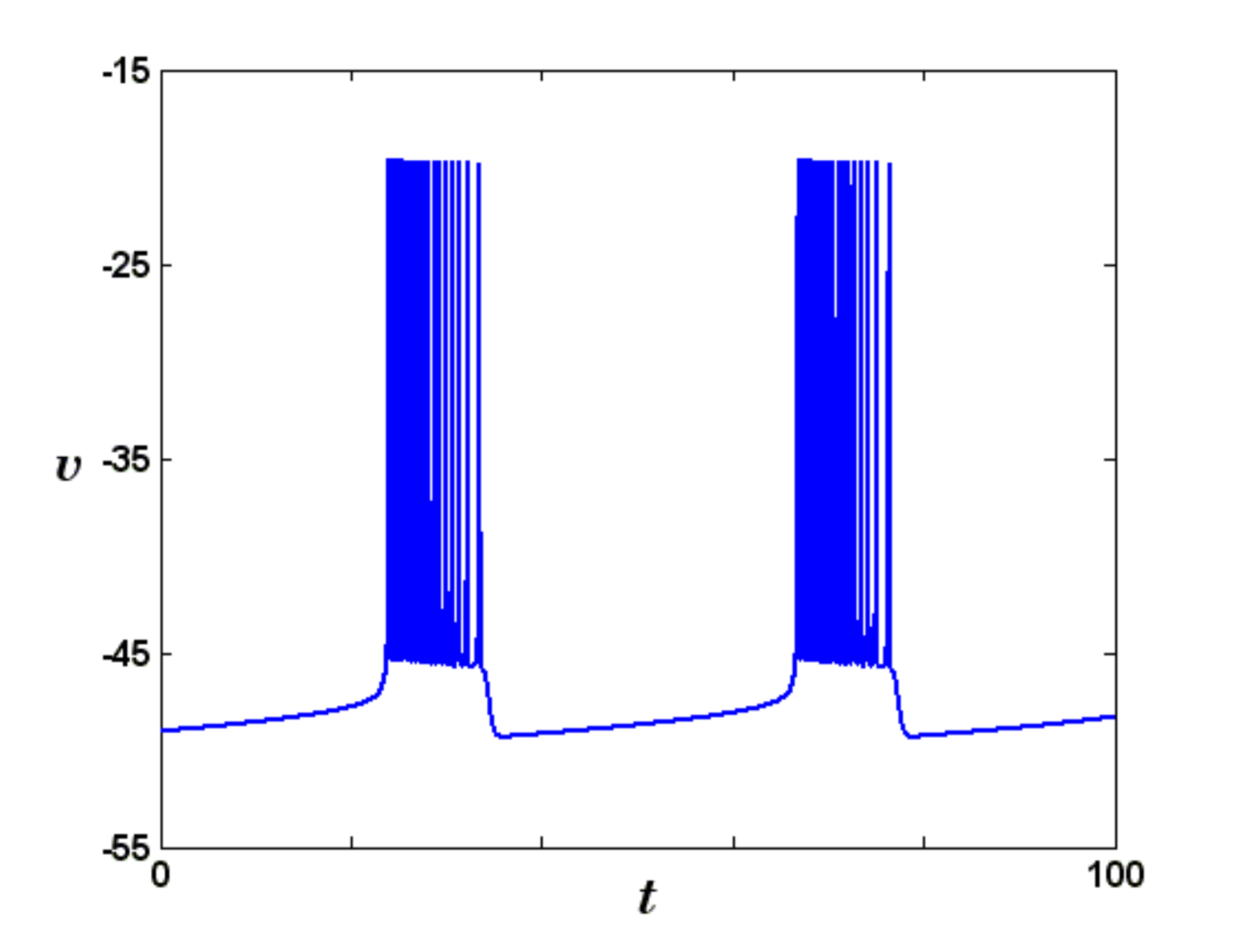, height=1.6in, width=2in}
{\bf e}\epsfig{figure=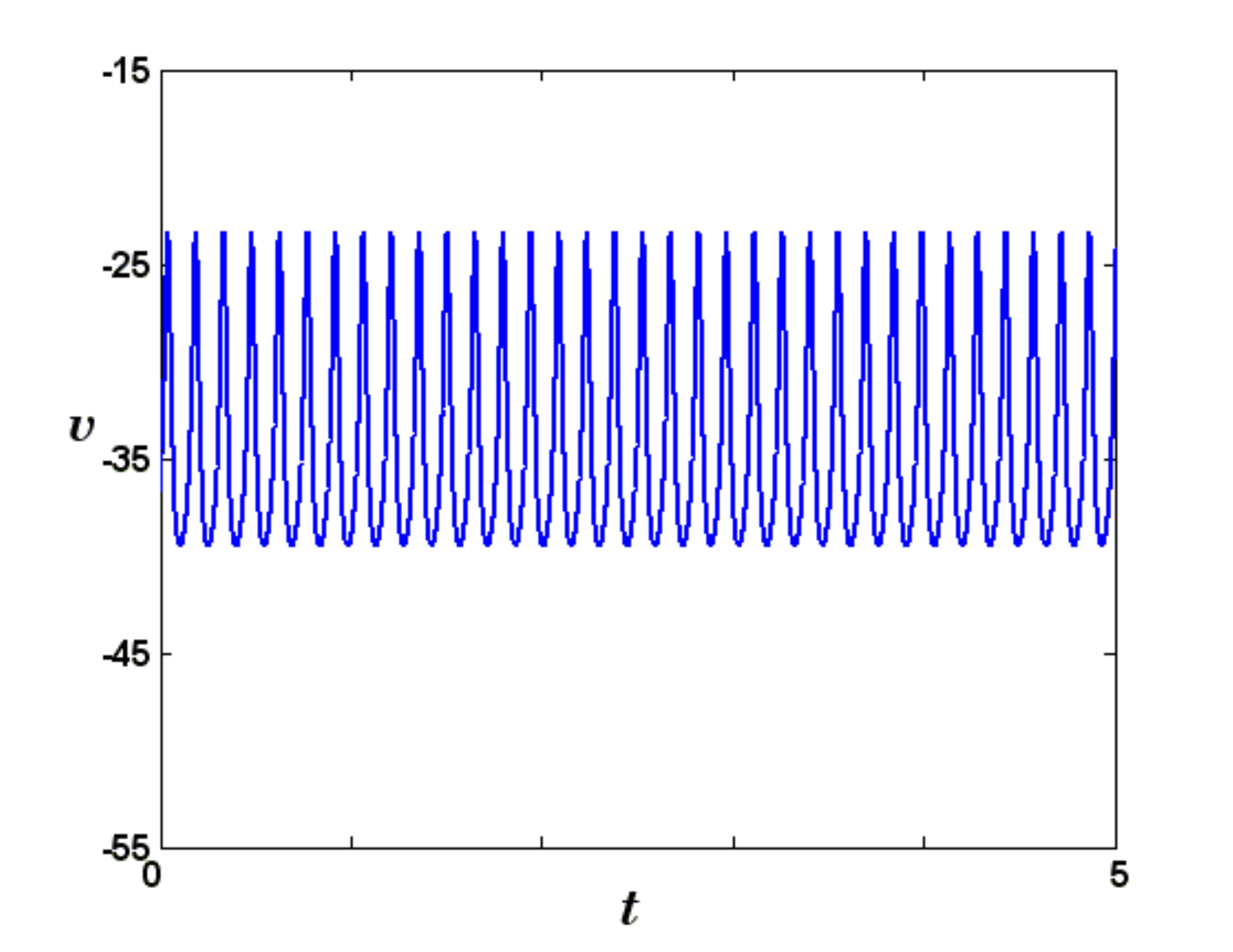, height=1.6in, width=2in}
\end{center}
\caption{
The bifurcation diagram of the fast subsystem (\ref{2.3}).
Depending on the location of the null surface $S$, the 
full deterministic system (\ref{c.1}$)_0$ and (\ref{c.2})
can be in one of the three regimes:
bursting (a), spiking (b), or excitable (quiescence) (c).
The timeseries in (d) and (e) illustrate bursting and
spiking respectively (see \S\ref{single-cell} and Appendix B 
for details).
}
\label{f.regimes}
\end{figure}

\subsection{The single cell model}\lbl{single-cell}
In numerical experiments throughout this paper, we are
going to use a conductance based model of a pancreatic $\beta-$cell
due to Chay \cite{chay85}. The analysis in the following sections does not
depend on any specific details of this model and more general
assumptions will be formulated below. However, 
we believe that it is instructive 
to start from a concrete model to make the biological 
interpretation of the analysis that follows transparent.
With this in mind, following \cite{chay85}, we introduce the
a system of differential equations modeling electrochemical
dynamics in the $\beta-$cells:
\begin{eqnarray}\lbl{ch.1}
C_m\dot v &=& -I_{ion}(v, n, u)+\sigma_1 \dot w_1,\\
\lbl{ch.2}
\dot n &=& {n_\infty(v)-n\over \tau(v)}+\sigma_2 \dot w_2,\\
\lbl{ch.3}
\dot u &=& \epsilon (I_{Ca}(v)-ku).
\end{eqnarray}
Here $v, n,$ and $u$ stand for the cell membrane potential,
gating variable, and the concentration of calcium respectively.
$I_{ion}(v,n,u)$, $I_{Ca}(v)$, $n_\infty(v)$ and $\tau(v)$ denote  
nonlinear functions, which are used for modeling ionic currents. 
$C_m$ denotes   membrane capacitance.
The small parameter $\epsilon>0$ multiplying the right hand side
of the third equation reflects the separation of the
timescales of the calcium dynamics and the fast variables $v$ and $n$.
The right hand sides of the first two equations also contain
independent copies of Gaussian white noise $\dot w_{1,2}$, which
account for the deviations from  the deterministic dynamics due
to various fluctuations \cite{smith, wrk}. For further details of 
(\ref{ch.1})-(\ref{ch.2}) including the values
of parameters, we refer the reader to Appendix A and \cite{chay85}. 

To describe the structure of (\ref{ch.1})-(\ref{ch.3}), it is convenient 
to rewrite it in a more general form
\begin{eqnarray}\lbl{2.1}
\dot x &=& f(x,y)+\Sigma \dot w,\\
\lbl{2.2}
\dot y &=& \epsilon g(x,y),
\end{eqnarray}
where $x=(x_1,x_2)^\t:=(v,n)^\t$, $y:=u$, and 
$\Sigma=\mbox{diag}(\sigma_1,\sigma_2)$. 
Let us first consider the deterministic model (\ref{2.1}$)_0$ and (\ref{2.2}), 
where the zero subscript indicates that the stochastic perturbation is set to zero,
$\Sigma=0$.  The {\it fast} subsystem associated with (\ref{2.1}$)_0$ 
and (\ref{2.2}) 
is obtained by setting $\epsilon=0$ in (\ref{2.2}) and treating $y$ as a 
parameter:
\be\lbl{2.3}
\dot x = f(x,y).
\ee

Under the variation of $y$, the fast subsystem has the bifurcation structure
as shown schematically in Fig. \ref{f.regimes}a. Specifically,
\begin{description}
\item[(PO)]
There exists $y_{hc}\in\R$ such that for each $y<y_{hc}$, 
Equation (\ref{2.3}) has an 
exponentially stable limit cycle of period $T(y)$:
\be\lbl{2.4}
L_y=\{x=\phi(s,y): \;0\le s< T(y)\}.
\ee
The family of the limit cycles, $L=\bigcup_{y<y_{hc}}L_y$, forms a cylinder 
in $\R^3$, which terminates at a homoclinic loop at $y=y_{hc}$ \cite{Kuz98} (Fig. \ref{f.regimes}a).  

\item[(EQ)]
There is a branch of asymptotically stable equilibria of (\ref{2.3}),
$E=\bigcup_{y>y_{sn}} E_y, E_y=\{x=\psi(y)\}$, which terminates at
a saddle-node bifurcation at $y=y_{sn}< y_{hc}$ (Figure \ref{f.regimes}a).

\item[(LS)] For each $y\in\R$, the $\omega-$limit set of almost 
all trajectories of (\ref{2.3})
belongs to $L_y\bigcup E_y$.
\end{description}

Deterministic models of bursting are well understood (see, e.g.,\cite{IZH00, LT, M05, RIN87}).
For small $\epsilon>0$, (\ref{2.1}$)_0$ and (\ref{2.2}) features three main regimes:
bursting, spiking, and quiescence (or excitable). In the former, the trajectory alternates between 
drifting along the cylinder $L$ foliated by periodic orbits of the fast subsystem 
and the curve of equilibria (see Fig.~\ref{f.regimes}a,d). Alternatively,
(\ref{2.1}$)_0$ and (\ref{2.2}) may have a stable limit cycle near $L$ (spiking)
or a stable fixed point near $E$ (excitable). The latter two regimes are
illustrated in Fig.~\ref{f.regimes} (b,e) and (c) respectively. 
The analytical conditions for bursting, spiking, and excitable regimes
are given in Appendix B.

\subsection{The electrically coupled network} \lbl{coupled-network}
Next, we consider a gap-junctionally coupled ensemble of $n$ cells,
whose dynamics is generated by (\ref{ch.1})-(\ref{ch.3}).
In the coupled network, Cell~$i$, $i\in [n]:=\{1,2,\dots,n\}$ receives current
\be\lbl{3.1}
I_c^{(i)}= g\sum_{j=1}^N a_{ij}( v^{(j)}-v^{(i)}),
\ee
from other cells in the network.
Conductance $a_{ij}>0$ if Cell~$i$ and Cell~$j$ are connected and
$a_{ij}=0$, otherwise. Without loss of generality, we set
$a_{ii}=0$, $i\in [n]$, and denote $A=(a_{ij})$. Nondimensional parameter
$g>0$ is used to control the coupling strength.

By including the coupling current (\ref{3.1}) into the models
of individual cells (\ref{ch.1})-(\ref{ch.3}), 
we obtain a differential equation model of the electrically
coupled network
\begin{eqnarray}\lbl{net.1}
C_m\dot v^{(i)} &=& -I_{ion} (v^{(i)}, n^{(i)}, y^{(i)})+ 
g\sum_{j\ne i}  a_{ij} (v^{(j)}-v^{(i)}) 
+\sigma_1 \dot w^{(i,1)},\\
\lbl{net.2}
\dot n^{(i)} &=& {n_\infty(v^{(i)})-n^{(i)}\over \tau(v^{(i)})}+\sigma_2 \dot w^{(i,2)},\\
\lbl{net.3}
\dot y^{(i)} &=& \epsilon (I_{Ca}(v^{(i)})-ky^{(i)}), \quad i=1,2,\dots,n, 
\end{eqnarray}
where $W^{(i)}=(w^{(i,1)}, w^{(i,2)})^\t$ are independent copies of $2D$ Brownian motion.
Using the notation, which we adopted for the single cell model in (\ref{2.1}) and
(\ref{2.2}), we rewrite the coupled system in the following more general form:
\begin{eqnarray}\lbl{c.1}
\dot X &=& F(X,Y)-g(L\otimes J_1)X+(I_n\otimes \Sigma)\dot W,\\
\lbl{c.2}
\dot Y &=& G(X,Y),
\end{eqnarray}
where
\be\lbl{L-and-J}
(L)_{ij}=\left\{\begin{array}{ll}
-a_{ij},& \;i\neq j \\
\sum_{j=1}^n a_{ij}, & \; i=j,
\end{array}
\right.
\quad\quad J_1=\begin{pmatrix} 1 & 0 \\ 0 & 0 \end{pmatrix},
\ee
$$
X=(x^{(1)}, x^{(2)},\dots, x^{(n)})^\t, Y=(y^{(1)}, y^{(2)}, \dots, y^{(n)})^\t,
W=(W^{(1)}, W^{(2)},\dots, W^{(n)})^\t,
$$
$$
F(X,Y)=\left(f(x^{(1)},y^{(1)}),f(x^{(2)},y^{(2)}),\dots,f(x^{(n)},y^{(n)})\right),
$$
$$
G(X,Y)=\left(g(x^{(1)},y^{(1)}),g(x^{(2)},y^{(2)}),\dots,g(x^{(n)},y^{(n)})\right),
$$
and $\otimes$ stands for the Kronecker product.

\subsection{The graph of the network}
\lbl{graph}
Dynamics of the coupled system depends on the spectrum
of matrix $L$ appearing in the coupling operator in (\ref{c.1}).
The eigenvalues of $L$ in turn depend on the structure of the graph
of the network. Throughout this paper, we will repeatedly use the relation between 
the spectrum of $L$ and
the structural properties of the network to study how the network topology
affects its dynamics.  To this end, we will need certain constructions
and results from the algebraic graph theory, which we review below following
\cite{Biggs}.

Let $\mathcal{G}=(V(\mathcal{G}),E(\mathcal{G}))$ denote the  graph of 
interactions between the cells 
in the network. Here, $V(\mathcal{G})=\{v_1, v_2, \dots, v_n \}$ and 
$E(\mathcal{G})=\{e_1, e_2, \dots,e_m\}$
denote the sets of vertices (cells) and  edges 
(pairs of connected cells), respectively.
Throughout this paper, we assume that $\mathcal{G}$ is an undirected connected graph.

It is instructive to consider first the case when all nonzero conductances are 
equal to $1$:
\be\lbl{isotropic}
a_{ij}=\left\{ \begin{array}{ll}
1 & \;\mbox{Cell}~i\; \mbox{and Cell}~j\; \mbox{are connected},\\
0 & \; \mbox{otherwise}.
\end{array}
\right.
\ee
As before, we set the diagonal elements of $A$ to zero, $a_{ii}=0$.  Matrix $A=(a_{ij})$ in (\ref{isotropic})
is called the adjacency matrix of $\mathcal{G}$ and
\be\lbl{def-Lap}
L=\mathrm{diag}(\mathrm{deg}(v_1),\mathrm{deg}(v_2),\dots,\mathrm{deg}(v_n))-A, 
\ee
is called the Laplacian of $\mathcal{G}$. By $\mathrm{deg}(v_i)$ we denote
the degree of $v_i$, i.e., the number of  edges incident to $v_i$.
Alternatively, the graph Laplacian can be defined by
\be\lbl{Lap-1}
L=H^\t H,
\ee
where $H\in\R^{m\times n}$ is the coboundary matrix of $\mathcal{G}$ \cite{Biggs}. 
The  definition  of $H$ uses an orientation of the edges of $\mathcal{G}$.
For each edge $e_j=(v_{j_1},v_{j_2})\in V(\mathcal{G})\times V(\mathcal{G})$, 
we specify the positive and  negative ends; e.g., let $v_{j_1}$ be the negative
end of $e_{j_2}$. Then the  coboundary 
matrix of $G$ is defined as follows (cf. \cite{Biggs}) 
\be\lbl{incidence}
H=(h_{ij})\in \R^{m\times n},\quad
h_{ij}=\left\{ \begin{array}{cl}
1, & v_j\;\mbox{ is a positive end of}\; e_i,\\
-1, & v_j\;\mbox{ is a negative end of}\; e_i,\\
0, &\;\mbox{otherwise}.
\end{array}
\right.
\ee
Definitions (\ref{def-Lap}) and (\ref{Lap-1}) are equivalent
(cf. \cite{Biggs}). From either of these definitions, it is easy
to see that $\lambda_1(L)=0$ is an eigenvalue of $L$. If $\mathcal{G}$ is 
a connected graph then the zero eigenvalue is simple and all
other eigenvalues are positive (cf.~\cite{Fiedler73})
\be\lbl{EVs}
0=\lambda_1(L) < \lambda_2(L)\le \dots\le \lambda_n(L).
\ee
\begin{rem}\lbl{convention}
Following a common in the algebraic graph theory convention, 
we will refer to the eigenvalues
of $L$ as the eigenvalues of $\mathcal{G}$.
\end{rem}

The second eigenvalue $\mathfrak{a}(\mathcal{G})=\lambda_2(L)$ is called the algebraic 
connectivity of $\mathcal{G}$, because it yields a lower bound for the edge and the vertex connectivity of 
$\mathcal{G}$ \cite{Fiedler73}. 
The algebraic connectivity is important  for a variety
of combinatorial, probabilistic, and dynamical  aspects of the network analysis.
In particular, it is used in the studies of the graph expansion \cite{Hoory06}, 
random walks \cite{Bollobas98}, and synchronization of dynamical networks
\cite{Jost07, medvedev10b, MZ}. Below, we show that $\mathfrak{a}(\mathcal{G})$ determines
the rate of convergence to synchrony in the coupled model (\ref{c.1}) and (\ref{c.2}).  

Another spectral function of $\mathcal{G}$, which will be  useful in the analysis of the 
coupled system is the total effective resistance of $\mathcal{G}$ (cf. \cite{klein93, Gutman03})
\be\lbl{resistance}
\mathcal{R}(\mathcal{G})=n\sum_{j=2}^n \lambda_j^{-1}(L).
\ee
For electrical and graph-theoretic interpretations of $\mathcal{R}(\mathcal{G})$ 
as well as for numerous applications, we refer the reader to 
\cite{Bollobas98, Boyd08}.

All definitions and constructions, which we reviewed above, naturally extend to cover
nonhomogeneous networks with different conductances $a_{ij}$. 
To this end, we define conductance matrix
\be\lbl{conductance}
C=\mbox{diag}(c_1, c_2, \dots c_m),
\ee
where $c_i\ge 0$ is the conductance of edge $e_i, i\in [m].$
The graph Laplacian of the weighted graph $\mathcal{G}=(V,E,C)$
is defined by
\be\lbl{weightedLap}
L(\mathcal{G})=H^\t CH.
\ee
The algebraic connectivity and the total effective resistance of $\mathcal{G}$ are defined
through the EVs of $L(\mathcal{G})$ as before.

\begin{figure}
\begin{center}
{\bf a}\includegraphics[height=2in,width=2in]{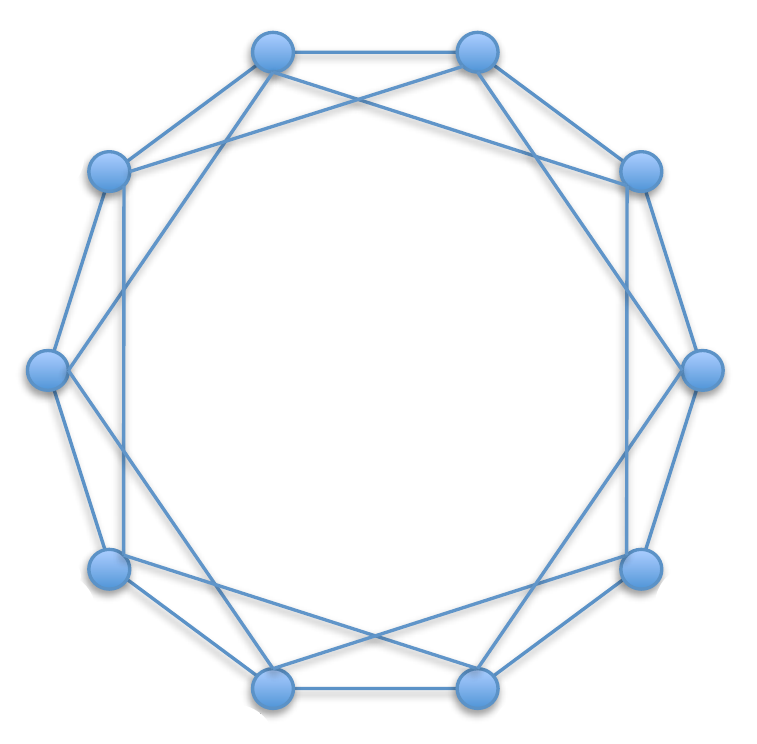} \hspace{0.8in}
{\bf b}\includegraphics[height=2in,width=2in]{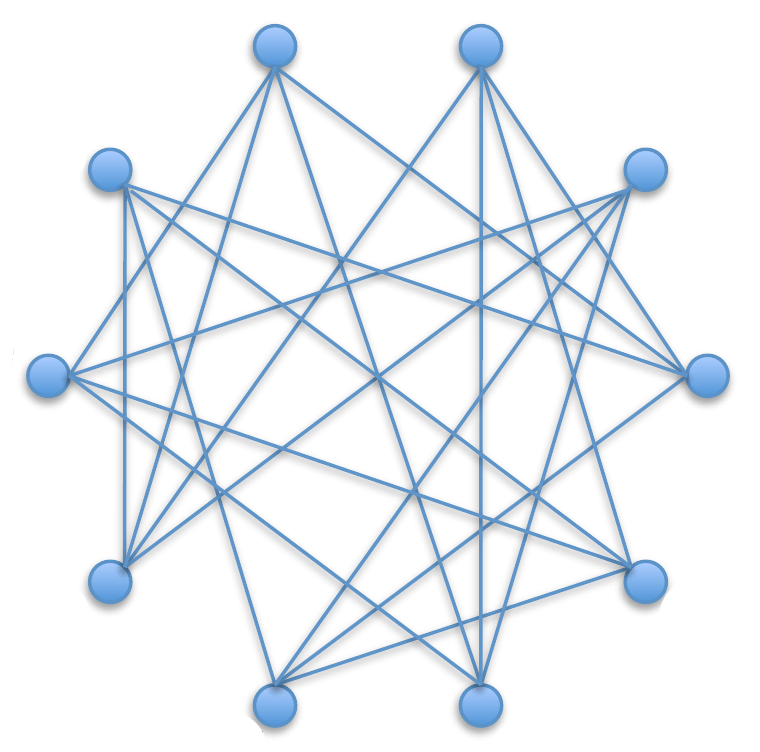}
\end{center}
\caption{ 
Symmetric (a) and  random (b) degree$-4$ graphs defined in Example~\ref{ex.4}.
}
\lbl{f.connectivity}
\end{figure}

\subsection{Examples of the network connectivity}
\lbl{examples}
In this subsection,  we present several examples of  network connectivity, which 
will be useful for illuminating the role of the network topology in the dynamical
phenomena analyzed in this paper.

\begin{ex}\lbl{ex.1} 
One of the most common examples of local connectivity is a $1D$
nearest neighbor coupling. It has been used in many studies of the
coupled ensembles of $\beta-$cells (see, e.g.,\cite{SRK88, SR91}). 
We use the $1D$ nearest-neighbor coupling in most numerical 
experiments throughout this paper.

In this configuration,
each cell in the interior of the array is coupled
to two nearest neighbors. This leads to the following expression for the 
coupling current:
$$
I_c^{(j)}=g(v^{(j+1)}-v^{(j)})+g(v^{(j-1)}-v^{(j)}),\;
j=2,3,\dots,n-1.
$$
The coupling currents for the cells on the boundary
are given by
$$  
I_c^{(1)}=g(v^{(2)}-v^{(1)})\quad\mbox{and}\quad 
I_c^{(n)}=g(v^{(n-1)}-v^{(n)}).
$$
The corresponding graph Laplacian is 
\be\lbl{1.6}
L=\left(\begin{array}{cccccc}
1 &-1 & 0& \dots &0&0 \\
-1 & 2& -1 & \dots& 0&0\\
\dots&\dots&\dots&\dots&\dots&\dots\\
0 &0 & 0& \dots &-1 &1
\end{array}
\right).
\ee
The undirected graph corresponding to the $1D$ nearest neighbor coupling 
scheme is called a path and is denoted by $P_n$ \cite{Biggs}.
\end{ex}

Clearly, the total number of edges in the graph (connections in the network)
is one of the important factors shaping the dynamics of the coupled system.
However, the way how these connections are distributed,
i.e. the connectivity of the network,
is also important as the analysis of the following example will show.

\begin{ex}\lbl{ex.4}
Consider two graphs on $n$ nodes of degree $d=4$
(i.e., each node in each of these graphs has precisely four edges 
incident to it). Such graphs are referred to as $(n,d)-$graphs. 
Thus, each of these graphs has $2n$ edges. To illuminate the role
of connectivity, we assign to these graphs two  
different connectivity patterns as shown schematically
in Fig.~\ref{f.connectivity}. The graph in Fig.~\ref{f.connectivity}a 
has symmetric connections. The edges of the graph in  
Fig.~\ref{f.connectivity}b were selected randomly.

Specifically, the second graph was generated using so-called permutation
model of a $(n,d)$ random graph of even degree $d=2\tilde d$ In this model, one chooses
at random $\tilde d$ permutations 
$$
\pi_1,\pi_2,\dots,\pi_{\tilde d}
$$
in the symmetric group $S_n$. Then the edges between $n$ vertices 
$v_1,v_2,\dots, v_n$ of $\mathcal{G}$ are generated as follows
$$
E=\{ (v_j, \pi_i(v_j)):\quad j\in [n], \; i\in [\tilde d]\}.
$$
\end{ex}

Spectral properties of the random graphs similar to the one constructed in the previous 
example have important
implications for analyzing dynamical phenomena like synchronization and 
denoising in large networks. In the remainder of this subsection, we will
discuss several facts about the spectra of random graphs that are particularly
relevant to the analysis that follows. Our review is very brief and we refer 
an interested reader for more information and extensive bibliography to 
an excellent survey by Hoory et al \cite {Hoory06}.

First, we note that the random graph constructed in Example~\ref{ex.4} is
a (spectral) expander, which means that for some positive $\alpha$
\be\lbl{expander}
\lambda_2(\mathcal{G}_n)\ge \alpha,\quad\mbox{uniformly in}\quad n\in\N,
\ee
where $n$ stands for the power of the set of vertices \cite{Hoory06, Sar04}. 
For those readers who have not seen this condition before, we note that this property 
fails to hold for any family of lattices (see Section~5 in \cite{medvedev10b}
for a related discussion). In particular, for a path on $n$ vertices
(cf.~Example~\ref{ex.1}), the second eigenvalue 
\be\lbl{expand-path}
\lambda_2(P_n)=4\sin^2\left({\pi\over 2n}\right)=O(n^{-2})
\ee
tends to zero as $n\to\infty$. Thus, the existence of a uniform
bound for the second eigenvalue postulated by \ref{expander} is 
a special, if not counter-intuitive, property. Nonetheless, it holds
for random graphs. There are also known explicit (nonrandom) algorithms 
generating expanders, including the 
celebrated Ramanujan graphs \cite{Margulis88, LPS88}.

In many applications, it is desirable to have a large bound on the expansion constant 
$\alpha$. However, for $(n,d)$ graphs, $\lambda_2(\mathcal{G}_n)$ can not exceed
the Alon-Bopana bound $g(d)=d-2\sqrt{d-1}$ \cite{Hoory06}. A remarkable property of the
family of $(n,d)$
random graphs is that for $n\gg 1$ they posses nearly optimal 
expansion constant with overwhelming probability.
In particular, it is known that for any $\epsilon>0$
\be\lbl{Friedman}
\mathrm{Prob}\left\{ \lambda_2(\mathcal{G}_n)\ge d-2\sqrt{d-1}-\epsilon\right\}=
1-o_n(1) \; \forall\epsilon>0,
\ee
where $\mathcal{G}_n$ stands for the family of $(n,d)$ random
graphs of degree $d\ge 3$ and $n\gg 1$ \cite{Fri08}.

\section{Transitions to bursting in the coupled model}\lbl{transitions}
\setcounter{equation}{0}
Having reviewed bursting in deterministic systems and the definition of the 
coupled network, we now turn to the main theme of this work - the roles of
noise and electrical coupling in shaping firing patterns of the coupled system
(\ref{c.1}) and (\ref{c.2}). Fig.~\ref{f.intro} shows transitions in the
coupled system's dynamics observed over long intervals of time
upon increasing the coupling strength. In the first case, which in the sequel we will
refer to as Scenario A, irregular spiking patterns for weak 
coupling are transformed into fairly regular bursting patterns for sufficiently
strong coupling (see Fig.~\ref{f.intro}a,b). In the second case, Scenario B,
robust very irregular bursting becomes synchronous spiking when coupling strength
is increased. 

We show that at the heart of both transitions lies denoising, the mechanism responsible
for greatly diminishing the effects of noise on network dynamics. Below, we discuss 
both scenarios illustrating them with numerical results. For Scenario A, we also
develop analytical estimates characterizing denoising. Our goal is to show how 
statistical features of the firing patterns depend on the coupling strength, excitability,
and network topology. The analysis of Scenario B can be developed along
the same lines, but it requires certain additional techniques for dealing with
the analysis of trajectories near a limit cycle. These extensions will be considered
in the future work.

\begin{figure}
\begin{center}
{\bf a}\epsfig{figure=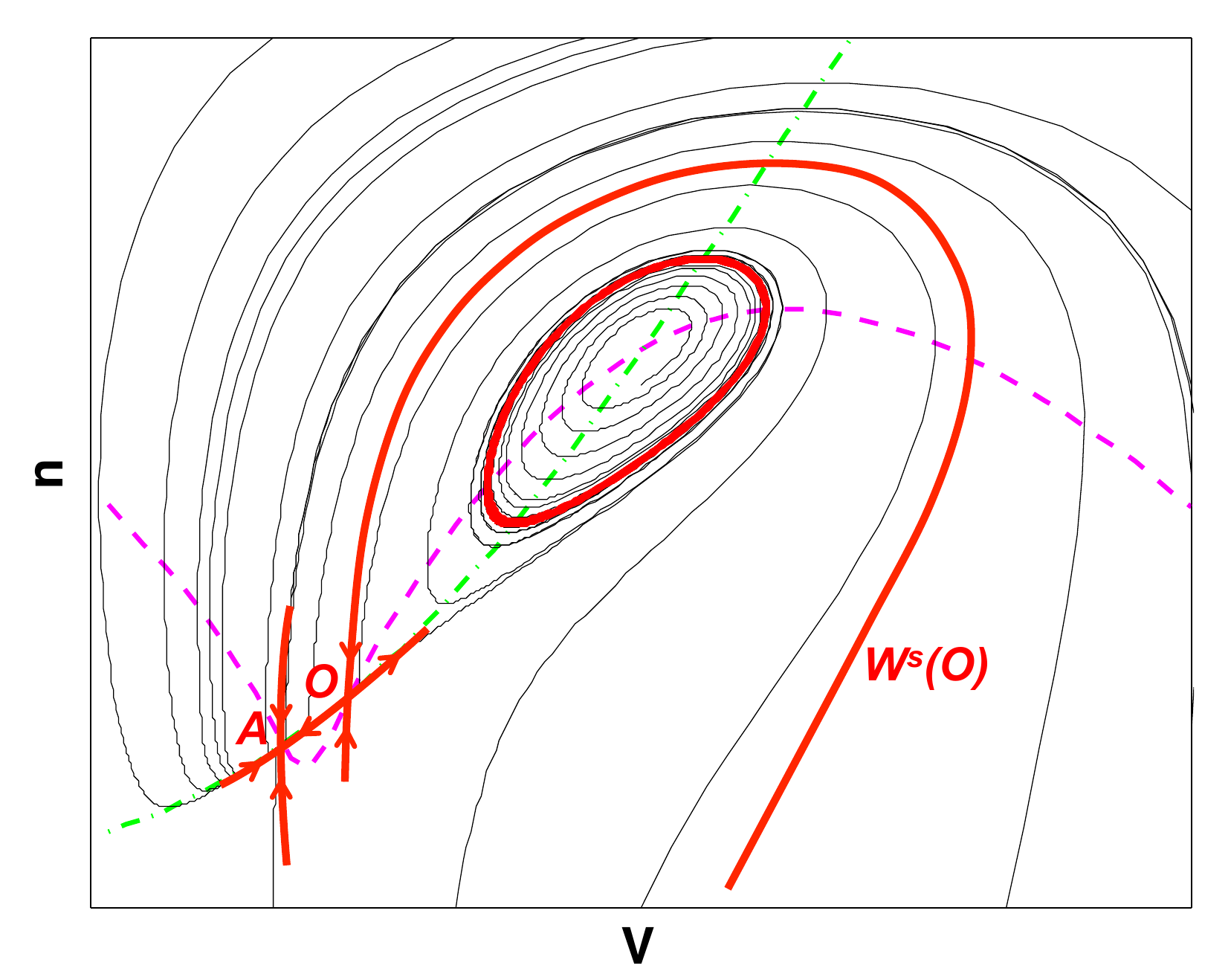, height=2.0in, width=2.2in}\qquad
{\bf b}\epsfig{figure=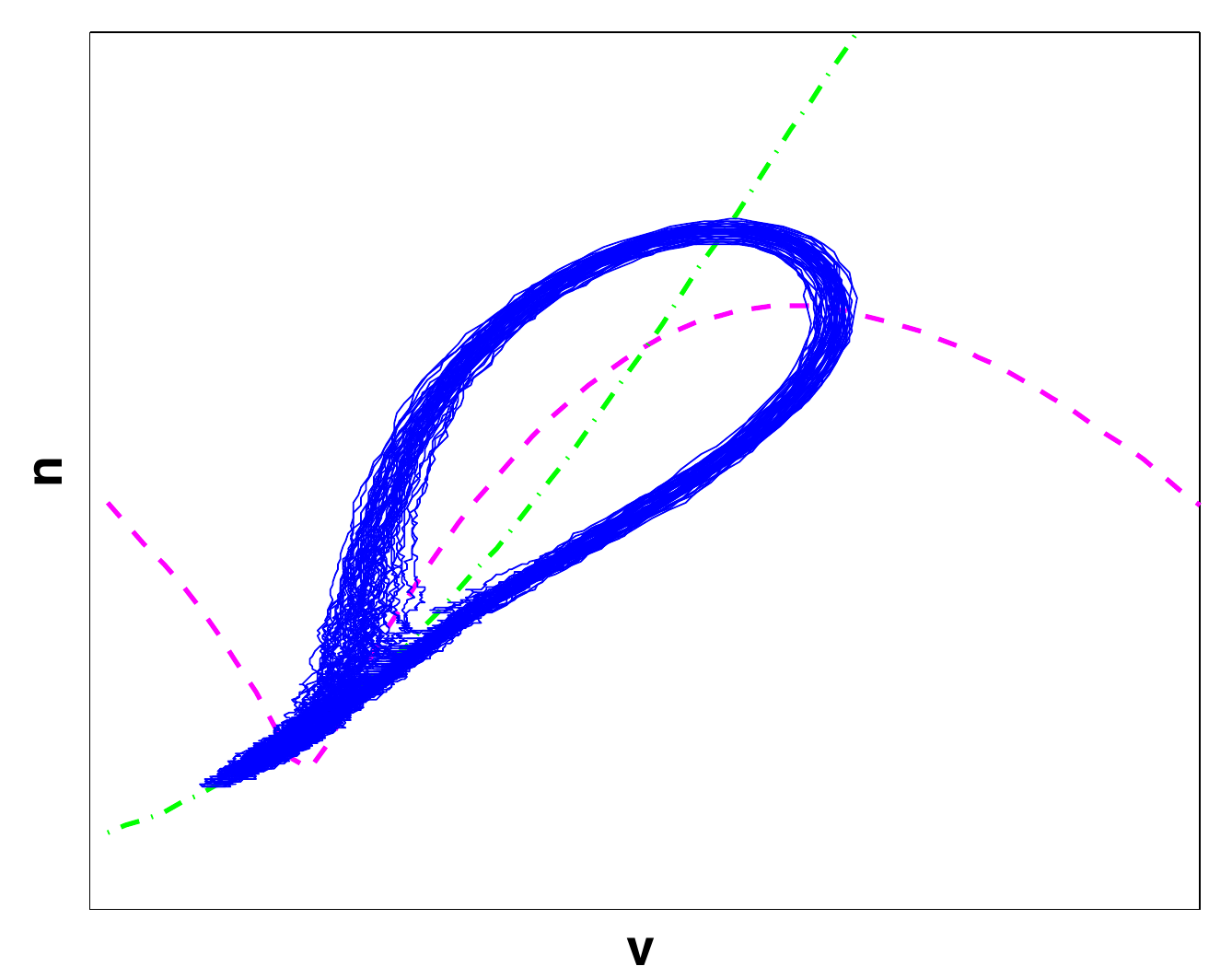, height=2.0in, width=2.2in}
\end{center}
\caption{The phase portrait of the fast subsystem. (a) The basins
of attractions of the stable fixed point $E_y$ and  limit cycle $L_y$
are separated by the stable manifold of the saddle fixed point.
(b) The trajectory of the randomly perturbed model 
(\ref{c.1}) and (\ref{c.2}) switches between the neighborhoods 
of the two attractors. Note that the transitions take place
along the unstable manifold of the saddle. This shows
that the center manifold reduction used in 
Sections~\ref{transitions} and \ref{synchronization} adequately
describes the dynamics observed numerically.
}\lbl{f.pplane}
\end{figure}

\subsection{Scenario A: The single-cell model}
In this and in the following subsections, we discuss Scenario A in more detail. 
We start with the single-cell model (\ref{2.1}) and (\ref{2.2}).

 Assume that the deterministic model (\ref{2.1}$)_0$ and
(\ref{2.2}) is in the bursting regime, i.e., a typical trajectory alternates between
drifting along a curve of stable equilibria of the fast subsystem, $E$, and a cylinder
of limit cycles, $L$ (see Fig.~\ref{f.regimes}a). Let us focus now on a slow evolution
along $E$. Since $\dot y=O(\epsilon)$ (cf.~(\ref{2.2})), on time intervals $o(\epsilon^{-1})$
long, the dynamics of the slow-fast system (\ref{2.1})-(\ref{2.2}) is approximated by
the fast subsystem 
\be\lbl{fast-loc}
\dot x=f(x,y)+\Sigma \dot w,
\ee
where $y\in (y_{sn}, y_{hc})$ is fixed,  $x=(x_1,x_2)^\t\in \R^2, 
\Sigma=\mbox{diag}(\sigma_1,\sigma_2),\; \sigma_{1,2}\ge 0$, 
$f:\R^2\times\R\mapsto\R^2$ is a smooth function, and 
$w=(w_1, w_2)$ is a standard Brownian motion in $\R^2$.

The frozen system (\ref{fast-loc}$)_0$ is bistable. It has two co-existing
attractors: the stable equilibrium $E_y$ and the limit cycle $L_y$ 
(see Fig.~\ref{f.pplane} a). A trajectory of the randomly perturbed system
(\ref{fast-loc}) alternates between the basins of $E_y$ and $L_y$,
$\mathcal{B}(E_y)$ and $\mathcal{B}(L_y)$, separated by the stable
manifold of the saddle point $O$ (see Fig.~\ref{f.pplane}b).
In fact, most of the time it spends in small neighborhoods of 
$E_y$ and $L_y$. The plot in Fig.~\ref{f.pplane}b shows that a typical
trajectory  leaves the neighborhood of $E_y$
along the weak stable manifold of the sink, i.e., the dynamics 
near $E_y$ is effectively one-dimensional. We will use this observation
to simplify the analysis of the coupled system by reducing it (via the center
manifold theorem \cite{CH82}) to a simpler system.

The following exit problem is instrumental for understanding the effects of
noise on the dynamics of the bistable system (\ref{fast-loc}). For a trajectory
of (\ref{fast-loc}), which starts from a deterministic initial condition
$x(0)=x_0\in\mathcal{B}(E_y)$, define the first exit time from $\mathcal{B}(E_y)$
by
\be\lbl{first-exit}
\tau(E_y,x_0) =\inf \{ t>0:~ x(t) \notin\mathcal{B}(E_y)  \}.
\ee
Below we show that one can choose $\sigma(\epsilon)>0$ such that with overwhelming
probability for small $\epsilon>0$,
$$
\tau(E_y, x_0)=o(\epsilon^{-1})
$$ and at the same time $\sigma(\epsilon)\to 0$ as $\epsilon\to 0$ (see Lemma~\ref{easy}).
Therefore, with arbitrarily small noise one can make the trajectory of the frozen
system (\ref{fast-loc}) leave $\mathcal{B}(E_y)$ in time insufficient for $y$
to change by $O(1)$ amount thus destroying bursting.

In Fig.~\ref{f.sceA}a, we show a typical trajectory of
(\ref{ch.1})-(\ref{ch.3}) superimposed on the bifurcation diagram of the
fast subsystem. Note that the trajectory can not advance far enough
along the curve of stable equilibria $E$, because it is forced to jump
to the vicinity of $L$ by noise. Thus, it lands on $L$ near $y_{hc}$ every time
and does not have enough room to generate many spikes before it reaches
the right end of $L$ and jumps back to $E$. This results in very irregular bursting
patterns with very few spikes in one burst (see Fig.~\ref{f.sceA}b,c). In theory,
for sufficiently small $\epsilon>0$ and small $\sigma>0$ one can make the probability
of clusters of $2$ and more spikes in the timeseries of (\ref{ch.1})-(\ref{ch.3})
arbitrarily low. This means, that one can  transform bursting into irregular spiking
by adding small noise. Showing this numerically, however, requires integrating
stiff stochastic differential equations over very long time. Thus, we did not strive
to achieve irregular spiking in our experiments resorting to irregular bursting
patterns with very few spikes in Fig.~\ref{f.sceA}b, which already illustrate this
effect.    

The noise-induced irregular firing pattern shown in Fig.~\ref{f.sceA}b is characterized
by the approximetelly geometric distribution of spikes in one cluster (see Fig.~\ref{f.sceA}c).
The geometric distribution  has its origins in the exponential distribution
of the first exit time $\tau(E_y, x_0)$ \cite{Day83}, which implies roughly that the distance 
from the landing point on $L$, $y_l$, to the right hand end of 
$L$, $y_{hc}$, $y_l-y_{hc}$ is distributed
approximately exponentially. The exponential distribution of $y_l-y_{hc}$ 
translates into the geometric distribution of the number of spikes in one burst. Later we will
see that this distribution is qualitatively different for the coupled system 
(see Fig.~\ref{f.sceA}f). The distinct probability distributions in Fig.~\ref{f.sceA}c
and Fig.~\ref{f.sceA}f corresponding to
 different values of the coupling strength show  that the transition in the network dynamics 
has taken place.
  
In conclusion, we note that while the fact that we were able to destroy deterministic
bursting with noise may not seem very surprising, the possibility of doing this with
small noise is far from obvious. Note that the equilibria of the fast subsystem, $E_y$,
for $y$ near $y_{hc}$ are extremely stable. The slow-fast structure of the vector field
is the key to this important effect. Extremely slow evolution along $E$ gives the trajectory
of the frozen system enough time to develop large deviations 
(necessary to leave $\mathcal{B}(E_y)$), which are highly unlikely on time intervals
$O(1)$ long. This is a general mechanism by which adding noise to slow-fast systems
may create qualitatively new dynamical regimes \cite{FR01}. 

Before we move on to discuss the coupled system, we prove the following lemma,
which provides an estimate of the noise intensity sufficient for 
destroying bursting.

\begin{lem}\lbl{easy}
Let $k>0$ and $\Sigma=\mbox{diag}~(\sigma, k\sigma)$ be a matrix defining
the noise intensities in (\ref{fast-loc}).
Then there exists $C_1>0$ such that for every $C_2\ge C_1$
and
\be\lbl{sigma-kill}
\sigma = {C_2\over \sqrt{|\ln\epsilon|}}
\ee
a trajectory of (\ref{fast-loc}) with initial condition
$x(0)\in\mathcal{B}\left(E_y\right)$ leaves 
$\mathcal{B}\left(E_y\right)$ in time $o(\epsilon^{-1})$
with probability converging to $1$ as $\epsilon\to 0$.
\end{lem}
\pf\;
By rescaling $x_2$ in (\ref{fast-loc}), one can always achieve $k=1$,
which we assume without loss of generality.

Thanks to the large deviation theory, we have  the following estimate for the 
first exit time $\tau(E_y,x_0)$
(cf. Theorem 4.4.2, \cite{FW}): for any $h>0$,
\be\lbl{r.7}
\lim_{\epsilon\to 0} \P_{x_0} \left\{\exp\{(V_y-h)\sigma^{-2}(\epsilon)\} <\tau (E(y),x_0) 
< \exp\{(V_y+h)\sigma^{-2}(\epsilon)\}\right\} =1,
\ee
where positive constant
$V_y$ is the minimum of the quasipotential associated with the randomly perturbed system 
(\ref{fast-loc}).  The definition and the  properties of the quasipotential can 
be found in \cite{FW}.

Fix $0<\alpha<1$ and take
$$
 C_1^2 = {V_y+h\over 1-\alpha}.
$$ 
The combination of (\ref{r.7}) and (\ref{sigma-kill}) implies
\be\lbl{r.9}
\tau (E(y),x_0)< \epsilon^{-V_y+h\over C_1^2} < \epsilon^{-1+\alpha}
\ee
with probability tending to $1$ as $\epsilon\to 0$, provided $0<h<V_y$. \\
$\qed$

\begin{figure}
\begin{center}
{\bf a}\epsfig{figure=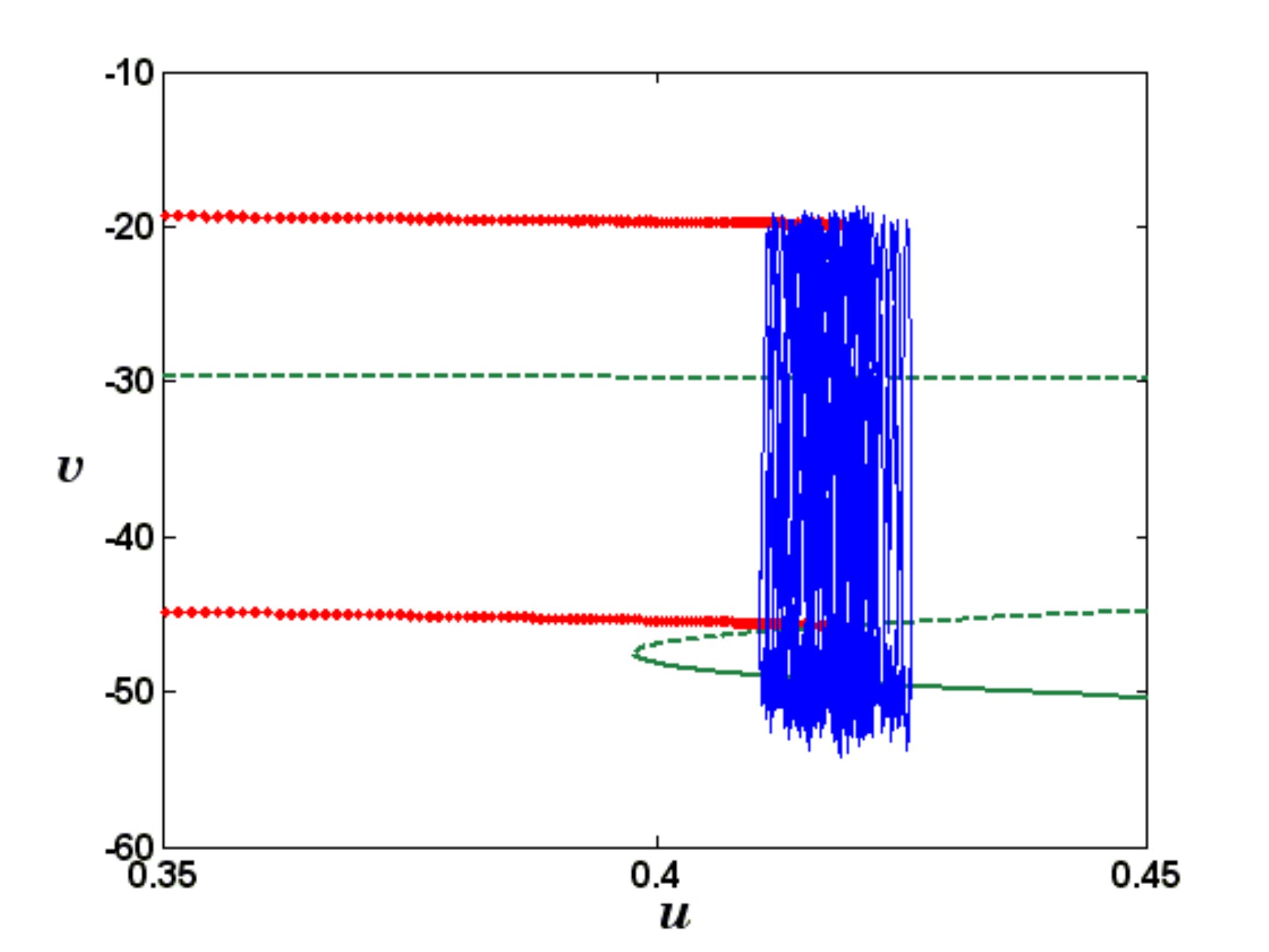, height=2.0in, width=2.0in}
{\bf b}\epsfig{figure=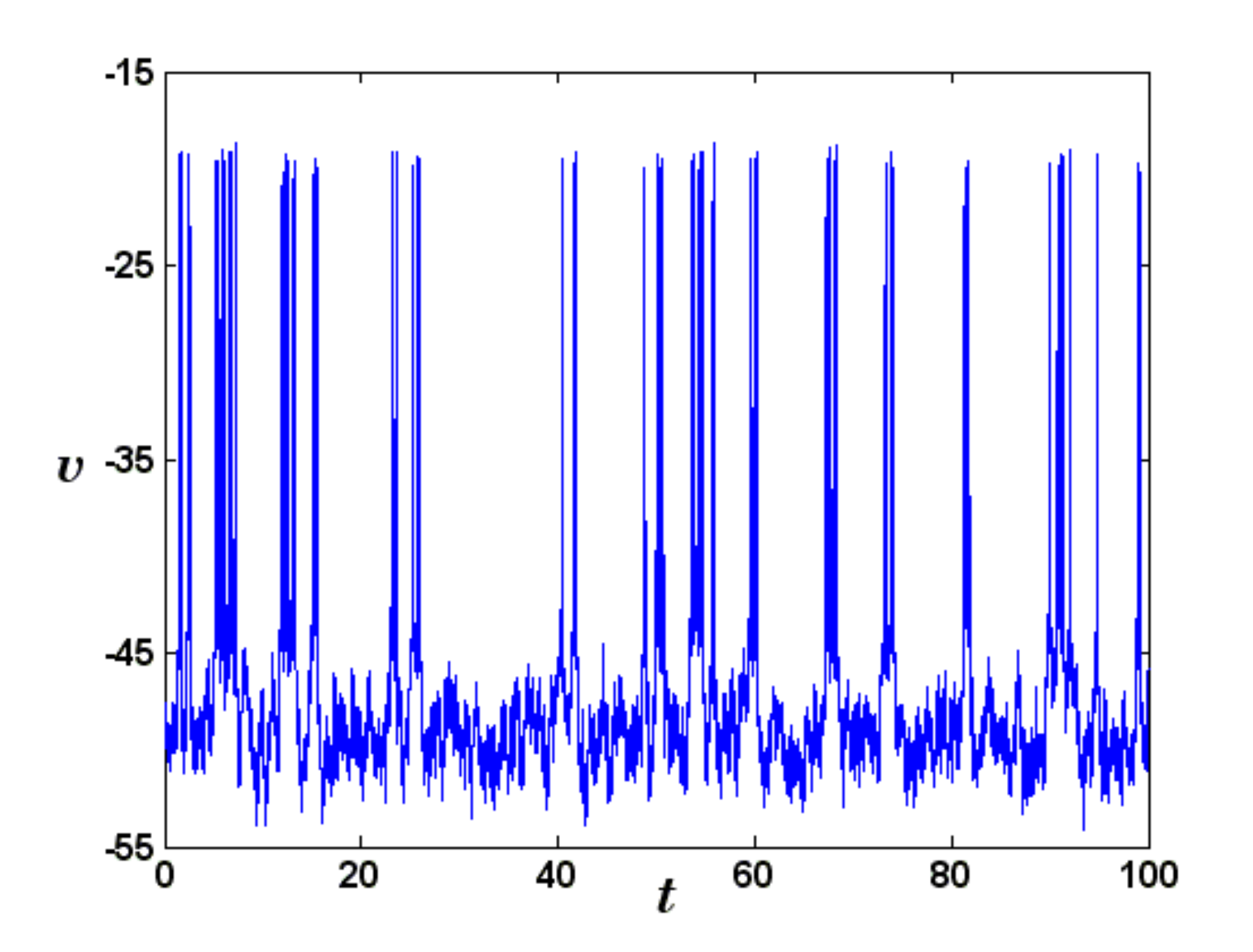, height=2.0in, width=2.0in}
{\bf c}\epsfig{figure=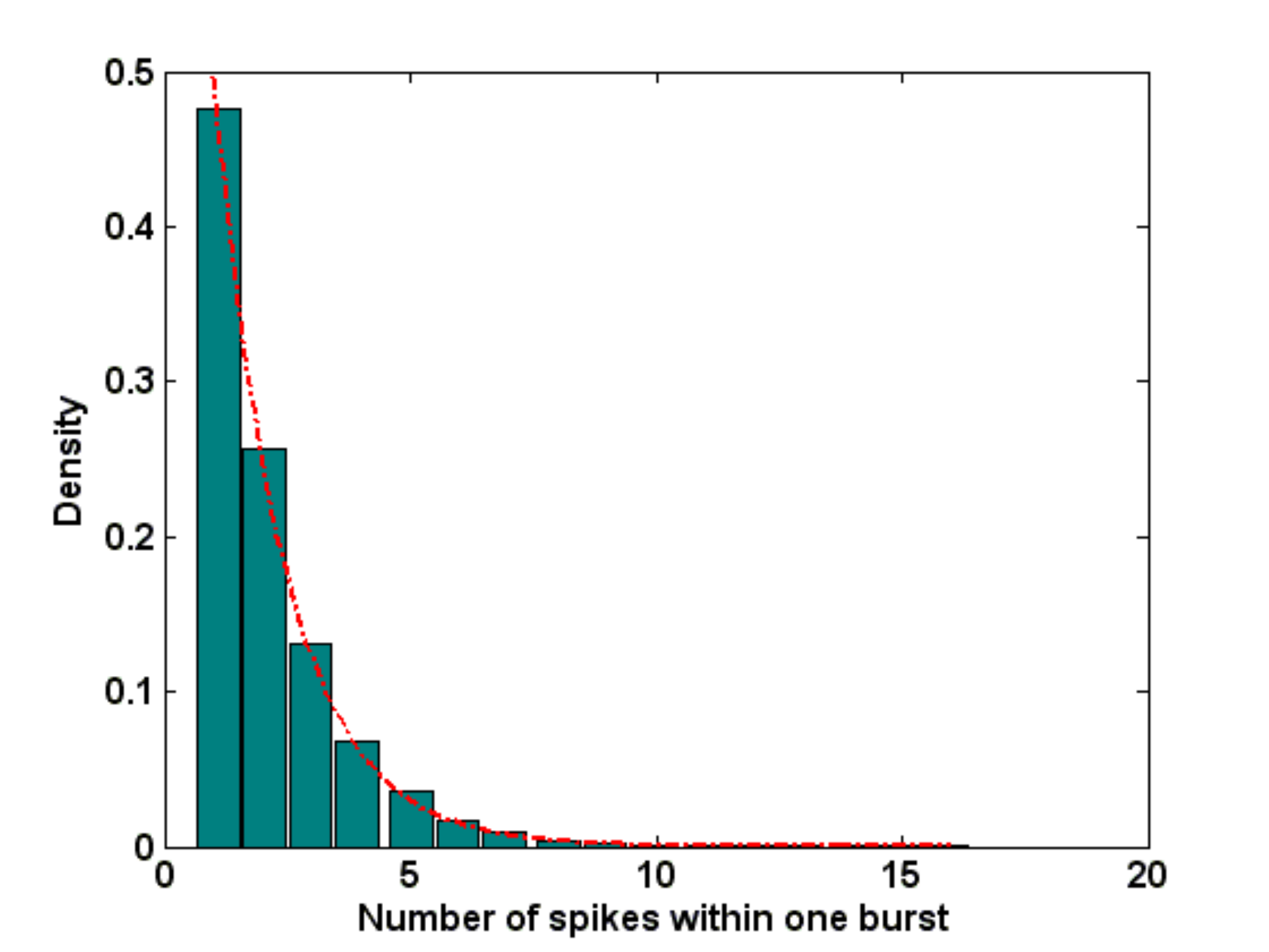, height=2.0in, width=2.0in}\\
{\bf d}\epsfig{figure=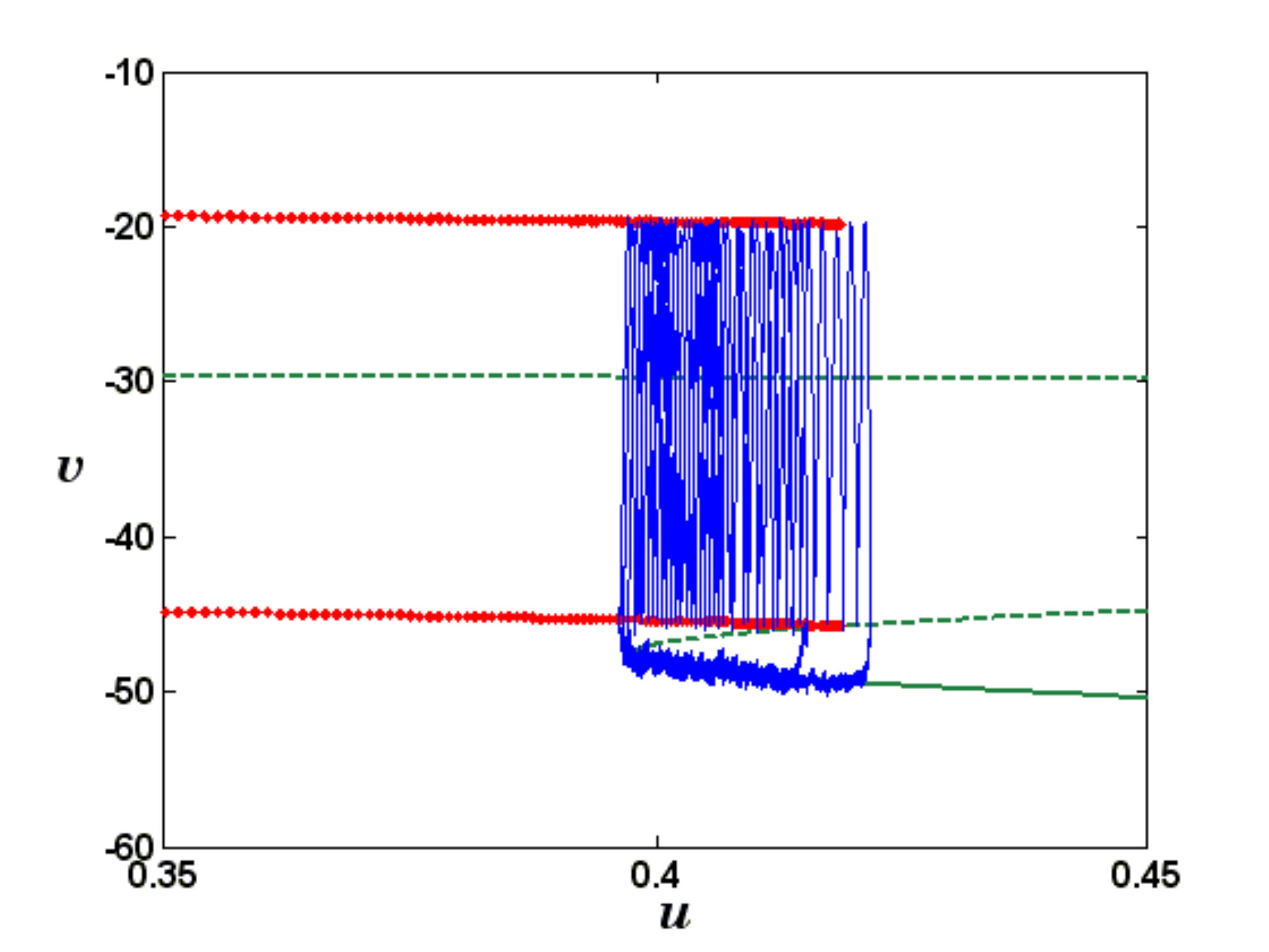, height=2.0in, width=2.0in}
{\bf e}\epsfig{figure=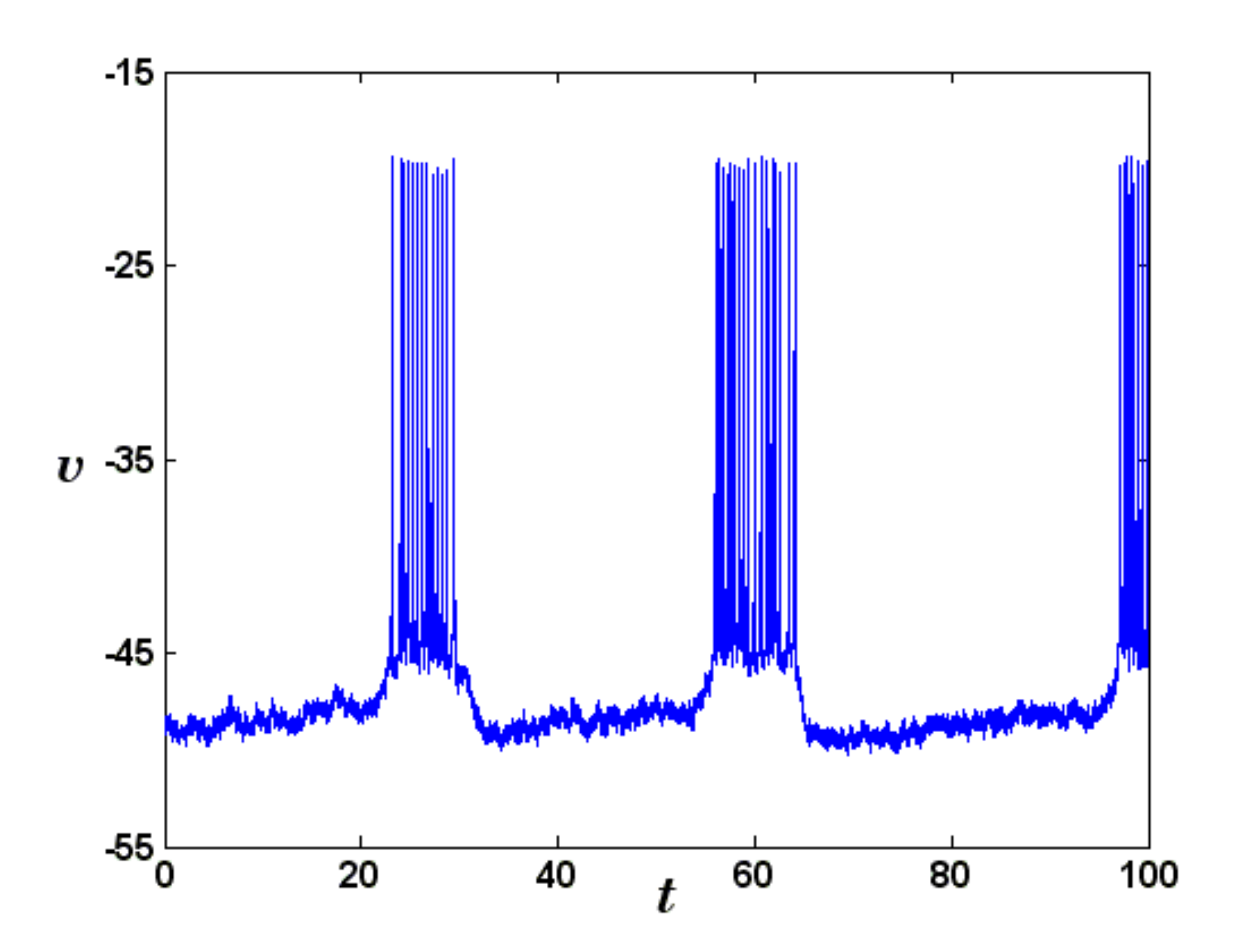, height=2.0in, width=2.0in}
{\bf f}\epsfig{figure=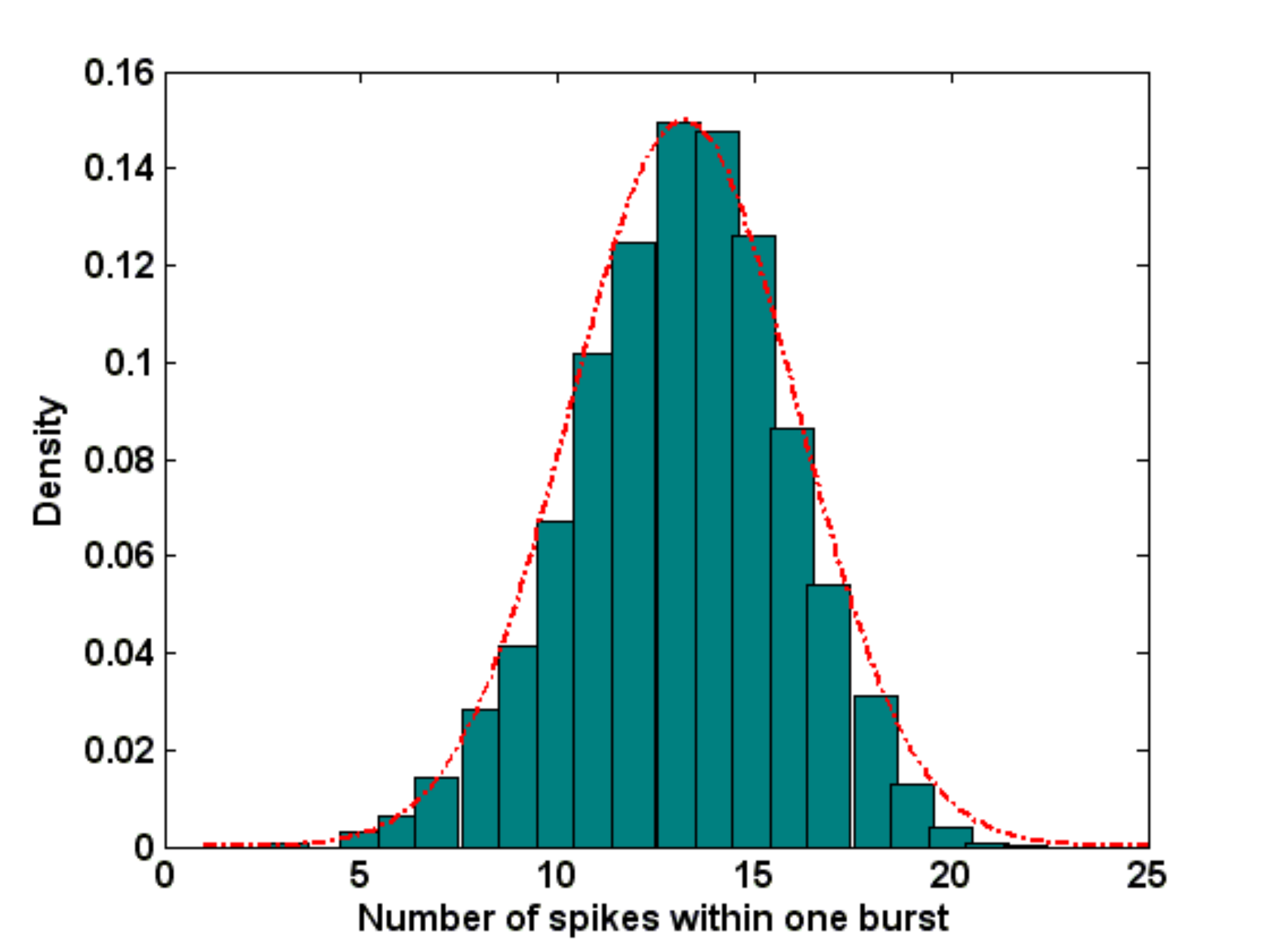, height=2.0in, width=2.0in}
\end{center}
\caption{Scenario A. The deterministic model
(\ref{ch.1})-(\ref{ch.3}) is tuned to the bursting
regime (a,b). Weak noise can destroy bursting in the single
cell model creating irregular spiking pattern
(cf.~Lemma~\ref{easy}). 
The plots of trajectory of the coupled system superimposed
on the bifurcation diagram (d) and the corresponding timeseries
(e) show that bursting is recovered in the network thanks to denoising.
The transition to bursting can be clearly seen
from the normalized histograms of the number of spikes
in one cluster or burst. The geometric distribution
corresponding to (a,b) is transformed to the Gaussian
distribution in (c,d).  In these numerical experiments, we used 
system (\ref{ch.1})-(\ref{ch.3}) with $50$ oscillators coupled 
through the nearest neighbor coupling 
(see Example~\ref{ex.1}) with the  coupling strength $g=5000$
and other parameters specified in Appendix A.
}
\lbl{f.sceA}
\end{figure}

\subsection{Scenario A: The coupled system}

Next, we turn to the analysis of the coupled system.
We want to understand how bursting is recovered for
larger values of the coupling strength (see Fig.~\ref{f.sceA} d-f). 
Thus, we consider the coupled system (\ref{c.1}) and (\ref{c.2}).
Below, we use two simplifying asuumptions, which let us
avoid certain technical details, which are peripheral to the mechanism 
analyzed below,
First, we focus on the fast subsystem ignoring $O(\epsilon)$
changes in the slow variables:
\be\lbl{fast-coup}
\dot X = F(X,Y) -g(L\otimes J_1)X +(I_n\otimes \Sigma)\dot W,
\ee
where $X=(x^{(1)},x^{(2)},\dots, x^{(n)})^\t\in\R^d\times\dots\times\R^d\cong\R^{nd}.$
Furthermore,  
it can be shown that with the diffusive coupling like 
in (\ref{fast-coup}), $y_i'$s synchronize and remain close after some initial transients, 
provided the coupling is sufficiently strong (see \cite{medvedev10a, medvedev10}). 
Thus, we set the frozen slow variables for all subsystems 
to the same value $y\in (y_{sn}, y_{hc})$ so that
$$
F(X, Y)=\left(f(x^{(1)},y),f(x^{(2)},y),\dots, f(x^{(n)}, y)\right)^\t.
$$

For the coupled system (\ref{fast-coup}), we consider the problem
of exit from the basin of the rest state 
$\mathbf{E_y}=E_y\times\dots\times E_y\in\R^{2n}$. To this
end, we define
$$
\tau (\mathcal{B}(\mathbf{E}_y), X_0)=
\inf\{t>0:~ X(t)\notin\mathcal{B}(\mathbf{E}_y)\}.
$$

The key difference between the coupled system 
(\ref{fast-coup}) and the individual subsystems (\ref{fast-loc}) is 
that the former is much less susceptible to the effects of noise. 
The robustness of the coupled system to noise manifests itself  
in the disparity of the exit times:
$$
\tau (\mathcal{B}(\mathbf{E}_y), X_0)\gg \tau (\mathcal{B}(E_y), x_0).
$$
More precisely, the asymptotic relation between the exit times
corresponding to the single cell and coupled models is given in the 
following theorem.

\begin{thm} 
Suppose that (\ref{fast-loc}) is close to a nondegenerate 
saddle-node bifurcation \footnote[1]{The nondegeneracy conditions
are specified in the proof of the theorem.}
Then for some $\sigma_0>0$ and $g_0>0$, and for all 
$0<\sigma<\sigma_0$ and $g>g_0$ the following asymptotic relations
hold:
\be\lbl{exit-asymptotics}
\tau (\mathcal{B}(E_y), x_0)\asymp \exp\left\{ {C_3\over\sigma^2}\right\}
\quad\mbox{and}\quad
\tau (\mathcal{B}(\mathbf{E}_y), X_0)\asymp 
\exp\left\{ {C_3 n\over\sigma^2}\right\},
\ee
where $\asymp$ denotes logarithmic asymptotics (cf. (\ref{r.7})), $n$ is the number of
the cells in the network, and $C_3$ is a positive constant independent
of $\sigma$ and $n$.
\end{thm}
\begin{rem}
Relations in (\ref{exit-asymptotics}) show that for the level of noise
chosen in Scenario A, by taking sufficiently strong coupling with
sufficiently many cells in the network, one can make 
$\tau (\mathcal{B}(\mathbf{E}_y), x_0)$ longer than the time necessary
for $y$ to reach the vicinity of $y_{sn}$. This means that 
the level of noise, which prevents a single cell model
from bursting, does not affect bursting in the coupled system
(compare Fig.~\ref{f.sceA} a and d).
\end{rem}
\pf
The proof of the theorem follows from the analysis of a closely
related model in \cite{MZ}. For completeness, we outline the main steps 
of the proof and refer the interested reader to \cite{MZ} for further details. 
The proof consists of several steps,
the main of which are
the center manifold reduction of the single cell and coupled
models near the saddle-node bifurcation (cf.~(\ref{rescale})) and the variational interpretation
of the reduced systems (cf.~(\ref{grad})).
\begin{enumerate}
\item
By our assumptions on the fast subsystem, the 
Jacobian $Df(0,0)$ has a $1D$ kernel. Denote
\be\lbl{kerA}
e\in\ker Df(0,0)/\{0\} \;\mbox{and}\; 
p\in\ker (Df(0,0))^\t\;\mbox{such that}\; 
p^\t e=1.
\ee
In addition to standard nondegeneracy and transversality 
conditions for a saddle-node bifurcation
\begin{eqnarray}
\lbl{a1}
a_1 &=& {1\over 2} {\partial^2\over \partial u^2} p^\t f(ue,0)\left|_{u=0}\right.\neq 0,\\
\lbl{a2}
a_2 &=& {\partial \over \partial y} p^\t f(0,\mu)\left|_{\mu=0}
\right.\neq 0,
\end{eqnarray}
we assume that
\be\lbl{a3}
a_3 = p^\t Je\neq 0,
\ee
where $J_1$ is the matrix involved in the coupling operator (see (\ref{fast-coup})).
Condition (\ref{a3}) guarantees that the projection of the coupling 
onto the center subspace is not trivial. 
Without loss of generality, we assume that nonzero
coefficients $a_{1,2,3}$ are positive.
\item
Under the conditions in 1., near the bifurcating equilibrium
the coupled model can be reduced to an $n-$dimensional slow manifold.
The reduced system (after appropriate rescaling of the dependent and 
independent variables and dropping higher order terms) has 
the following form:
\be\lbl{rescale}
\dot z=z^2-\mathbf{1_n} - \gamma Lz+ \sigma\dot W,
\ee
where $\gamma$ is the rescaled coupling strength, 
$\mathbf{1_n}=(1,1,\dots,1)^\t\in\R^n$, and
$W$ is the standard Brownian motion in $\R^n$
\footnote[1]{
Throughout this paper, we use $W$ to denote multidimensional Brownian motion 
in Euclidean spaces of different dimensions associated with the coupled
systems. We reserve $w$ to denote Brownian motions used to perturb single
cell models. The dimension of the space for stochastic processes will be clear 
from the context and should not cause any confusion.}.

Similarly, the single cell model is reduced to the following
$1D$ equation:
\be\lbl{single}
\dot\zeta=\zeta^2 -1+\sigma\dot w_t.
\ee
\item 
We recast (\ref{single}) and (\ref{rescale}) as randomly
perturbed gradient systems. The former is rewritten as
\be\lbl{grad-1d}
\xi=-\Phi^\prime(\xi)+\sigma \dot w,\quad 
\Phi(\zeta)={2\over 3}+\zeta-{\zeta^3\over 3}.
\ee
For the latter, we use the structure of the coupling matrix (\ref{weightedLap}),
to reduce it to the following form
\be\lbl{grad}
\dot z=-{\partial\; \over\partial z} U_\gamma (z)+\sigma\dot W,
\quad
U_\gamma(z) = {\gamma\over 2} \langle \tilde L z, \tilde L z\rangle 
+\sum_{i=1}^n\Phi(z_i),
\ee
where $\tilde L=\sqrt{C}H$ and 
$\sqrt{C}=\mbox{diag}(\sqrt{c_1}, \sqrt{c_2},\dots,\sqrt{c_m})$
is a square root of the nonnegative definite conductance matrix
$C$ (cf.~\ref{conductance}).
\item The large deviation estimates yield the logarithmic asymptotics
for the exit times associated with the stable fixed point of 
(\ref{grad-1d}) (cf. Theorems 2.1 \& 3.1 in \cite{FW}):
\be\lbl{asymptotics-1}
\lim_{\sigma\to 0} \P_{\zeta_0}\left\{ \exp\{(2\Phi(1) - h)\sigma^{-2}\}\le 
\tau (\mathcal{B}(-1),\zeta_0)\le 
\exp\{(2\Phi(1) + h)\sigma^{-2}\}\right\}=1, \;\forall h>0,
\ee
where $\Phi(1)=4/3$ is the value of the potential at the barrier at $\zeta=1$.
Thus, 
\be\lbl{var-as}
\tau(\mathcal{B}(-1),\zeta_0)\asymp \exp\left\{{8\over 3\sigma^2}\right\}.
\ee
which shows the first relation in (\ref{exit-asymptotics}).
\item 
The (deterministic) coupled system (\ref{rescale}$)_0$ has a stable fixed point at 
$z=-\mathbf{1_n}$.
The basin of attraction of this fixed point is bounded by
\be\lbl{boundary}
\partial D=\bigcup_{i=1}^n D_i, \; D_i=\{ z=(z_1,z_2,\dots, z_n)\in \R^n:\; 
(\exists i\in [n]: z_i=1)\; \&\; (z_k \le 1, k\in [n]) \}
\ee
The estimate of the exit times from $D$ follows from the analysis
of the minima of the potential function $U_\gamma(\cdot)$  on 
$\partial D$. In \cite{MZ}, we prove that 
for $\gamma \ge 2\lambda_1(L^1)^{-1}$,
$U_\gamma (z)$ achieves its minimal value on $\partial D$ at $z=\mathbf{1_n}$:
\be\lbl{u-lambda-min}
u_\gamma:=U_\gamma(\mathbf{1_n})={4n\over 3}.
\ee
(Here, $\lambda_1(L^1)$ denotes the smallest eigenvalue of matrix $L^1$,
obtained from $L$ by deleting the first row and the first column.)
Knowing the minimum of the potential function on the boundary 
of the basin of attraction of the stable fixed point, the 
large deviation estimates for the randomly perturbed gradient 
systems yield 
\be\lbl{var-as-coup}
\tau (\mathcal{B}(-\mathbf{1}_n), z_0)\asymp \exp\left\{ {8n\over 3\sigma^2}\right\}.
\ee
\end{enumerate}
$\qed$

\begin{figure}
\begin{center}
{\bf a}\epsfig{figure=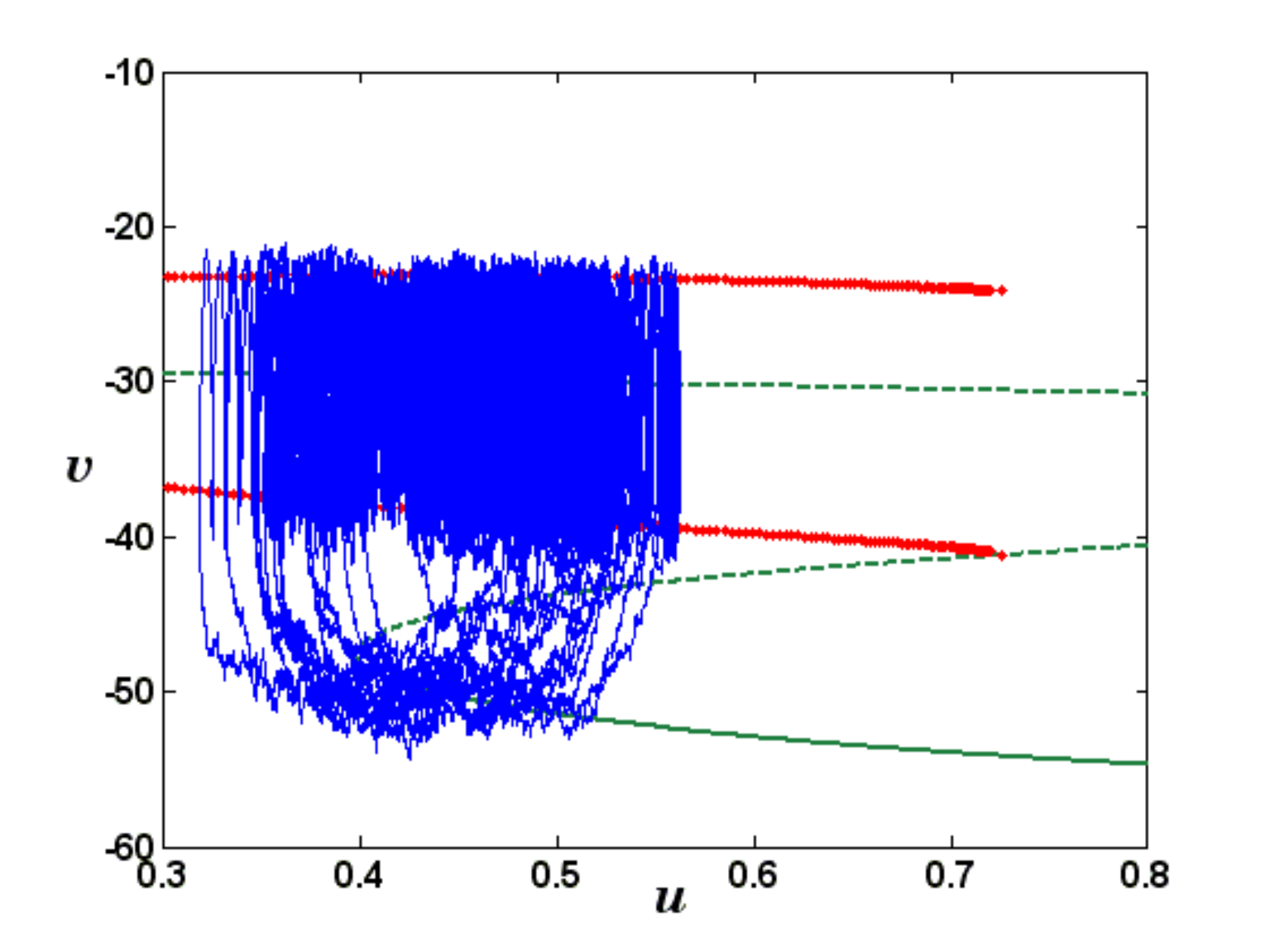, height=2.0in, width=2.0in}
{\bf b}\epsfig{figure=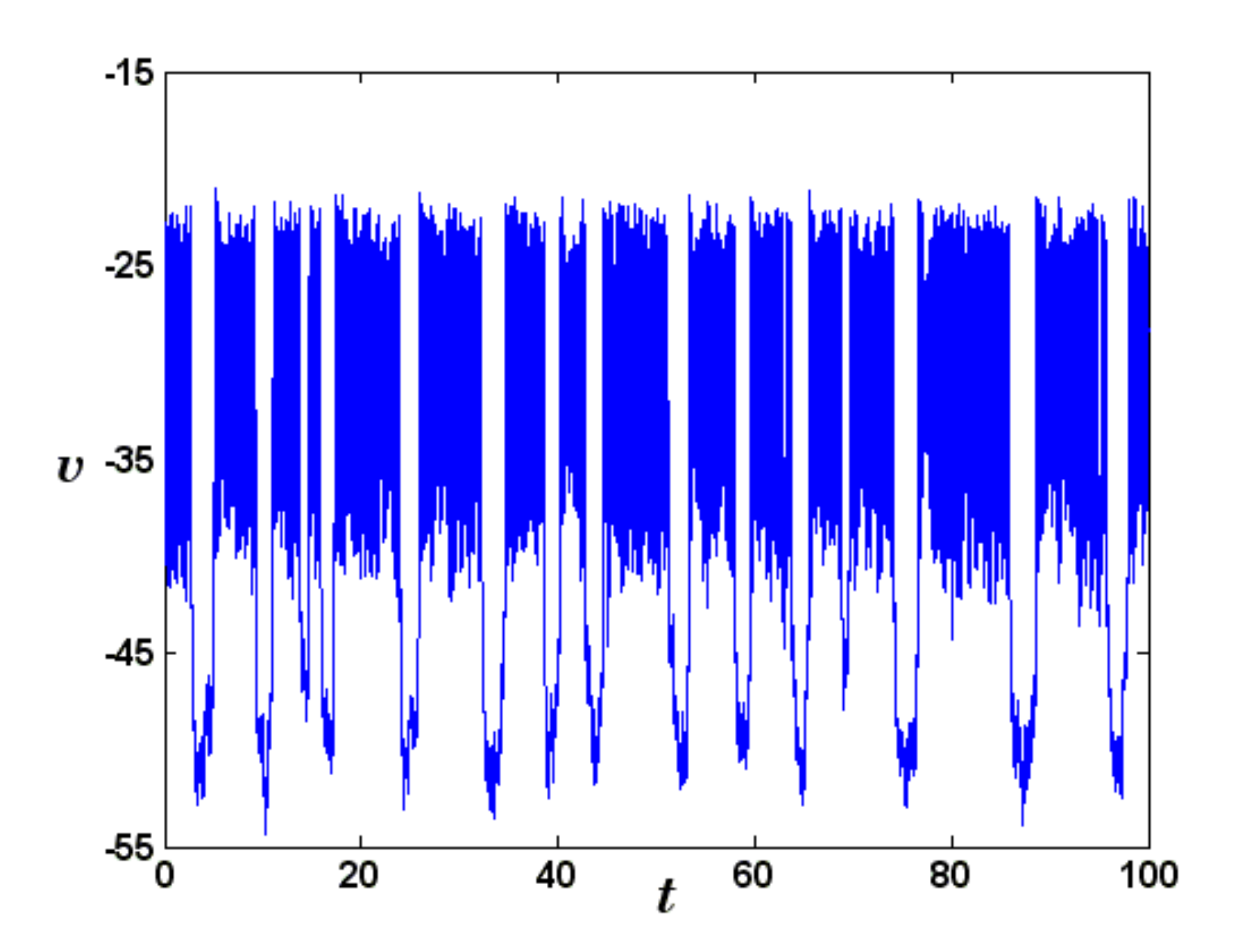, height=2.0in, width=2.0in}
{\bf c}\epsfig{figure=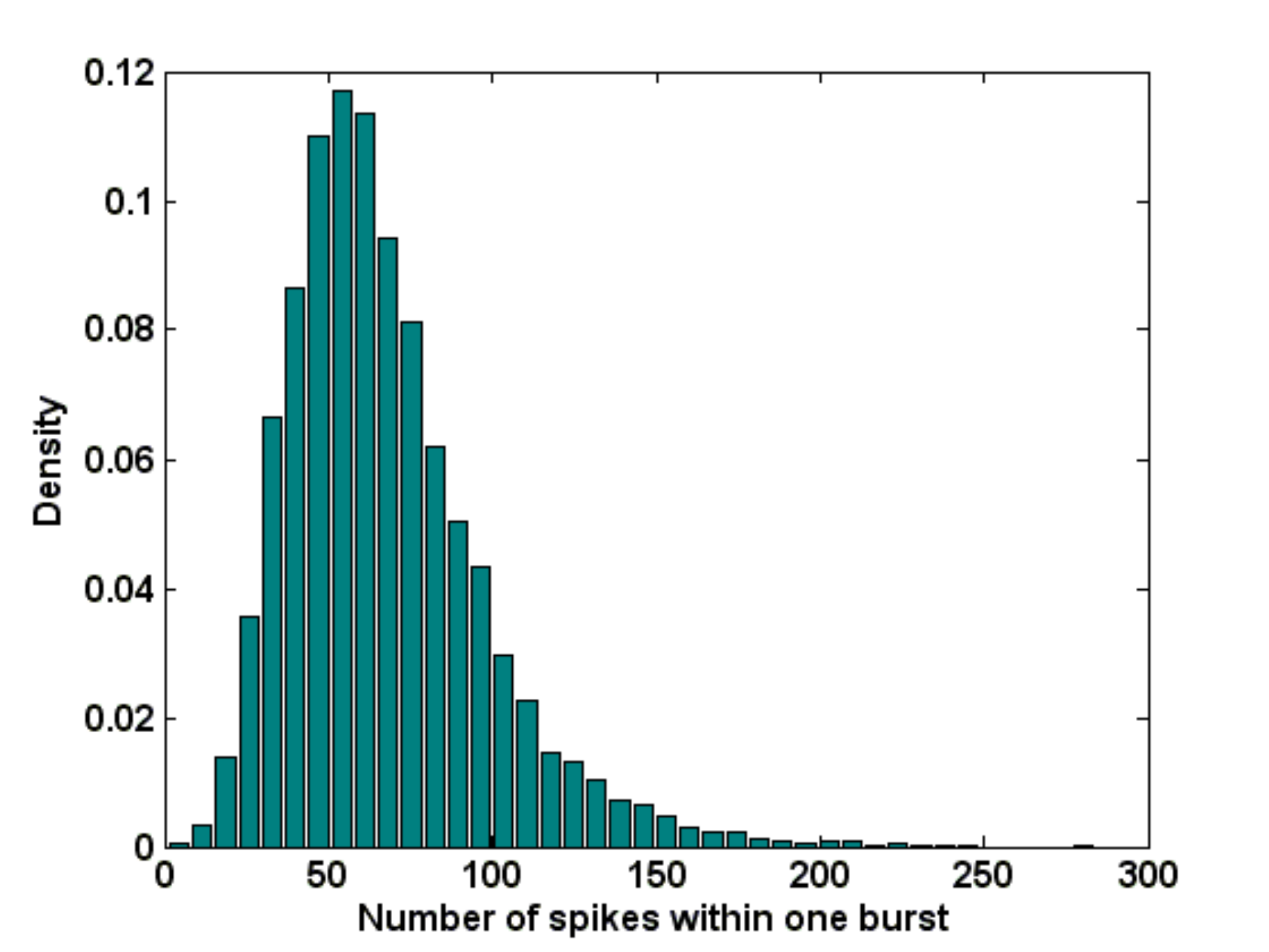, height=2.0in, width=2.0in}\\
{\bf d}\epsfig{figure=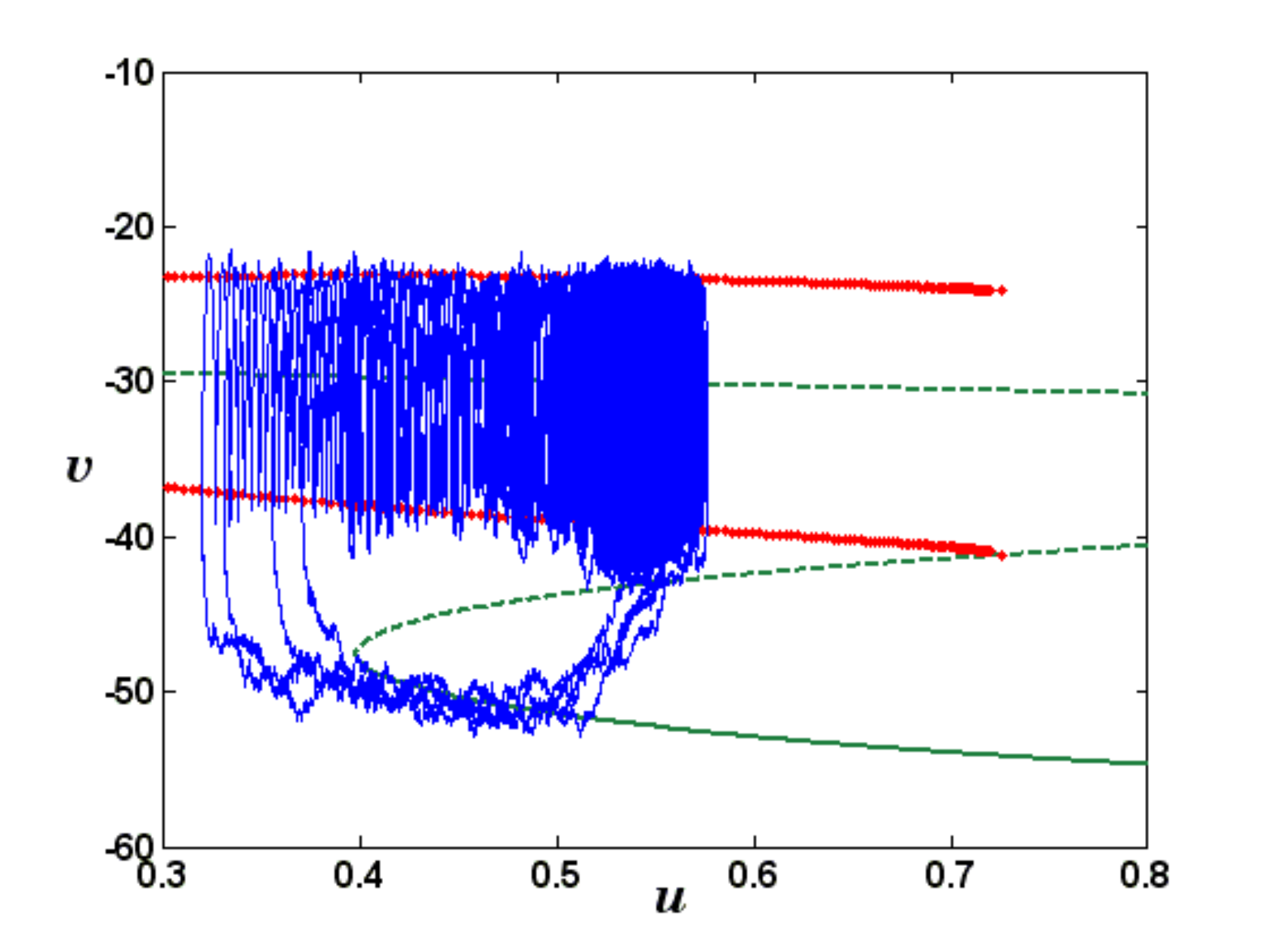, height=2.0in, width=2.0in}
{\bf e}\epsfig{figure=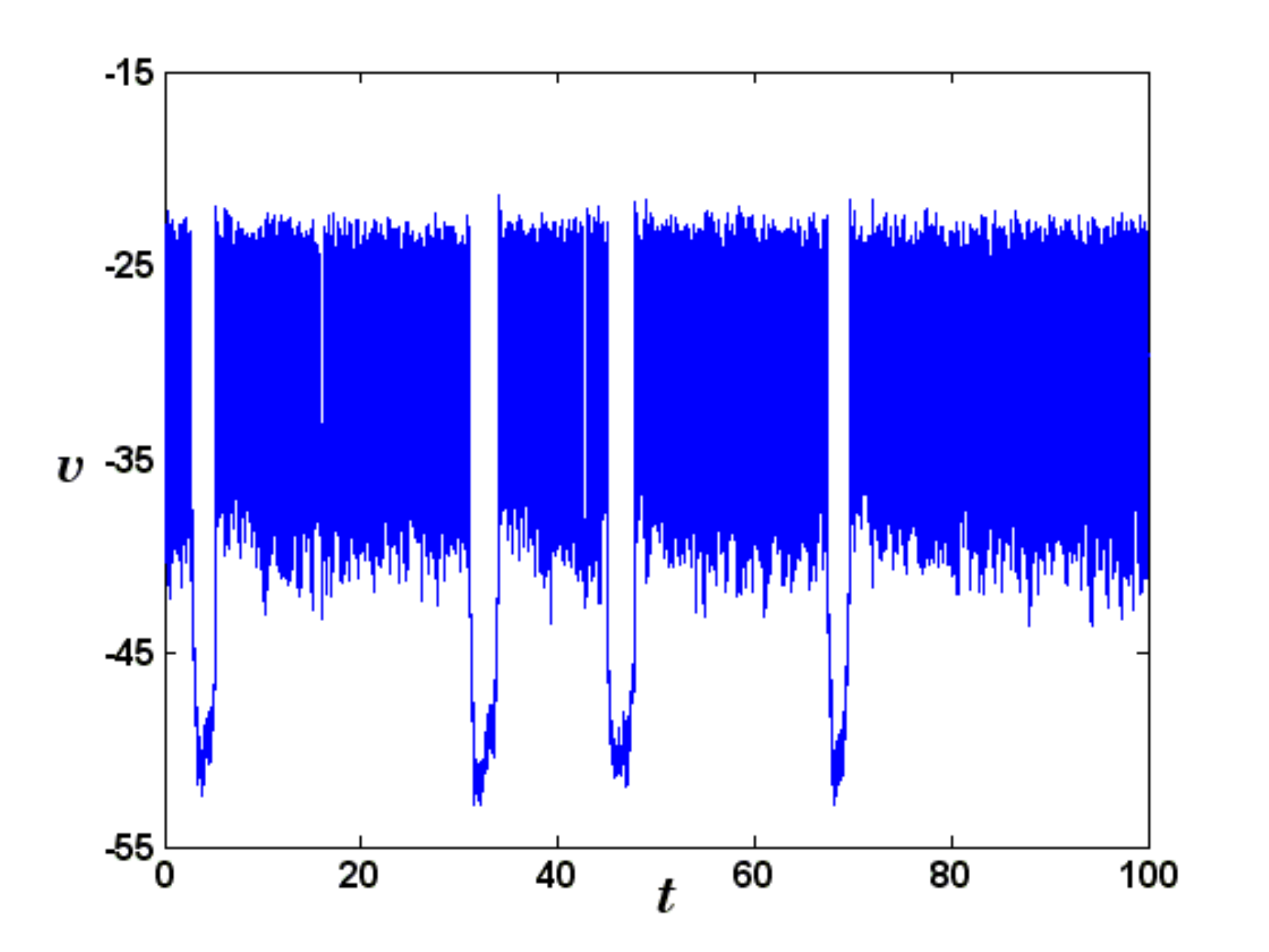, height=2.0in, width=2.0in}
{\bf f}\epsfig{figure=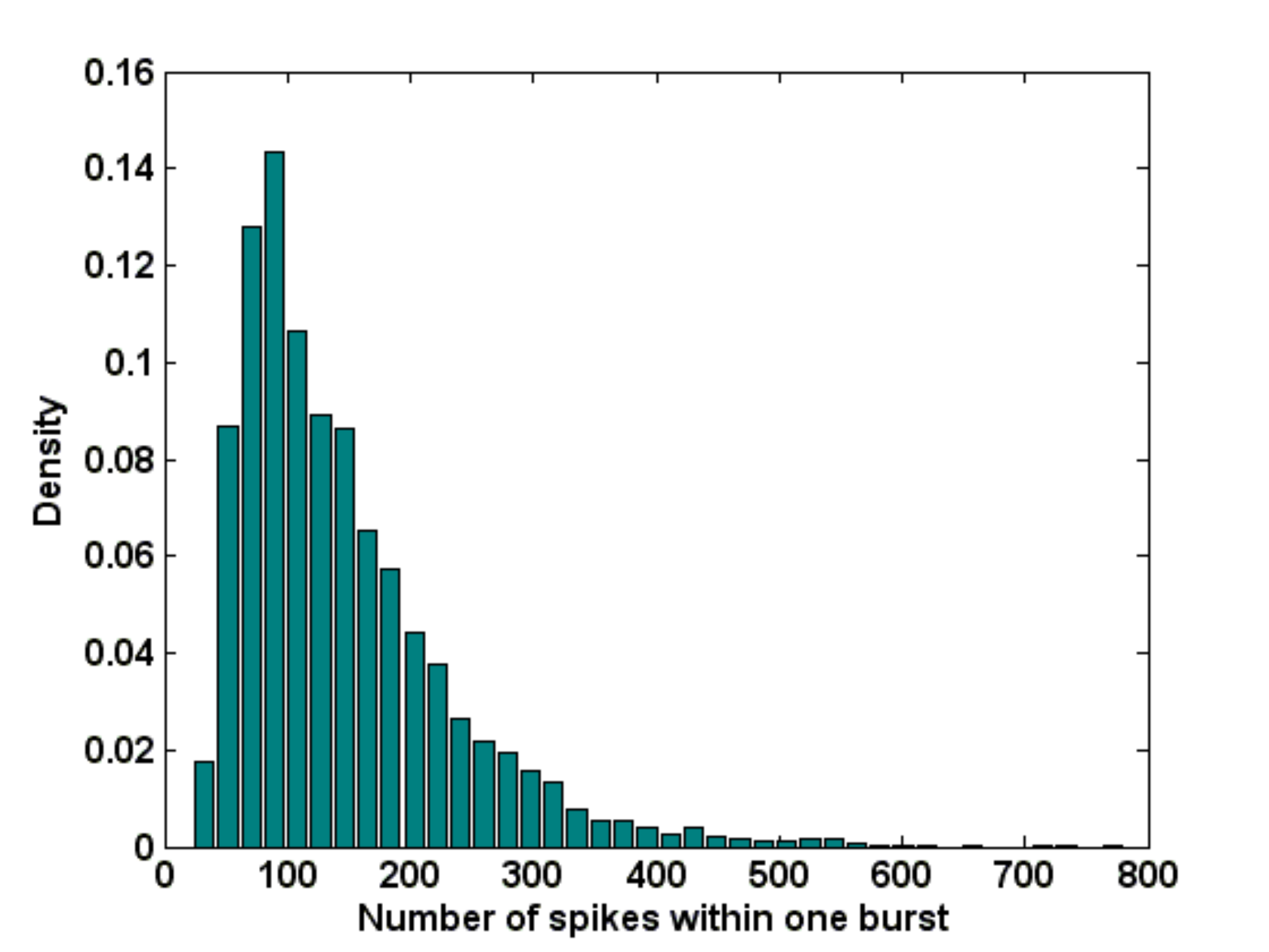, height=2.0in, width=2.0in}\\
{\bf g}\epsfig{figure=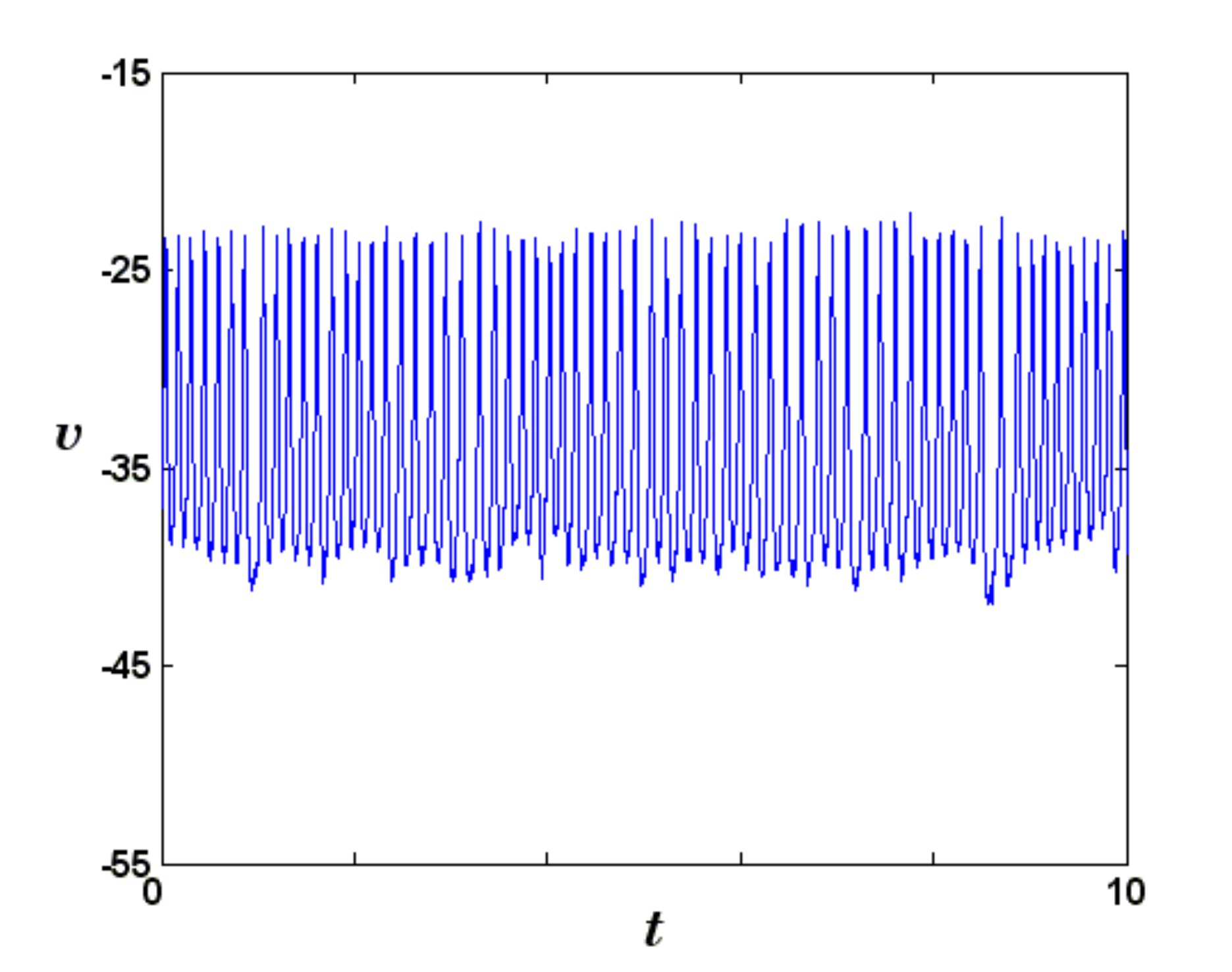, height=2.0in, width=2.0in}
{\bf h}\epsfig{figure=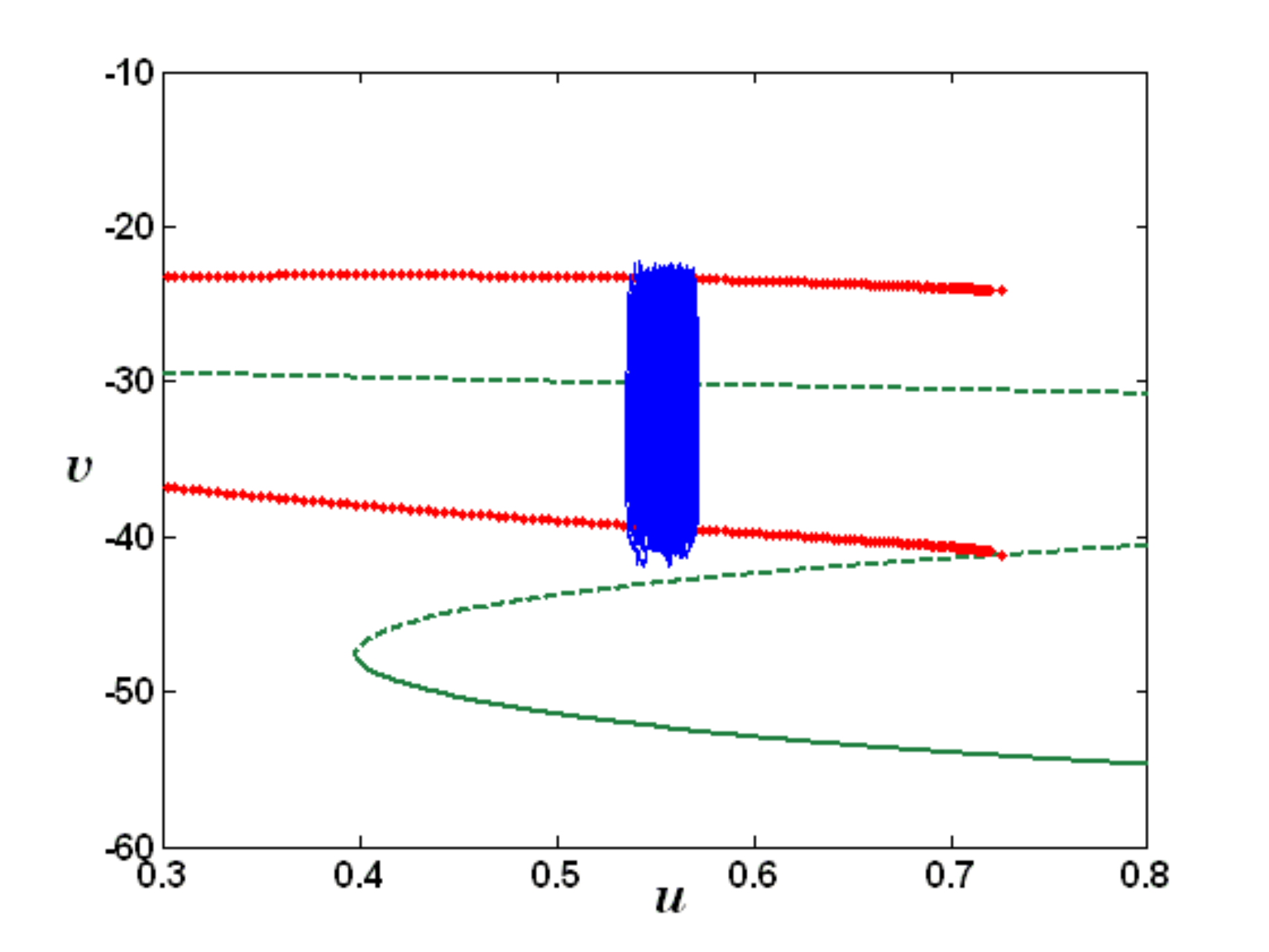, height=2.0in, width=2.0in}
{\bf i}\epsfig{figure=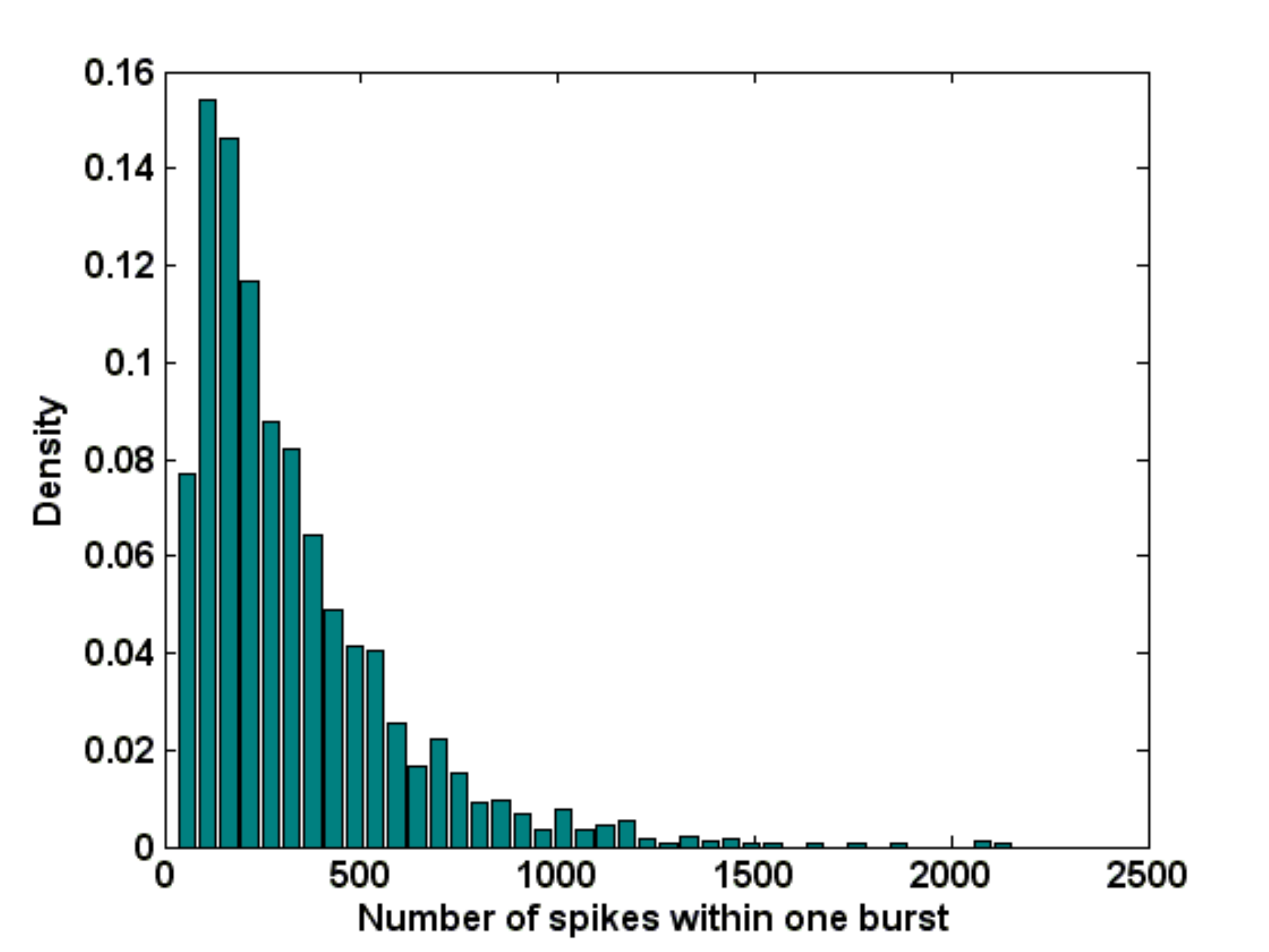, height=2.0in, width=2.0in}
\end{center}
\caption{Scenario B. The underlying deterministic system is 
in the spiking regime (see Fig.~\ref{f.regimes}b). 
Adding small noise transforms spiking into irregular bursting
shown in plots (a,b). Coupling two cells together already increases
the duration of bursts significantly (d,e). For ten coupled cells
the duration of typical bursts is extremely long so that for
all practical purposes (g,h), it can be considered as a transition to
spiking.
The normalized histograms for the number of spikes in
one burst are plotted for a single cell, two, and three 
coupled cells in c, f, and i respectively. We were unable
to numerically detect any bursts for the $10$ coupled cells
due to their extremely long duration. All histograms
show approximately geometric distribution confirming the large deviations
nature of the bursting patterns.  
The parameters of the geometric distribution (i.e., the
expected value of the number spikes per burst) increases
significantly as we go from one cell to three cell network (c,f,i).
}
\lbl{f.sceB}
\end{figure}

\subsection{Scenario B}
In this scenario, the deterministic single cell model is tuned to be in 
the spiking regime, i.e.,
it has a stable limit cycle $L_{y_c},\; y_c< y_{hc}$ in the vicinity of $L$ 
(see Fig.~\ref{f.regimes}b,e).
When noise is added to the modeling equations (\ref{ch.1})-(\ref{ch.3}), it forces the
trajectory to leave the basin of attraction of the limit cycle $\mathcal{B}(L_{y_c})$
once in a while (see Fig.~\ref{f.sceB}a), producing irregular bursting (Fig.~\ref{f.sceB}).
This mechanism of irregular bursting was studied in detail in \cite{HM}. 
In particular, it is shown in \cite{HM}
that the number of spikes in one burst has asymptotically geometric distribution with
parameter $p\approx \exp\{C_4\sigma^{-2}\}$ for some positive constant $C_4$.

The normalized histogram  in Fig.~\ref{f.sceB}c shows that with the level of noise
chosen for this experiment, the system generates very long irregular bursts. In the second
row in Fig~\ref{f.sceB}d-f, we show the results for two coupled cells and the coupling 
strength $g=50$. Taking just two cells already changes the statistics of bursting significantly
(compare  Fig.~\ref{f.sceB} c and f). For three coupled cells, the bursts become even longer
preserving the geometric distribution (see Fig. ~\ref{f.sceB} i). For ten coupled cells
and the coupling strength $g=500$, we were unable to detect a single 
termination of spiking activity. 
While the coupled system is technically still in the
bursting regime, the bursts become so long that for all practical purposes, one can speak   
about effective transition to spiking (see Fig.~\ref{f.sceB} g,h).

The changes in statistical properties of the firing patterns generated by the coupled 
system for increasing values of the coupling strength are due to denoising.
Here, electrical coupling diminishes the effects of noise
on synchronous limit cycle oscillations and results in longer times that the trajectory
of the coupled system spends in $\mathcal{B}(\mathbf{L}_{y_c})$ for larger values of $g$.
Qualitatively, this is the situation that we have already analyzed for Scenario A.
However, the study of the Scenario B involves certain
additional technical details needed for extending the analysis to systems of coupled
limit cycle oscillators (see \cite{medvedev10, medvedev10a}). We will consider the
analysis of Scenario B in the future work.

\section{Synchronization}
\lbl{synchronization}
\setcounter{equation}{0}

In this section, we study synchronization in electrically coupled
network (\ref{c.1}) and (\ref{c.2}). Our motivation is twofold. 
On the one hand, synchronization is an important aspect of the
network dynamics, because for sufficiently strong coupling
the network activity becomes synchronized. In particular, all
numerical results shown in this paper for the coupled system feature
synchronous activity. On the other hand, we show that the mechanism of 
synchronization is closely related to that of denoising.
In particular, the estimates of stochastic stability of synchrony,
which we derive below, reveal the contribution of the
network topology and the strength of connections distribution
to synchronization and 
denoising. We show that networks with larger algebraic 
connectivity and smaller total effective resistance are more
effective for implementing denoising. To illustrate this point,
we use random and symmetric degree-$4$ graphs defined in 
Example~\ref{ex.4}. Our analysis and numerics show that 
random graphs and, more generally, expanders have  
good synchronization and denoising properties. 
The analytical approach, which we develop in this section, complements
that of the previous section.

\subsection{The new coordinates}

To study synchronization, we introduce a
special system of coordinates, which to leading order decouples
the two principal modes of the system's dynamics:
fast synchronization and ultra slow excitation due to noise.
Remarkably, the latter is captured by a scalar stochastic 
ordinary differential equation. The slow-fast decomposition is
used to show that synchronization takes place for sufficiently 
strong coupling
and to quantify various aspects of the synchronized dynamics.

The new coordinate system takes into account the 
structure of the network. To this end, we will need 
a spanning tree of 
$\mathcal{G}$,
$\tilde{\mathcal{G}}=(V,\tilde E)$, i.e., a subgraph
of $\mathcal{G}$ with $n=|V(\mathcal{G})|$ vertices,
$n-1$ edges, and containing no loops \cite{Biggs}.
Having chosen the spanning tree $\tilde{\mathcal{G}}$, 
let $\tilde H\in\R^{(n-1)\times n}$
denote the coboundary matrix corresponding to  $\tilde{\mathcal{G}}$. 
Matrix $S$, which we introduce in the next lemma will be
useful for constructing the new coordinates. 

\begin{lem}\lbl{S}
Let 
\be\lbl{define-S}
S=(\tl H{\tl H}^\t)^{-1/2}\tl H.
\ee
Then
\be\lbl{study-S}
SS^\t=I_{n-1}\quad\mbox{and}\quad S\mathbf{1_n}=0.
\ee
\end{lem}
\pf\; Properties (\ref{study-S}) follow from the definition
(\ref{define-S}).\\
$\qed$

The synchronization subspace spanned by $\mathbf{1_n}$
coincides with the kernel of $S$, while the columns of 
$S^\t$ form an orthonormal basis of $\mathbf{1_n}^\perp$.
These two subspaces are important for studying synchronization,
which motivates the following coordinate transformation
\be\lbl{coordinates}
\R^n\ni z\mapsto (\xi,\eta)\in \R^{n-1}\times \R,
\ee
where
\be\lbl{xi-eta}
\xi= Sz\quad \mbox{and} \quad \eta=n^{-1}\mathbf{1_n}^\t z.
\ee
Note that $|\xi|=|P_{\mathbf{1_n}^\perp} z|$ measures
the distance of the solution of the coupled system
(\ref{rescale}) to the synchronization subspace
corresponding to $\xi=0$. Here, $P_{\mathbf{1_n}^\perp}$ stands
for the orthogonal projector onto $\mathbf{1_n}^\perp$.

\subsection{The slow-fast system}
Throughout this section, we work with the reduced (rescaled) 
system (\ref{rescale}), which we derived using the center
manifold approximation of (\ref{fast-coup}) near the excitable
equilibrium. In Section~\ref{transitions}, we analyzed
(\ref{rescale}) using its gradient structure and
the large deviations estimates. This time, we use 
a complementary approach by first identifying and then exploiting
the slow-fast structure of (\ref{rescale}).  

By projecting (\ref{rescale}) onto $\mathbf{1_n}^\perp$ 
and $\mathbf{1_n}$, we obtain
\begin{eqnarray}\lbl{xi}
\dot \xi &=& (-\gamma \hat L +2\eta I_{n-1})\xi +\sigma S\dot W +O(|\xi|^2),\\
\lbl{eta}
\dot \eta &=& f(\eta) +O(|\xi|^2) +{\sigma\over\sqrt{n}}\dot w,
\end{eqnarray}
where matrix 
$
\hat L= SLS^\t
$
is the unique solution of the matrix equation 
$$
SL=\hat LS.
$$
Here, $\dot W$ and $\dot w$ denote the Gaussian white noise
processes in $\R^n$ and $R^1$ respectively. In the derivation of 
(\ref{eta}), we use the fact that $\mathbf{1_n}^\t \dot W$
and $n^{-1/2}\dot w$ are identically distributed.
For details of the derivation of (\ref{xi}) and (\ref{eta}),
we refer the interested reader to Lemma 5.3 of \cite{MZ}.
About the reduced matrix $\hat L$,  the following is known:
$\hat L$ is a positive definite matrix, whose spectrum
consists of the nonzero eigenvalues of $L$ 
(cf. Lemma~2.6, \cite{medvedev10b}). Moreover, 
$S$ maps the generalized eigenspaces of 
the nonzero eigenvalues of $L$ bijectively onto those of $\hat L$ 
\cite{medvedev10b}.

Equation (\ref{xi}) captures the dynamics along the orthogonal
complement of the synchronization subspace $\mathbf{1_n}$.
Thus, it describes the process of synchronization.
On the other hand, Equation (\ref{eta}) tracks the motion along 
the synchronization subspace. Since $\hat L$ is positive
definite, in the strong coupling regime ($\gamma\gg 1$), it follows 
from (\ref{xi}) that the trajectory of the full system (\ref{xi})
and (\ref{eta}) relaxes to an $O(\sigma)$ neighborhood
of the synchronization subspace, $\mathbf{1_n}$, at $O(\gamma)$ rate. 
Equation (\ref{eta}) to leading order (with $\xi\approx 0$) is 
a standard model of a particle in a potential well forced by noise. 
On the time intervals not exceeding the Kramer's time $O(\exp\{C_5\sigma^{-2}\})$
for some $C_5>0$, $\eta$ remains close to $-1$ for the most of the time. 
This sums up the qualitative dynamics of (\ref{xi}) and (\ref{eta}).
In the remainder of this section, we study it in more detail.

\subsection{The fast subsystem: synchronization}

The analysis of the fast subsystem  elucidates
several important aspects of synchronization in electrically coupled networks.
In particular, we estimate the rate of synchronization in terms of the network
connectivity. We then study robustness of synchrony to noise.

The stability of the synchronization subspace is determined by the 
linear part of (\ref{xi}): 
\be\lbl{lead-xi}
\xi=-\gamma \hat L\xi +\sigma S\dot W.
\ee
Assuming for simplicity the deterministic initial condition $\xi(0)=\xi_0\in\R^{n-1}$, 
we compute the mean vector and the covariance matrix of $\xi(t)$:
\begin{eqnarray}\lbl{mean}
\E\xi(t)&=& \exp\{-\gamma t \hat L\} \xi_0,\\
\lbl{cov}
\cov \xi(t) &=& \sigma^2 \int_0^t \exp\{-2\gamma \hat L(t-u)\} du =
{\sigma^2\over 2\gamma}{\hat L}^{-1}
\left(I_{n-1}-\exp\{ -2t\gamma\hat L\}\right).
\end{eqnarray}

It follows from (\ref{mean}) and the geometric interpretation of $\xi$,
that the rate of convergence to the synchronization
subspace is set by the product of the strength of coupling
$\gamma$ and the algebraic connectivity $\mathfrak{a}(\mathcal{G})$:
\be\lbl{rate-sync}
\left|\E P_{\mathbf{1_n}^\perp} z(t)\right|\le 
C_6\exp\{-\gamma\mathfrak{a}(\mathcal{G})t\},
\ee
where $P_{\mathbf{1_n}^\perp}\cdot$ stands for the orthogonal projection 
onto $\mathbf{1_n}^\perp$ and $C_6$ is a positive constant independent
from $\gamma$ and $\mathfrak{a}$.
Therefore, networks with larger algebraic connectivity synchronize faster.
There are many fine results in the spectral graph theory relating algebraic
connectivity to various structural properties of the network (see \cite{Hoory06}
and references therein). These results can be used via (\ref{rate-sync})
to elucidate the contribution of the network topology to synchronization
properties of the coupled system.
In particular, (\ref{rate-sync}) shows that networks on random
graphs and on expanders in general (cf. Example~\ref{ex.4}) 
admit a lower bound on the synchronization rate, which is uniform
in the size of the network $n$. 

In the presence of noise, it is important to know how well the synchrony 
can withstand stochastic perturbations. This leads to the question of stochastic 
stability. There are many ways for measuring stability of synchrony to random
perturbations \cite{Hasminsky}. In this paper, we use the mean square stability, which provides
a natural metric for the problem at hand. Specifically, we are interested in transverse 
stability of the synchronization subspace. From (\ref{rate-sync}), we know that
$$
\E P_{\mathbf{1_n}^\perp} z(t) \to 0,\;\mbox{as}\; t\to\infty.
$$
Thus, we next look at the second moments, i.e., we estimate the dispersion of the
trajectories around $\mathbf{1_n}$:
\be\lbl{var}
\var P_{\mathbf{1_n}^\perp} z(t)=\sum_i^{n-1}\var \xi_i(t)=:\var\xi(t).
\ee
From (\ref{cov}), we find that 
$$
\var\xi(t)=\tr~\cov\xi(t)={\sigma^2\over 2\gamma} \tr\left\{{\hat L}^{-1}
\left(I_{n-1}-\exp\{ -2t\gamma\hat L\}\right)\right\}.
$$
Since $\hat L$ is positive definite, $\var\xi(t)$ has a finite asymptotic
value 
\be\lbl{lim-var}
\overline{\var} \xi:=\lim_{t\to\infty} \var \xi(t)=
{\sigma^2\over 2\gamma}\tr~{\hat L}^{-1}=
{\sigma^2\over 2n\gamma }\mathcal{R}(\mathcal{G}),
\ee
where $\mathcal{R}(\mathcal{G})$ is  the total effective resistance 
of the weighted graph of the network $\mathcal{G}$  (cf. \ref{resistance}).
Thus, $\overline{\var} \xi$ provides a convenient measure of stochastic
stability of the synchronization subspace. The smaller the value of 
$\overline{\var} \xi$, the more stable the synchrony is.
In the next section, we will
use the asymptotic value of the variance to estimate the 
effectiveness of denoising. 

Estimate (\ref{lim-var}) shows that the stochastic stability of the 
synchronization subspace
is fully determined by the strength of coupling and the total
effective resistance of the network. 
Similarly to the algebraic connectivity of the graph,
the value of the total effective resistance $\mathcal{R}(\mathcal{G})$ 
can be related to the structure of the graph and the weight distribution 
\cite{Boyd08}.
However, while the rate of convergence 
to the synchronization subspace depends only on the leading nonzero eigenvalue
of the graph Laplacian, the description of stochastic stability requires
the entire spectrum of $\mathcal{G}$. The information about higher eigenvalues of $\mathcal{G}$
in general is hard to obtain. 
However, as one can see from  the following crude estimate of 
$\mathcal{R}(\mathcal{G})$ 
\be\lbl{rough}
\mathcal{R}(\mathcal{G})\le {n^2\over\mathfrak{a}(\mathcal{G})},
\ee
graphs with larger algebraic connectivity enjoy
better bounds on the total effective resistance. In fact, 
for expanders, from (\ref{rough}) one gets 
a bound on $\mathcal{R}(\mathcal{G})=O(n^2)$,
which can not be improved (up to a multiplicative constant). 
Therefore, we expect that graphs with good bounds on algebraic
connectivity, like a random graph in Example~\ref{ex.4},
are robust against random perturbations. Below, we will illustrate
this point with numerical results.

Above we have used the analysis of the fast subsystem to gain  useful 
information about synchronization properties
of the coupled network. Further, with these  results at hand
one can easily understand the $1D$ slow equation (\ref{eta})
and get a  complete description of the slow-fast system (\ref{xi}) and
(\ref{eta}). This is our next step.

Since on time intervals not exceeding the Kramer's time, 
with high probability $\xi(t)=O(\sigma)$, we 
approximate (\ref{eta}) by
\be\lbl{lead-eta}
\dot \eta = f(\eta) +{\sigma\over\sqrt{n}} \dot w.
\ee 
From (\ref{lead-eta}), we estimate the time that a typical trajectory 
spends  in the basin of the rest state $\eta=-1$
\be\lbl{tau2}
\tau ({-1}, \eta_0)\asymp \exp\left\{ {8n\over 3\sigma^2}\right\}.
\ee
Comparison of (\ref{var-as-coup}) and (\ref{tau2}) shows that the results of 
the fast-slow analysis are consistent with  those of  the previous section. 
Furthermore, through the estimates of stability of the synchronization
subspace (\ref{rate-sync}) and (\ref{lim-var}) the analysis of this
section elucidates the contribution of the structural properties 
of the network to its synchronization properties. 

\section{The mechanism of denoising}
\lbl{denoising}
\setcounter{equation}{0}

In this section, we continue to study denoising, the mechanism responsible for
variability of activity patterns in electrically coupled network 
(\ref{net.1})-(\ref{net.2}). The analysis
of synchronization and denoising in the previous sections relied on the proximity
of the models of individual cells to the saddle-node bifurcation. To clarify to what extent
the mechanism of denoising depends on this assumption  and to elucidate
the scope of its applicability, in this section, we analyze the 
coupled system around the stable fixed point without assuming anything about the
bifurcation structure of the problem. 

Because the dependence of the single cell models on the slow variable is
not essential for the mechanism of denoising per se, we omit to indicate it explicitly 
to simplify notation.
Thus, in this section, we consider
\be\lbl{vect-loc}
\dot x= f(x)+\Sigma\dot w,\quad x\in\R^d,
\ee
where $f:~\R^d\to\R^d$ is a smooth function, $w$ is a standard Brownian
motion in $\R^d$, and $\Sigma\in \R^{d\times d}$ is a nonsingular
matrix.  We assume that 
(\ref{vect-loc}$)_0$ has a stable equilibrium at the origin,
i.e.,
\be\lbl{equil}
f(0)=0,\quad -A=Df(0),
\ee
where $A^s:=0.5(A+A^\t)>0$ is a positive definite matrix.
The coupled system has the following form
\be\lbl{vect-coup}
\dot X=F(X)-g(L\otimes J)X+(I_n\otimes\Sigma) \dot W,
\ee
where as before $X=(x^{(1)}, x^{(2)},\dots, x^{(n)})^\t\in\R^d\times\R^n\cong\R^{nd}$,
$F(X)=(f(x^{(1)}), f(x^{(2)}),\dots, f(x^{(n)}))^\t$, $L=H^\t CH$, and
$J$ is  $d\times d$ matrix, such that $J^s$ is positive definite. 
Finally, $W$ stands for the Brownian motion in $\R^{nd}$
The coupled system has an equilibrium at the origin in $\R^{nd}$. 
The linearization
of (\ref{vect-coup}) about this equilibrium is given by
\be\lbl{linearize}
\dot X =-gBX +(I_n\otimes\Sigma)\dot W,\quad
B=\hat L\otimes J+\delta I_n\otimes A,\; \delta=g^{-1}.
\ee

In the previous sections, to quantify the effect of denoising
we used the times of the first exit from the basins of stable
equilibria of the single cell and coupled models (cf.~(\ref{var-as-coup}) 
and (\ref{tau2})).
Since in this section we do not assume that the local 
systems are located near a saddle-node bifurcation,  we no longer can
rely on explicit estimation 
of the first exit times. Thus, we seek other means for measuring
the effectiveness of denoising. To this end, we recall that 
in the previous section we found that
the asymptotic value of  the variance of the trajectories near
the synchronization subspace to provide a convenient measure for
estimating stochastic stability. We thus adapt it to our current
purpose. Specifically, we use 
$$
\max_{i\in [n]} \overline{\var} x^{(i)}=
\max_{i\in [n]}\lim_{t\to\infty}\sum_{j=1}^d \var x^{(i)}_j(t) 
$$
to measure the variability of the trajectories of the
coupled system $X(t)=(x^{(1)}(t), x^{(2)}(t),\dots,x^{(n)}(t))$.
We estimate the effect of denoising by comparing 
$
\max_{i\in [n]} \overline{\var} x^{(i)}
$
to $\overline\var x$, where $x(t)$ is the solution of the local
subsystem (\ref{vect-loc}). We identify conditions which guarantee that
$$
\max_{i\in [n]} \overline{\var} x^{(i)}\ll \overline\var x,
$$
and show how the former quantity depends on the coupling strength,
network size and topology, as well as on the intrinsic properties
of the local system (\ref{vect-loc}).

The following lemma is the key for understanding the mechanism 
of denoising. 
\begin{lem}\lbl{var-asymp}
\be\lbl{triangle1}
\max_{i\in [n]}\overline{\var} x^{(i)} \le  {\sigma^2\over 2}
\left({1\over n \lambda_1(A^s)} +{n-1\over g\lambda_1(B^s)}\right),
\ee
where $\sigma=|\Sigma|_F$ stands for the Frobenius norm of $\Sigma$,
$B$ is defined in (\ref{linearize}), and
$\lambda_1(\cdot)$ denotes the smallest eigenvalue of a symmetric matrix.
\end{lem} 
In the process of proving Lemma~\ref{var-asymp}, we derive the 
following estimates, which are of independent interest.
\begin{cor}\lbl{cor-sync}
\begin{eqnarray}\lbl{triangle}
\overline{\var} X &\le & {\sigma^2\over 2}
\left({1\over \lambda_1(A^s)} +{n-1\over g\lambda_1(B^s)}\right),\\
\lbl{vect-sync}
|\E P_{\mathbf{1_n}^\perp} X(t)| &\le &C\exp\{-g\lambda_1(B^s)t\}\quad\mbox{and}\quad
\overline{\var} P_{\mathbf{1_n}^\perp} X(t)={(n-1)\sigma^2\over 2g\lambda_1(B^s)}.
\end{eqnarray}
\end{cor}
\pf\; The proof of Lemma~\ref{var-asymp} consists of the following steps.
\begin{enumerate}
\item
First, we separate the dynamics along the synchronization subspace from that
along its orthogonal complement. To this end, we 
switch to new coordinates
\be\lbl{vect-new}
X\mapsto (Y,Z), \; Y=(S\otimes I_d)X\;\mbox{and}\; Z=(n^{-1}\mathbf{1_n}^\t\otimes I_d) X.
\ee
By multiplying both sides of (\ref{linearize}) by $S\otimes I_d$ and 
$(n^{-1}\mathbf{1_n}\otimes I_d)^\t$,
we obtain the equations for $Y$ and $Z$
\begin{eqnarray}\lbl{Y}
\dot Y &=& -gBY+(S\otimes\Sigma)\dot W,\; B= \hat L\otimes J+\delta (I_{n-1}\otimes A), \\
\lbl{Z}
\dot Z &=& -AZ +n^{-1/2}\Sigma \dot w.
\end{eqnarray}

\item
Recall that $A^s$ is a positive definite matrix and  the smallest EV of $A^s$ denoted
by $\lambda_1(A^s)$. For the mean vector of the solution (\ref{Z})
subject to the deterministic initial condition $Z_0$
\be\lbl{mean-Z}
|\E Z|=|\exp\{-tA\}Z_0|\le C\exp\{-\lambda_1(A^s)t\} |Z_0|,
\ee
for some positive constant $C$. Next, we estimate
\begin{eqnarray}\nonumber
\var Z &=& n^{-1}\tr~\left(\int_0^t \exp\{(t-s)A\} \Sigma\Sigma^\t\exp\{(t-s)A^\t\} ds\right)\\
\lbl{var-Z}
       &\le &  n^{-1}\tr~(\Sigma\Sigma^\t) \int_0^t\exp\{2\lambda_1(A^s)s\}ds\to 
(2n|\lambda_1(A^s)| )^{-1}\sigma^2\;\;\mbox{as}\;\; t\to\infty.
\end{eqnarray}

\item 
Similarly,
\begin{eqnarray}\lbl{mean-Y}
|\E Y|&=&|\exp\{-tgB\}Y_0|\le C\exp\{-g\lambda_1(B^s)t\} |Y_0|\;
\mbox{and}\;\\
\lbl{var-Y}
\overline{\var} Y &=& {|S\otimes \Sigma|^2_F\over 2g\lambda_1(B^s)}=
{(n-1)\sigma^2\over 2g\lambda_1(B^s)}. 
\end{eqnarray}
\item
Using (\ref{vect-new}), we express $X$ in terms of $Y$ and $Z$  
\be\lbl{sum}
X=(S^\t\otimes I_d)Y + (\mathbf{1_n}\otimes I_d)Z=: M_1 Y + M_2 Z.
\ee
The definition of $S$ implies
\be\lbl{M1M2}
M_1^\t M_1=I_{(n-1)d}\;\mbox{and}\; M_2^\t M_2 =nI_d.
\ee
By (\ref{sum}) and (\ref{M1M2}), we have
\begin{eqnarray}\nonumber
\tr~\E XX^\t &=& \tr\{M_1\E(YY^\t)M_1^\t + M_2\E(YY^\t)M_2^\t\}\\
\lbl{linearity}
&=&\tr\{M_1^\t M_1\E(YY^\t) + M_2^\t M_2\E(YY^\t)\}\\
\nonumber
& =& \tr\{\E(YY^\t)+ n\E(ZZ^\t)\}.
\end{eqnarray}
Using the elementary properties of $\cov$, (\ref{mean-Z}), (\ref{mean-Y}),
and (\ref{linearity}), we have
\be\lbl{var-X}
\overline{\var} X=\overline{\var}Y +n\overline{\var}Z.
\ee
The combination of (\ref{var-Z}), (\ref{var-Y}), and (\ref{var-X}), shows (\ref{triangle}).

\item 
Since $X=(x^{(1)},x^{(2)},\dots, x^{(n)})$, from (\ref{sum}) we have
\be\lbl{sumi}
x^{(i)}=z+(\mbox{Row}_i(S^\t)\otimes I_d)Y=:z+N_i Y.
\ee
By noting
$$
N_i^\t N_i=I_d,\; \mbox{and}\; zY^\t=0,
$$
from (\ref{sumi}) we have
$$
\tr~\E x^{(i)}{x^{(i)}}^\t= \tr~\E\{zz^\t + YY^\t\}.
$$
Thus,
\be\lbl{var-xi}
\overline{\var} x^{(i)}= \overline{\var}z + \overline{\var}Y.
\ee
Using (\ref{var-xi}), (\ref{var-Z}), and (\ref{var-Y}), we obtain (\ref{triangle1}).
\end{enumerate}
$\qed$

We want to understand how the variability of the coupled system can be smaller
than that of a local subsystem. Estimate (\ref{triangle1}) suggests a possible
scenario. Note that the first term on the right hand side of (\ref{triangle1})
can be made arbitrarily small by increasing $n$. The second term can be controlled
by $g$ provided that $\lambda_1(B^s)$ remains $O(1)$ as $g\to\infty$. 
Thus, we need to understand the behavior of $\lambda_1(B^s)$ for increasing
values of $g$. In the following lemma, we show that the latter depends 
on the coupling architecture and identify two cases of full and partial coupling,
which are important in the context of denoising.
\begin{lem}\lbl{perturb}
Let
$$
B=\hat L\otimes J +\delta (I_{n-1}\otimes A),
$$
where $J^s\ge 0, A^s>0,$ and $0<\delta\ll 1$ (cf.~(\ref{linearize})). If $J^s$ is a full rank
matrix then
\be\lbl{correction}
\lambda_1(B^s)=\lambda_1(\hat L)\lambda_1(J^s) + O(\delta).
\ee
Otherwise, let $k=\dim\ker J^s>0$, choose $\{ q_1,q_2,\dots, q_k\}$
an orthonormal basis for $\ker J^s$, and define an $k\times k$ matrix
\be\lbl{define-G}
(G)_{ij}=q_j^\t A^sq_i, \;\; (i,j)\in[k]\times [k].
\ee
Then
\be\lbl{correction1}
\lambda_1(B^s)=\delta\lambda_1(G)+O(\delta^2).
\ee
\end{lem}
\pf\;
For small $\delta>0$, the EVs of $B$ perturb smoothly from the EVs of
$\hat L\otimes J^s$ (cf.~\cite{gelfand}). If $J^s>0$ then 
$\lambda_1(\hat L\otimes J^s)=\lambda_1(\hat L)\lambda_1(J^s)$
and (\ref{correction}) follows.  Otherwise, $0$ is an EV of   $\hat L\otimes J^s$
of multiplicity $k(n-1)$.  We construct an orthonormal basis for the 
$0-$eigenspace of $\hat L\otimes J^s$
\be\lbl{basis}
h_i=e_{i_1}\otimes q_{i_2},\; i=(i_2-1)k+i_1, \; i_1\in [n-1], i_2\in [k],
\ee
where $e_i=(\delta^i_1,\delta^i_2,\dots,\delta^i_{n-1})$ and $\delta^i_j$
stands for the Kronecker delta.
By the perturbation results for the multiple eigenvalues of symmetric matrices
(cf. Appendix, \cite{gelfand}), we have
$$
\lambda_1(B^s)=\delta \lambda_1 (\mathbf{G})+O(\delta^2),
$$ 
where $\mathbf{G}$ is an $(n-1)d\times (n-1)d$ matrix 
$$
(\mathbf{G})_{ij}= h_j^\t (I_{n-1}\otimes A) h_i.
$$
It is easy to see that $\mathbf{G}=I_{n-1}\otimes G$ with $G$ defined in (\ref{define-G}).
Therefore, the EVs of $\mathbf{G}$ and $G$ coincide. 
This shows (\ref{correction1}).\\
$\qed$

In conclusion, we discuss the implications of the analysis of this section.
There are two effects that contribute to denoising: averaging and synchronization.
The former can be interpreted  as a manifestation of the law of large numbers: 
the combined effect of independent stochastic processes acting on individual cells 
vanishes as the size of the network goes to infinity. It is captured by the 
first term on the right hand side of (\ref{triangle1}). From (\ref{triangle1})
it follows that averaging depends on the network size, $n$, and the dissipativity
of the local dynamics through $\lambda_1 (A^s)$ but is independent from the 
properties of the coupling operator. The second term on the right hand side of 
(\ref{triangle1}), which captures synchronization, depends on the dissipativity 
of the coupling operator through
$\lambda_1(B^s)$, the coupling strength $g$, and the size of the network.
Denoising takes place when the variance of the trajectories of the coupled system
can be controlled by $n$ and $g$. In fact, we can always make the
first term in (\ref{triangle1}) arbitrarily small by taking $n$ large enough.
To control the second term on the right hand side of (\ref{triangle1}),
we can use $g$ provided  $\lim_{g\to\infty} \lambda_1(B^s)>0$. Thus, one
can make the right hand side of (\ref{triangle1}) arbitrarily small by increasing 
$n$ and $g$ provided $\lambda_1(B^s)$ is bounded away from $0$ as $g\to\infty$.

Lemma~\ref{perturb} identifies two cases important in the context
of denoising and synchronization. Depending on the rank of 
$J^s$, the leading eigenvalue
of $B^s$, $\lambda_1(B^s)$, is either $O(1)$ or $O(\delta)$ for $g\gg 1$.
If the coupling is full rank then $\lambda_1(B^s)=O(1)$ and (\ref{triangle1}) shows 
that the variability of large coupled systems can be effectively
controlled by the coupling strength and the network size. However, if the 
coupling  is partial ($\mbox{rank}~J^s<d$), as in the model of gap-junctionally
coupled network (\ref{net.1})-(\ref{net.3}), 
(\ref{triangle1}) becomes
\be\lbl{worse}
\max_{i\in [n]} \overline{\var} x^{(i)} \le {\sigma^2\over 2}
\left( {1\over n\lambda_1( A^s)}+ {n-1\over\lambda_1 (G^s) +O(\delta)}\right)
\le {\sigma^2n\over 2\lambda_1( A^s)} +O(\sigma^2\delta n)
\footnote[1]{The second inequality in (\ref{worse}) follows from 
(\ref{define-G})
and the variational properties of the EVs of symmetric matrices
$$
\lambda_1(G)=\min_{x\in\ker J^s,\; |x|=1} x^\t A^sx\ge 
\min_{x\in\R^d} x^\t A^sx =\lambda_1(A^s).
$$
}
\ee
This shows that in the partial coupling case the mechanism of denoising may fail.
Our numerical experiments and the analysis of 
(\ref{net.1})-(\ref{net.3}) show that denoising is at work despite the fact 
that the coupling is partial in this model.  We are able to observe
denosing  in (\ref{net.1})-(\ref{net.3}) because of the additional structure of this problem -
the proximity of the models of individual cells to the saddle-node bifurcation. This additional
feature of the problem affords center manifold reduction, and the coupling of the
reduced system is already full rank. This reconciles the results of this section
with the previous analysis and our numerical results.  
This also highlights
the importance of the bifurcation structure of the fast subsystem
of the bursting cell model (\ref{ch.1})-(\ref{ch.3}) for the mechanism
of denoising. More generally, this discussion implies that 
while the mechanism of denoising can be already studied at the level of the linearized system, 
the structure of the nonlinear system is nonetheless important for the realization of denosing 
in concrete (biophysical) models.

\begin{figure}
\begin{center}
{\bf a}\epsfig{figure=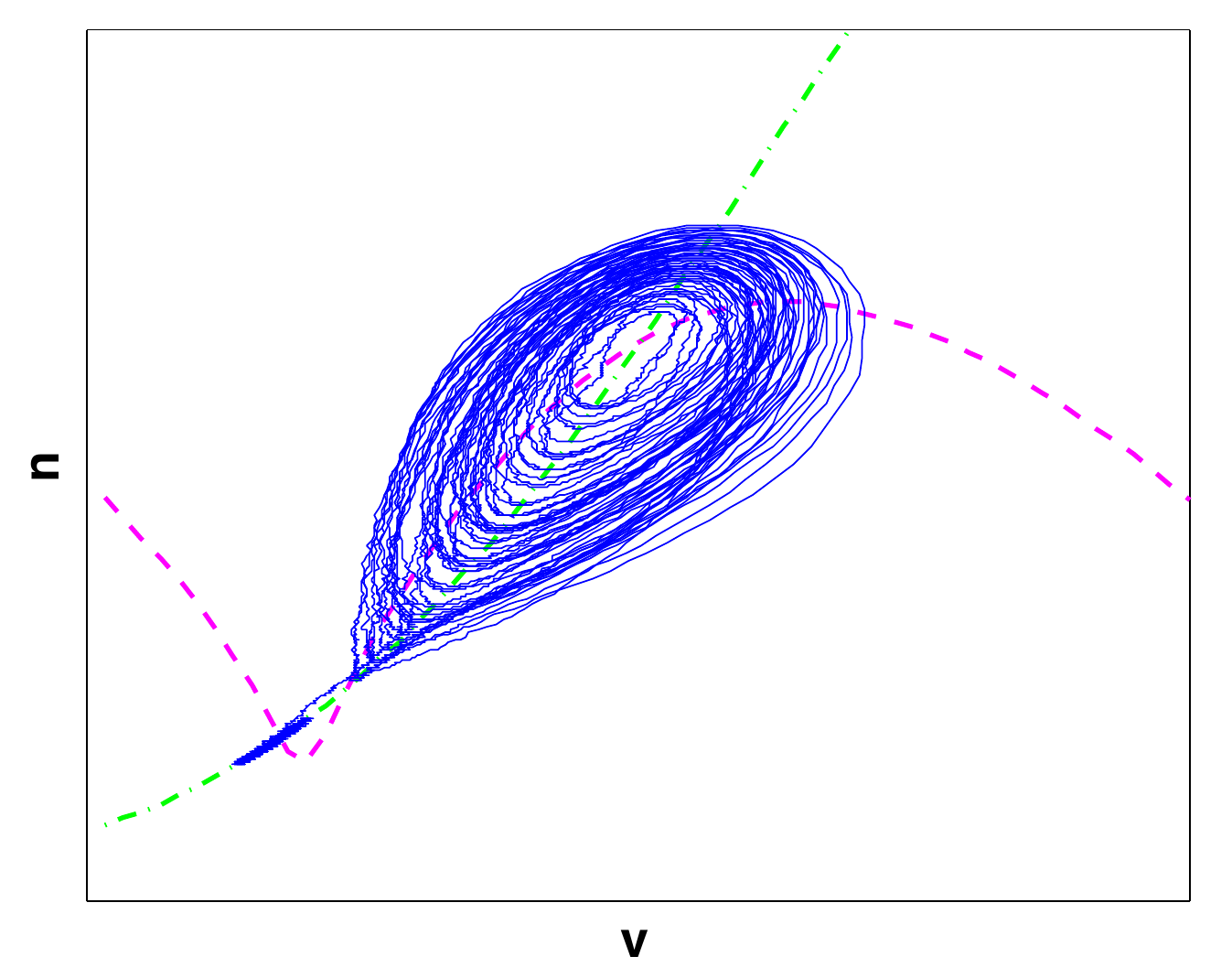, height=2.0in, width=2.5in}\qquad
{\bf b}\epsfig{figure=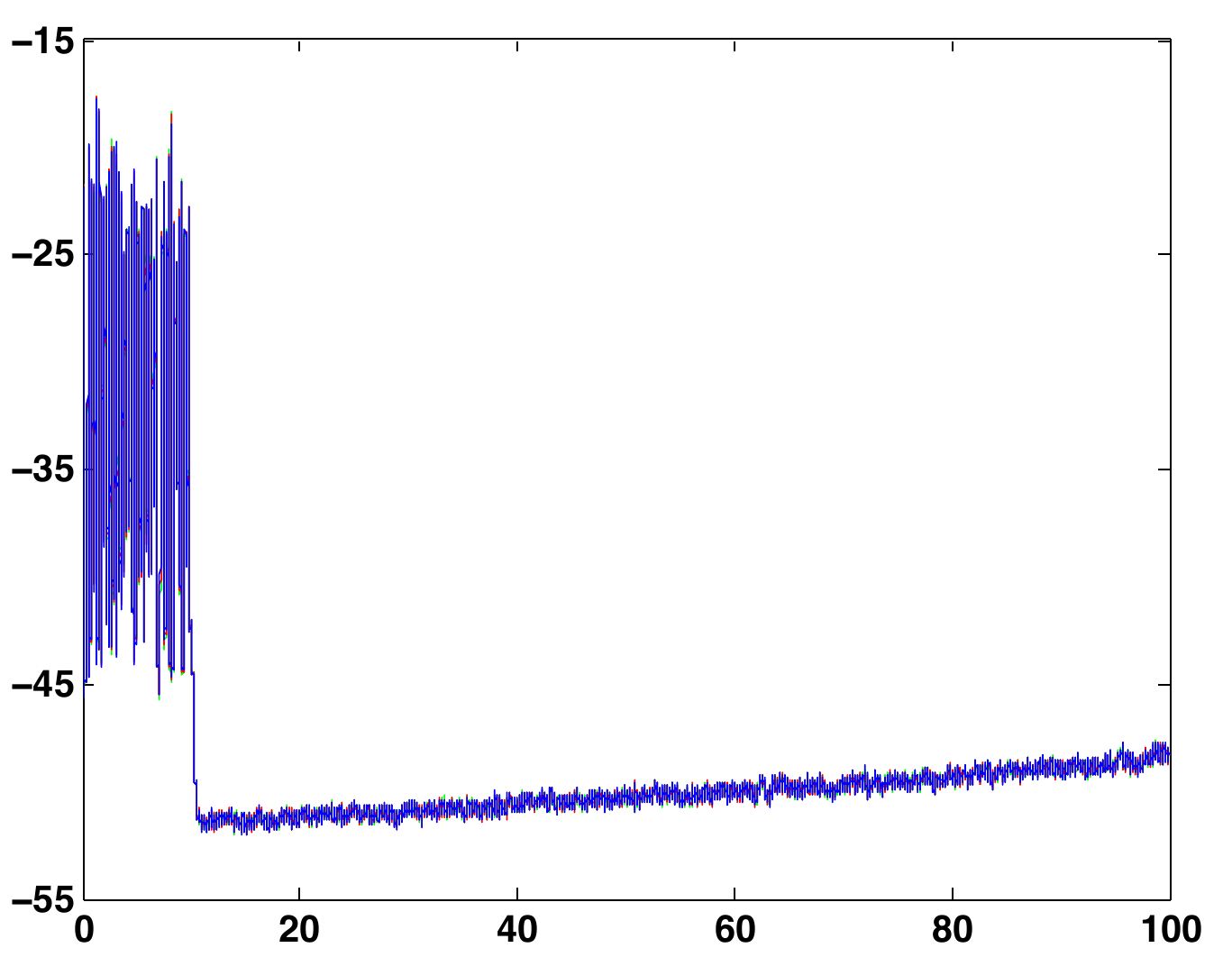, height=2.0in, width=2.5in}\\
{\bf c}\epsfig{figure=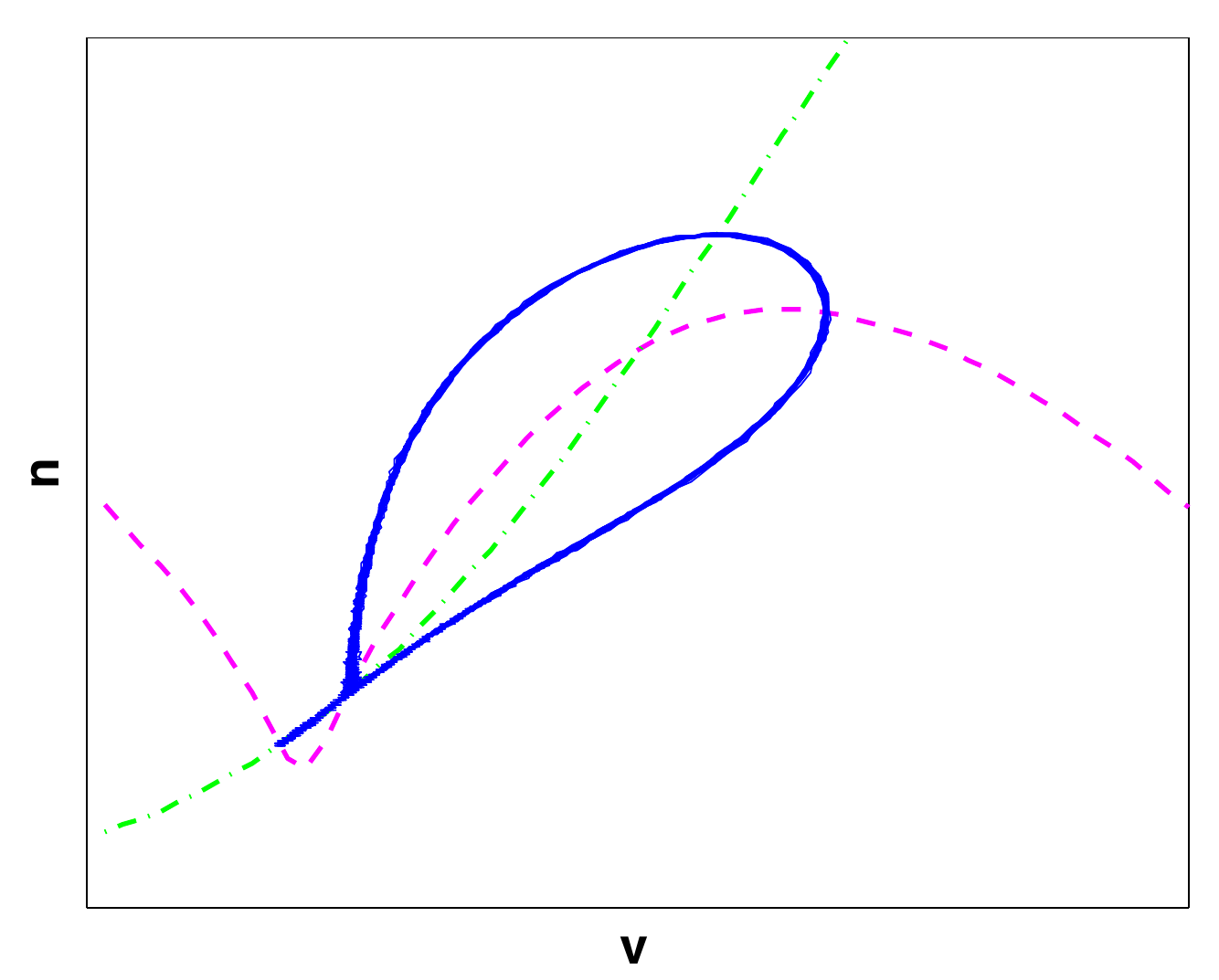, height=2.0in, width=2.5in}\qquad
{\bf d}\epsfig{figure=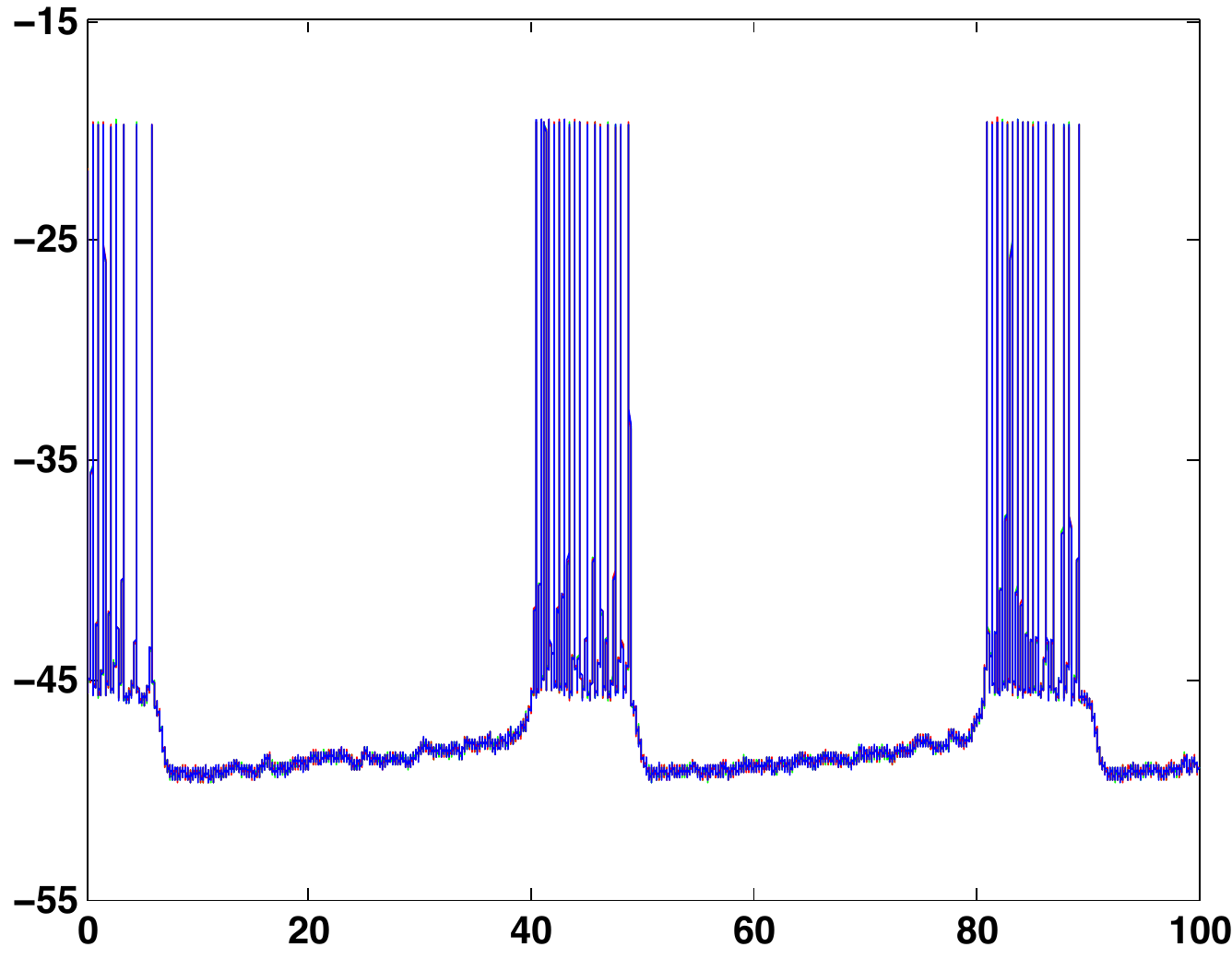, height=2.0in, width=2.5in}\\
{\bf e}\epsfig{figure=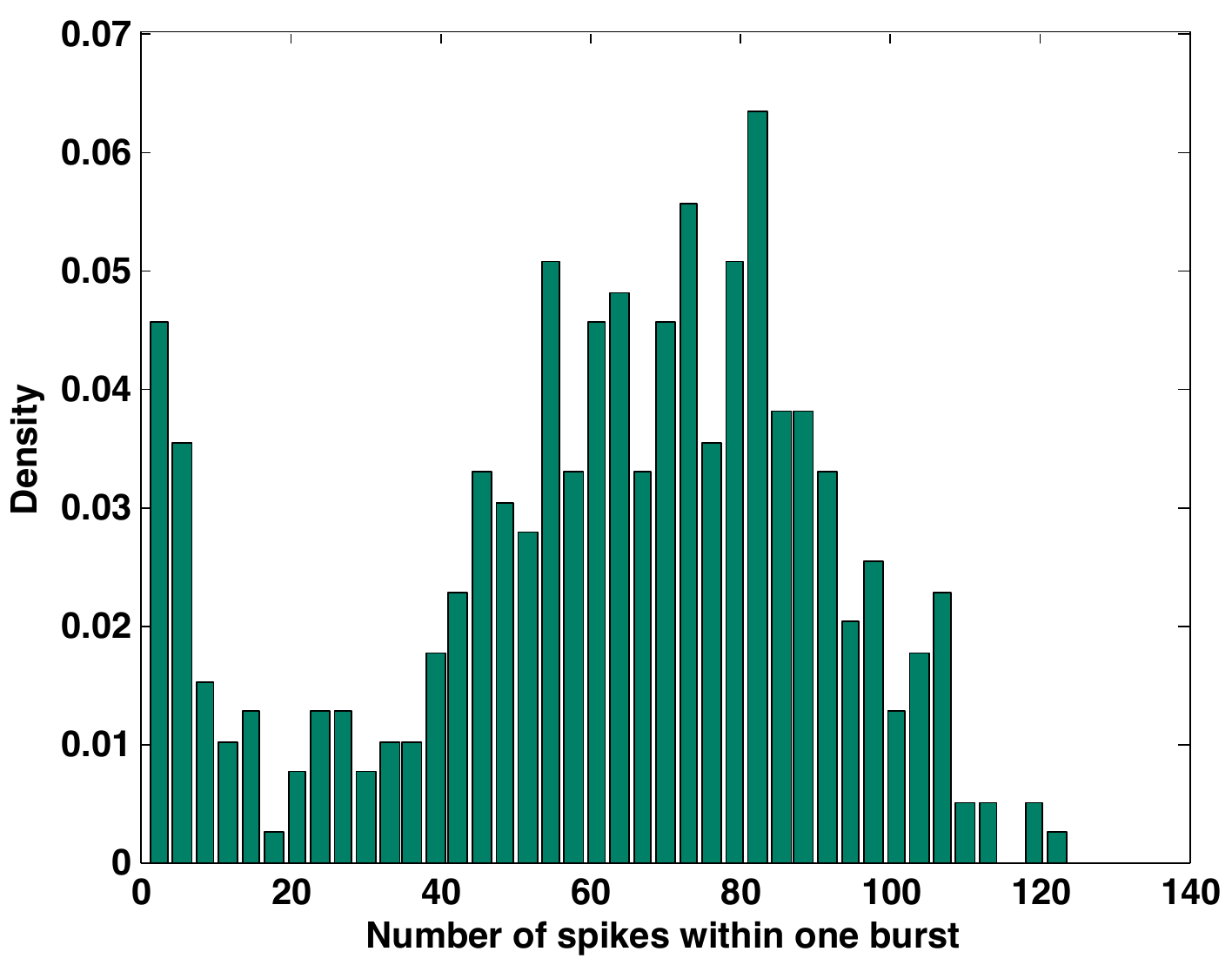, height=2.0in,width=2.5in}
\qquad
{\bf f}\epsfig{figure=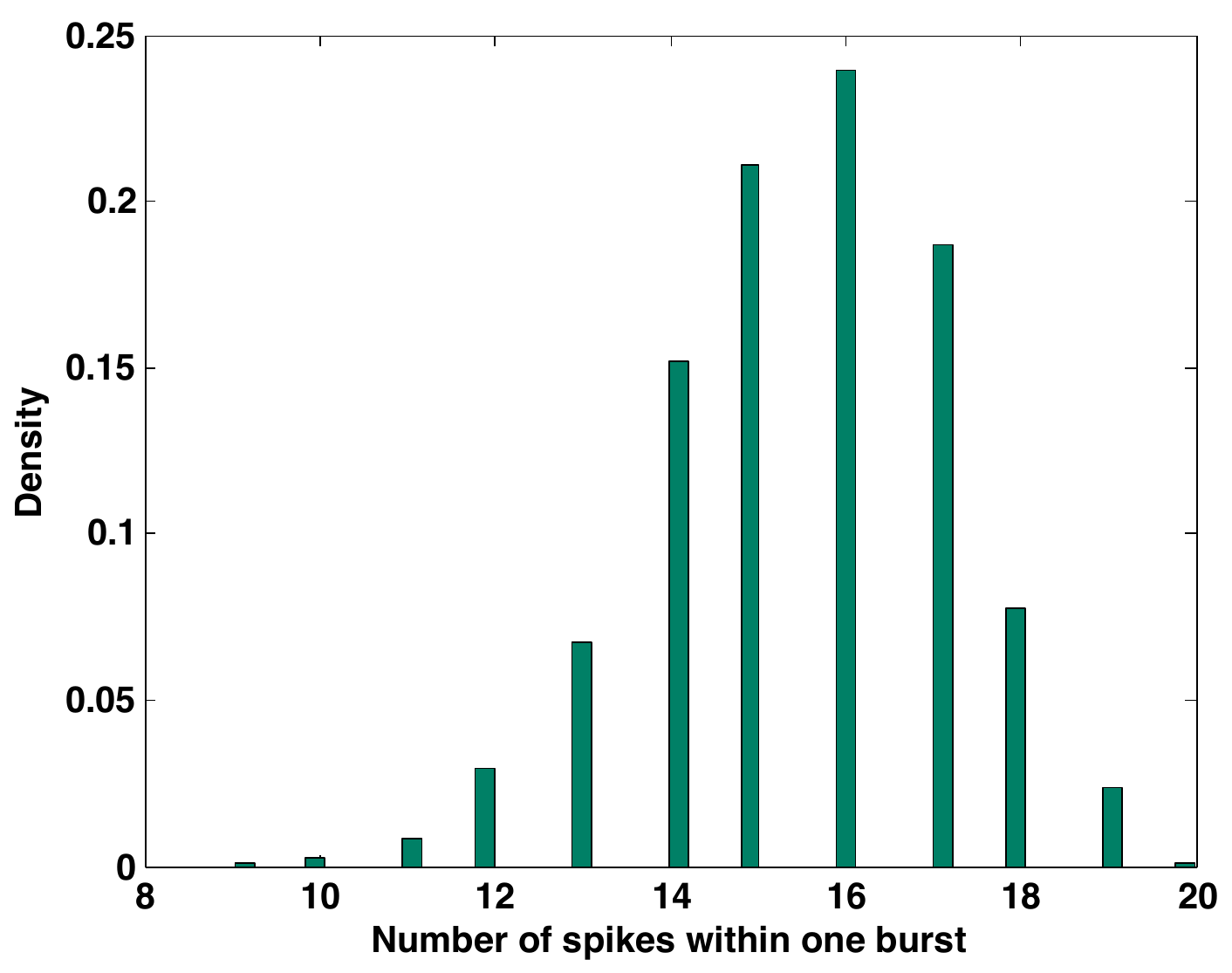, height=2.0in, width=2.5in}
\end{center}
\caption{Network connectivity and denosing. The trajectories of the
coupled systems of $200$ bursting cell models on a symmetric (a) and
random (b) degree$-4$ graphs respectively (see Example~\ref{ex.4}).
The trajectories in (a) fill a larger area as in (b), which suggests
that denoising is more effective in a network on a random graph.
The corresponding timeseries are shown in (c) and (d); and the histograms
for the number of spikes in one burst - in (e) and (f). The latter 
shows that the number of spikes in one burst generated by the random network
has approximately Gaussian distribution tightly localized around $15$. This is in a stark
contrast with a much broader distribution corresponding to the symmetric
network in (e). 
}
\lbl{f.topology}
\end{figure}

\section{Discussion}\lbl{discuss}
\setcounter{equation}{0}

In this work, we have analyzed synchronous regimes in electrically coupled
networks of bursting capable excitable cells. The individual cells in the network
can be tuned to one of the three main activity states: periodic spiking, bursting,
or quiescence. Strong electrical coupling synchronizes activity across the network.
However, in the presence of noise, the synchronous patterns generated by the network
can differ qualitatively from the patterns that the cells comprising the network exhibit 
in isolation. We have identified two scenarios of such behavior:
Scenario A and Scenario B. In the former, the network of
irregularly spiking cells generates very regular bursting provided the coupling is
sufficiently strong. In the latter, the network formed from irregularly bursting 
cells is effectively switched to spiking once the coupling strength and the size
of the network become sufficiently large.

In constructing these scenarios featuring the disparity of the firing patterns 
generated by the single cell and coupled models, we used the large deviations
type mechanisms to generate irregular patterns in the single cell models
and denoising for shaping the patterns of collective behavior in the network.
In particular, for irregular
spiking in Scenario A, we used noise to perturb the system from slowly evolving
stable equilibrium. A closely related mechanism was studied in the context
of emerging regular dynamics in randomly perturbed slow-fast systems \cite{FR01}
and self-induced stochastic resonance \cite{MVE, DMV}. 
The irregular bursting in Scenario B was organized by perturbing a system with 
a stable limit cycle via the mechanism proposed in \cite{HM}. In both cases, 
the slow-fast structure of the single cell models was essential. The emerging firing
patterns in both scenarios are very sensitive to the variations in the intensity of noise. 
Utilizing the ability of electrical coupling  to synchronize the activity and
to reduce the effects of noise on network dynamics \cite{medvedev09, medvedev10a, medvedev10}, 
for the coupled system we achieved effective
control of the firing patterns by varying the strength of coupling and the size 
of the network.

In this paper, we analyzed in detail the mechanism of denoising for coupled systems
with (slowly evolving) stable equilibria (Scenario A). We expect that the analysis
of Scenario B can be done along the same lines by combining the techniques 
of Section~\ref{denoising} and local analysis near the synchronous limit cycle
of the coupled system as in \cite{medvedev10a, medvedev10}. We will address this
problem in the future work. For Scenario A, we considered two cases: one -
when the individual
subsystems are close the saddle-node bifurcation as in the network of model 
bursting neurons (\ref{c.1}) and (\ref{c.2}), our motivating example, 
and another 
without assuming the proximity to the saddle-node bifurcation.  For systems near the bifurcation, 
we adapted 
two complementary
approaches from \cite{MZ}. The former of which utilizes the gradient structure 
of the reduced system to yield accurate estimates of  the first exit times
from the basins of the stable equilibria of the single cell and coupled models;
while the latter relies on the algebraic graph theory techniques \cite{medvedev10b}, 
which  elucidate the contribution of the network topology to synchronization and denoising. 
Furthermore, we extended the analysis of denoising 
in systems of coupled integrate and fire models in \cite{medvedev09} to conductance
based models with multidimensional phase spaces. We have shown that in this setting
the coupling architecture is very important for implementing denoising. In particular, 
for the common in applications
partial coupling case, we have shown that denosing generically does not take place,
in contrast to what one might expect from the analysis of the coupled one-dimensional 
systems \cite{medvedev09}. This result highlights
the significance of the bifurcation structure of the biophysical models of square wave
bursting neurons for the realization of denoising. The proximity to the saddle node
bifurcation makes the dynamics of the individual subsystems effectively one-dimensional
and the coupling effectively - full rank, thus circumventing the problems for implementing
denosing in partially coupled systems identified in Section~\ref{denoising}.
Therefore, the results of this study extend and complement the existing results
characterizing denoising in electrically coupled networks 
\cite{medvedev09, TSP10} and highlight the importance of the bifurcation
structure of the local dynamical systems comprising the network for implementing
denoising.

In analyzing the coupled system, we paid special attention to the role of the
network topology in shaping the network dynamics. We have found that
two spectral functions: the algebraic connectivity and the total effective resistance,
feature prominently in the quantitative descriptions of synchronization and denoising
in electrically coupled networks. The algebraic connectivity of the weighted graph
of the network sets the rate of convergence to the synchronization subspace 
(cf.~(\ref{rate-sync})),
while the total effective resistance is involved in the estimates of stochastic
stability of the synchronous regime (cf.~(\ref{lim-var})).
These analytical estimates allow one to use many known results relating 
the spectra of the graphs and their structural properties (see \cite{Chung97, Hoory06} and references
therein) to elucidate the contribution of the network structure to its dynamics.
As an example of such application, we used the spectral properties of random 
graphs \cite{Hoory06} to show that networks on random graphs feature
fast synchronization and robustness of synchrony to noise. They are also more effective
for implementing denoising than their symmetric counterparts (see Example~\ref{ex.4}),
as shown in numerical experiments in Fig.~\ref{f.topology}.

Understanding dynamical mechanisms for different patterns of electrical activity
in excitable cells and transitions between them has been long recognized as a 
fundamental problem in mathematical neuroscience. The results of this paper describe
the transition from irregular spiking to nearly periodic bursting in networks of
electrically coupled cells in the presence of noise. 
This transition was used in \cite{SRK88, SR91}
to explain why pancreatic $\beta-$cells burst in electrically coupled 
islets of Langerhans but not in isolation. Our results support and
extend the previous analysis in \cite{SRK88, SR91}. We believe that the 
results and techniques of this work will be useful for understanding dynamics in 
many other biophysical models of gap-junctionally coupled networks.  

\noindent {\bf Acknowledgments.} This work was partly supported by 
the NSF Award DMS 1109367 (to GM).

\renewcommand{\theequation}{A.\arabic{equation}}
\section*{Appendix A. The parameter values used in the 
conductance-based model of the beta cell } 
\setcounter{equation}{0}
\label{sec:A}
In the numerical experiments for this paper, we use a conductance
based model of a pancreatic $\beta-$cell due to Chay \cite{chay85}.
Below, we collect the expressions of the nonlinear functions and the 
parameter values used in (\ref{ch.1})-(\ref{ch.3}).

The first term on the right hand side of (\ref{ch.1}) models 
the combined effect of sodium and calcium currents, $I_{Na+Ca}$,
the calcium-dependent potassium current, $I_{KCa}$, delayed
rectifier $I_K$ and a small leak current, $I_l$
\be\lbl{apA.1} 
I_{ion}= I_{Na+Ca}+I_{KCa}+I_K+I_l,
\ee
where
\begin{eqnarray*}
I_{Na+Ca} &=& g_{Na+Ca}m_{\infty}(v)^3h_{\infty}(v)(E_I-v),\\
I_{KC} &=& {g_{KCa}u \over 1+u}(E_K-v),\\
I_K &=& g_Kn^4(E_K-v),\\
I_l &=& g_l(E_l-v).
\end{eqnarray*}
The steady state functions used to model the ionic currents
above are given by   
$$
f_\infty(v)=\frac{\alpha_f (v)}{\alpha_f(v)+\beta_f(v)},\quad f\in \{m,h,n\},
$$
where
\begin{eqnarray*}
\alpha_m = 0.1(v+25)(1-\exp\{-0.1(v+25)\})^{-1},  \beta_m=4\exp\{-(v+50)/18\}, & 
\alpha_h = 0.07\exp\{-0.05(v+50)\},\\
\beta_h=(1+\exp\{-0.1(v+20)\})^{-1},
\alpha_n = 0.01(v+20)(1-\exp\{-0.1(v+20)\})^{-1}, & 
\beta_n=0.125\exp\{ -(v+30)/80\}.
\end{eqnarray*}
The time constant of the delayed rectifier is given by
$$
\tau_n = (230(\alpha_n+\beta_n))^{-1}.
$$
The values of the remaining parameters are summarized in the following table.
\begin{center}
\textbf{Table A.1}
\end{center}
\begin{center}
\begin{tabular}{|l|c||l|c||l|c||l|c|}
\hline
$g_{Na+Ca}$    &   1800$s^{-1}$   & $E_{Na+Ca}$    & 100$mV$ & $g_K$    &   1700$s^{-1}$    & $E_K$    & -75$mV$ \\
$k_C$    &   $\frac{\{2,12\}}{18}$$mV$  & $g_l$    &   7$s^{-1}$    & $E_l$    & -40$mV$ & $g_{KC}$    &   12$s^{-1}$  \\
$E_{Ca}$    & 100$mV$ & $\epsilon$    &     0.03$mV^{-1}s^{-1}$ & $C_m$ & $1\mu F/cm^2$ &  $\sigma$ & 
10 $mVs^{-1}$\\
\hline
\end{tabular}
\end{center}

\renewcommand{\theequation}{B.\arabic{equation}}
\section*{Appendix B. Dynamical regimes of the conductance-based
model } 
\setcounter{equation}{0}
\label{sec:B}
The geometric theory for singularly perturbed systems implies the existence 
of the exponentially stable locally invariant manifolds $E_\epsilon$ and
$L_\epsilon$, which are $O(\epsilon)$ close to 
$E\bigcap\{(x,y): y>y_{sn}+\delta\}$ and $L\bigcap\{(x,y): y<y_{hc}-\delta\}$,
respectively, for arbitrary fixed $\delta>0$ and sufficiently small $\epsilon>0$
\cite{BG, jones}. Manifolds $E_\epsilon$ and $L_\epsilon$ are called 
{\it slow manifolds}. For small $\epsilon>0$, the dynamics of 
(\ref{2.1}) and (\ref{2.2}) on the slow manifolds is approximated by
\begin{eqnarray}\lbl{2.6}
L_\epsilon:&\qquad\qquad\qquad\dot y =\epsilon G(y),& y<y_{hc}-\delta,\\
\lbl{2.7}
E_\epsilon:&\qquad\qquad\qquad\dot y =\epsilon g(\psi(y),y),& y>y_{sn}+\delta,
\end{eqnarray}
where
\be\lbl{2.8}
G(y)={1\over T(y)} \int_0^{ T(y)} g\left(\phi(s,y),y\right) ds
\ee
and $T(y)$ stands for the period of the limit cycle $L_y$.
Depending on the location of the null surface $S=\{(x,y)\in\R^3:~ g(x,y)=0\},$
the slow-fast system (\ref{2.1}) and (\ref{2.2}) can be in one of the 
following states: {\em bursting}, {\em spiking}, and {\em quiescent} 
(see Fig.~\ref{f.regimes}). 

The following conditions on the slow subsystem yield bursting.
\noindent
For some $c>0$ independent of $\epsilon$,
\begin{description}
\item[(SE)]
\be\lbl{2.9}
  g(\psi(y),y)<-c \quad \mbox{for}\quad y>y_{sn},
\ee
\item[(SB)]
\be\lbl{2.9a}
G(y)>c \quad \mbox{for}\quad y<y_{hc}.
\ee
\end{description}
Under these assumptions, for sufficiently small $\epsilon>0$ a typical trajectory of 
(\ref{2.1}) and (\ref{2.2}) consists of the alternating segments
closely following $L_\epsilon$ and $E_\epsilon$ and fast transitions between them
(see Fig. \ref{f.regimes}a). The corresponding timeseries is shown in Fig.~\ref{f.regimes}d. 

Substituting (SB) with the condition that follows will switch (\ref{2.1}) and (\ref{2.2}$)_0$ 
to a spiking regime.
\begin{description}
\item[(SS)] $G(y)$ has a unique simple zero at $y=y_c\in (y_{sn},y_{hc})$:
\be\lbl{2.10}
G(y_c)=0 \quad\mbox{and}\quad G^\prime(y_c)<0.
\ee
\end{description}
In this case, the Pontryagin-Rodygin theorem \cite{MKKR} yields the existence of an
exponentially stable limit cycle $L_\epsilon(y_c)$ of period 
$ T(y_c)+O(\epsilon)$ lying in an $O(\epsilon)$ neighborhood of
$L_{y_c}$, provided (SS) holds and $\epsilon>0$ is sufficiently small (see Fig.~\ref{f.regimes}b,e).

Finally, the slow-fast system (\ref{2.1}$)_0$ and (\ref{2.2}) is said to be 
in the excitable regime if it has a stable fixed point lying on $E_\epsilon$ (see Fig.~\ref{f.regimes}c):
\begin{description}
\item[(Q)] $g(y):=g(\psi(y), y)$ has a unique simple zero at $y=y_q \in (y_{sn}, y_{hc})$:
\be\lbl{2.11}
g(y_q)=0 \quad\mbox{and}\quad g^\prime(y_q)<0.
\ee
\end{description}

\vfill
\break


\begin{thebibliography}{99}
\bibitem{BG}
N. Berglund and B. Gentz, {\it Noise-Induced Phenomena 
in Slow-Fast Dynamical Systems: A Sample-Paths Approach},
Springer, 2006.

\bibitem{Biggs}
N.~Biggs, {\it Algebraic Graph Theory}, second ed.,
Cambridge University Press, 1993.


\bibitem{Bollobas98}
Bela Bollobas,
{\it Modern graph theory}, Graduate Texts in Mathematics, vol. 184,
Springer, New York, 1998.


\bibitem{BRS99}
R. J. Butera, J. Rinzel, and J. C. Smith, 
Models of respiratory rhythm generation in the pre-Botzinger complex: 
I. Bursting pacemaker neurons, {\it Journal of Neurophysiology},
$\bf 82$, 382-397, 1999. 

\bibitem{chay85}
T.R. Chay,
Chaos in a three-variable model of an excitable cell,
{\it Physica D}, $\bf 16$, 233-242, 1985.

\bibitem{CK}
T.R. Chay and J. Keizer,
Minimal model for membrane oscillations in the pancreatic $\beta-$cell,
{\it Biophys. J.}, $\bf 42$, 181--190, 1983.

\bibitem{CK00}
C.C.~Chow and N.~Kopell, Dynamics of spiking neurons with electrical coupling, 
{\it Neural Comp.}, 12: 1643--1679 (2000).

\bibitem{CH82}
S.-N.~Chow and J.K.~Hale, 
{\it Methods Of Bifurcation Theory}, Springer-Verlag New York Inc,
New York, 1982.

\bibitem{Chung97}
F.R.K.~Chung,
{\it Spectral Graph Theory},
CBMS Regional Conference Series in Mathematics, No. 92,
1997. 




\bibitem{CCI} 
J.J.~Collins, C.C.~Chow, and T.T.~Imhoff,
Aperiodic stochastic resonance in excitable systems, {\it Phys. Rev. E},
$\bf 52$$(4)$, R3321--R3324, 1995.

\bibitem{CL04}
B.W.~Connors and M.A.~Long, 
Electrical synapses in the mammalian brain,
{\it Annual. Rev. Neurosci.}, 27:393--418, 2004.

\bibitem{COO08}
S. Coombes, Neuronal networks with gap junctions: 
A study of piece-wise linear planar
neuron models. {\it SIAM J. Appl. Dyn. Syst.}, vol. 7, 1101-1129, 2008.

\bibitem{Day83}
Martin V.~Day, On the exponential exit law in the small parameter
exit problem, {\it Stochastics,} $\bf 8$, 297--323, 1983.   


\bibitem{DMV}
R. E. Lee DeVille, Cyrill Muratov, and Eric Vanden-Eijnden, 
Two distinct mechanisms of coherence in randomly perturbed dynamical systems,
{\it  Physical Review E}, 72, 031105 (2005).

\bibitem{DVS00}
G.~De Vries and A.~Sherman, 
Channel sharing in pancreatic $\beta$-cells revisited: enhancement 
of emergent bursting by noise,
{\it J. Theor. Biol.}, $\bf 207$, 513-530, 2000.

\bibitem{DVZS98}  
G.~De Vries G, H.R.~Zhu, and A.~Sherman,
Diffusively coupled bursters: Effects of heterogeneity.
Bull. Math. Biol. 60: 1167-1200, 1998.





\bibitem{FR01}
M.I.~Freidlin, On stable oscillations and equilibriums induced by small noise,
{\it J. of Stat. Phys.}, $\bf 103$ (1-2), 283--300, 2001.


\bibitem{FW}
M.I.~Freidlin and A.D.~Wentzell, 
{\it Random perturbations of dynamical systems}, 
2nd ed., Springer, New York, 1998.

\bibitem{Fri08}
J.~Friedman,
{\it A Proof of Alon's Second Eigenvalue Conjecture and Related Problems},
Memoirs of the American Mathematical Society, vol. 195, 2008.

\bibitem{Fiedler73}
M.~Fiedler, Algebraic connectivity of graphs. 
{\it Czech. Math. J.} 23(98), 1973.

\bibitem{IZH00}
E.M. Izhikevich,
Neural excitability, spiking, and bursting,
{\it Int. J. of Bifurcation and Chaos},
$\bf 10$, 1171--1266, 2000.

\bibitem{jones}
C.K.R.T. Jones, Geometric singular perturbation theory, 
Lecture Notes in Mathematics, Vol. 1609, Springer, Berlin, pp. 44-118, 1995.

\bibitem{Jost07}
J.~Jost, Dynamical networks,
in  J.~Feng, J.~Jost, and M.~Qian, (Eds.)
{\it Networks: From Biology to Theory}, Springer, 2007 


\bibitem{gelfand}
I.M.~Gelfand, {\it Lectures on Linear Algebra},
Interscience Publishers, 1961.

\bibitem{Boyd08}
A.~Ghosh, S.~Boyd, and A.~Saberi,
Minimizing effective resistance of a graph,
{\it SIAM Rev.}, $\bf 50$(1), 37--66, 2008.

\bibitem{GP06}
D.S.~Goldobin and A.~Pikovsky,
Antireliability of noise-driven neurons, {\it Phys. Rev. E} $\bf 73$, 
061906 2006.

\bibitem{gra4}
A.A. Grace and B.S. Bunney,
The control of firing pattern in nigral dopamine neuron: single spike firing,
{\it J. Neurosci.}, $\bf 4$, 2866--2876, 1984.

\bibitem{GH}
J.~Guckenheimer and P.~Holmes,
{\it Nonlinear Oscillations,
Dynamical Systems, and Bifurcations of Vector Fields}, Springer, 1983.

\bibitem{GH07}
Juan Gao and  Philip Holmes,
On the dynamics of electrically-coupled neurons with inhibitory
synapses, {\it J. Comput. Neurosci.}, 22:39–61, 2007.





\bibitem{Hasminsky}
R.Z.~Has'minskii,
{\it Stochastic stability of differential equations,}
Sijthoff \& Noordhoff, Rockville, MD, 1980.

\bibitem{HM}
P.~Hitczenko and G.S.~Medvedev, 
Bursting oscillations induced by small noise, {\it SIAM J. Appl. Math.},
$\bf 69$(5): 1359-1392, 2009.


\bibitem{HI97}
F. C. Hoppensteadt and Eugene M. Izhikevich,
Weakly connected neural networks, Springer 1997.


\bibitem{Hoory06}
S. Hoory, N. Linial, and A. Wigderson, 
Expander graphs and their applications,
{\it Bulletin of the American Mathematical Society}, 
vol. 43, no. 4, pp 439–561, 2006.

\bibitem{KS}
I.~Karatzas and S.E.~Shreve, {\it Brownian Motion and Stochastic Calculus},
2nd ed., Springer, New York, 1991.

\bibitem{KE88}
N.~Kopell and G.B.~Ermentrout, Coupled oscillators and the design of 
central pattern generators, {\it Math. Biosci.} 90 (1-2), 87--109, 1988. 

\bibitem{KB09}
R. Kuske and P. Borowski, Survival of subthreshold oscillations: the interplay of noise, 
bifurcation structure, and return mechanism, {\it Discrete and Continuous Dynamical
Systems Ser. S}, 2(4):873--895, 2009.

\bibitem{Kuz98}
Yuri A. Kuznetsov,
{\it Elements of applied bifurcation theory}, Springer, 1998. 

\bibitem{LR03}
T.~Lewis and J.~Rinzel, 
Dynamics of spiking neurons connected by both inhibitory
and electrical coupling, {\it J. Comp. Neurosci.},
14:283--309, 2003.

\bibitem{LR10}
Sukbin Lim and John Rinzel, Noise-induced transitions in slow wave neuronal 
dynamics, {\it Journal of Computational Neuroscience}, 28(1): 1--17, 2010.   

\bibitem{klein93}
D.~Klein and M.~Randic, Resistance distance,
{\it J. Math. Chem.}, $\bf 12$, 81--95, 1993.

\bibitem{LT}
E.~Lee and D.~Terman, 
Uniqueness and stability of periodic bursting solutions, {\it Journal of 
Differential  Equations},
$\bf 158$, 48--78, 1999.

\bibitem{Lon97}
A.~Longtin,
Autonomous stochastic resonance in bursting neurons,
{\it Phys. Rev. E} $\bf 55$, 868 - 876, 1997.

\bibitem{LPS88}
A.~Lubotzky, R.~Phillips, and P.~Sarnak,
Ramanujan graphs, {\it Combinatorica}, $\bf 8$,
161--278, 1988.


\bibitem{Margulis88}
G.~Margulis,
Explicit group-theoretic constructions of combinatorial schemes 
and their applications in the construction of expanders and 
concentrators. (Russian) Problemy Peredachi Informatsii 24 (1988), 
no. 1, 51--60; (English translation in Problems Inform. Transmission 
24 (1988), no. 1, 39--46).

\bibitem{MS95}
Mainen Z.F. and Sejnowski T.J., 
Reliability of spike timing in neocortical neurons,
{\it Science}, 268, 1503-1506.

\bibitem{MMR10}
E. Manica, G.S. Medvedev, and J.E. Rubin, 
First return maps for the dynamics of synaptically coupled conditional 
bursters, {\it Biological Cybernetics}, 103:87-104, 2010. 

\bibitem{medvedev10b} 
G.S. Medvedev, Stochastic stability of continuous 
time consensus protocols,  submitted, arXiv preprint: 1007.1234. 

\bibitem{medvedev10a}
G.S.~Medvedev, Synchronization of coupled limit cycles, 
{\it J. Nonlin. Sci.},  $\bf 21$, 3, 441--464, 2011.

\bibitem{medvedev10}
G.S.~Medvedev, Synchronization of coupled stochastic limit cycle oscillators, 
{\it Physics Letters A} (374), 1712--1720, 2010.

\bibitem{medvedev09}
G.S.~Medvedev, Electrical coupling promotes fidelity of responses in the 
networks of model neurons, {\it Neural Computation},  $\bf 21$ (11), 
3057--3078, 2009.




\bibitem{M05}
G.S.~Medvedev, Reduction of a model of an excitable cell to a 
one-dimensional map, {\it Physica D}, 202(1-2), 37-59, 2005. 

\bibitem{M06} G.S.~Medvedev, Transition to bursting via deterministic chaos, 
{\it Phys. Rev. Lett.} 97, 048102, 2006. 

\bibitem{MC}
G.S.~Medvedev and J.E.~Cisternas, 
Multimodal regimes in a compartmental model of the dopamine neuron, 
{\it Physica D}, 194(3-4), 333-356, 2004.

\bibitem{MK}
G.S. Medvedev and N. Kopell, Synchronization and transient dynamics in the chains of electrically 
coupled FitzHugh-Nagumo oscillators, SIAM J. Appl. Math., 
vol. 61, No. 5, pp. 1762-1801.

\bibitem{MZ}
G.S. Medvedev and S. Zhuravytska, 
The geometry of spontaneous spiking in neuronal networks, submitted, 2011;
arXiv preprint: 1105.2801. 

\bibitem{MKKR}
E.F. Mishchenko, Yu.S. Kolesov, A.Yu. Kolesov, and N.Kh. Rozov,
{\it Asymptotic Methods in Singularly perturbed Systems}, 
Consultants Bureau, New York, 1994.

\bibitem{MVE}
C.B. Muratov, E. Vanden Eijnden, and W.~E, 
Self-induced stochastic resonance in excitable systems,
{\it Physica D} $\bf 210$, 227-240, 2005. 

\bibitem{RIN87}
J. Rinzel,
A formal classification of bursting mechanisms in excitable systems,
in A.M. Gleason, ed., Proceedings of the International Congress
of Mathematicians, AMS, pp. 135--169, 1987.

\bibitem{RE89}
J. Rinzel and G.B. Ermentrout,
Analysis of neural excitability and oscillations,
in C. Koch and I. Segev, eds
{\it Methods in Neuronal Modeling}, MIT Press,
Cambridge, MA, 1989.

\bibitem{RT00}
Rubin J, Terman D (2000) 
Geometric analysis of population rhythms in synaptically coupled 
neuronal networks. Neural Comp. 12: 597--645.

\bibitem{PS07}
M.G.~Pedersen and M.P.~Sorensen, 
The effect of noise on $\beta$-cell burst period,
{\it SIAM J. Appl. Math.}, $\bf 67$, 530--542  2007.

\bibitem{PMG}
B.~Pfeuty, G.~Mato, D.~Golomb, D.~Hansel, Electrical synapses and synchrony: 
the role of intrinsic currents, {\it  J. Neurosci.,} 23:6280--6294, 2003.

\bibitem{Sar04}
P. Sarnak, What is an expander?, 
{\it Notices of the American Mathematical Society}, 51, 762--763, 2004. 

\bibitem{SR91}
A.~Sherman and J.~Rinzel, 
Model for synchronization of pancreatic $\beta$-cells by gap junction coupling,
{\it Biophysical. J.}, $\bf 59$, 547--559, 1991.

\bibitem{SRK88}
A.~Sherman, J.~Rinzel, and J.~Keizer,
Emergence of organized bursting in clusters of pancreatic $\beta$-cells 
by channel sharing, 
{\it Biophysical. J.},
$\bf 54$, 411-425, 1988.

\bibitem{smith}
G.D.~Smith, Modeling of the stochastic gating of ion channels,
in C.P.~Fall et al., eds, {\it Computational cell biology},
Springer, New York, 2002. 

\bibitem{SRT}
Jianzhong Su, Jonathan Rubin, and David Terman, 
Effects of noise on elliptic bursters, {\it Nonlinearity}, 
17: 133--157, 2004.

\bibitem{TSP10}
Tabareau, N., Slotine, J.J.E., and Pham, Q.C., 
How synchronization protects from noise,
{\it PLoS Computational Biology},  6(1), 2010.

\bibitem{TR92}
D. Terman, 
The transition from bursting to continuous spiking in excitable 
membrane models,
{\it J. Nonl.Sci.}, $\bf 2$, 135--182, 1992.



\bibitem{WR95}
Wang XJ, Rinzel J (1995) 
Oscillatory and bursting properties of neurons. In:
MA Arbib, ed.
Handbook of Brain Theory and Neural Networks.
MIT Press, Cambridge, MA. pp. 686--691.

\bibitem{wrk}
J. White, J. Rubenstein, and A. Kay, Channel noise in neurons, {\it Trends in Neurosci.},
$\bf 23$(3), 131--137, 2000. 

\bibitem{UCS99}
M. Usher, J.D. Cohen, D. Servan-Schreiber, J. Rajkowski and G. Aston-Jones , 
The role of locus coeruleus in the regulation of cognitive performance,
{\it Science}, $\bf 283$ (1999), pp. 549--554.

\bibitem{Gutman03}
W.~Xiao and I.~Gutman, Resistance distance and Laplacian
spectrum, {\it Theor. Chem. Acc.}, $110:$ 284--289, 2003.

\end{thebibliography}
\end{document}